\documentclass[11pt,a4paper]{article}
\pdfoutput=1

\usepackage{a4wide}
\usepackage{graphicx} 
\usepackage{amsmath}
\usepackage{amssymb}
\usepackage{amsfonts}
\usepackage{cite}
\usepackage{color}
\usepackage{adjustbox}
\usepackage{pdflscape}
\usepackage{gensymb}
\usepackage{graphicx}
\usepackage{diagbox}
\usepackage{pifont}
\usepackage{bigstrut}
\usepackage[small,bf]{caption}
\usepackage{enumerate}
\usepackage{tabulary}
\usepackage{hyperref}
\usepackage{makecell}
\usepackage{multirow}
\usepackage{dcolumn}
\usepackage{textcomp}

\usepackage{ae} 
\usepackage{slashed}
\usepackage{aecompl}
\usepackage{bm}

\usepackage{epsfig}
\usepackage{fontenc}
\usepackage{graphicx}
\usepackage[lofdepth,lotdepth,caption=false]{subfig}
\usepackage{color}
\usepackage{booktabs}
\usepackage{tabularx}
\usepackage{multirow}
\usepackage[table]{xcolor}
\usepackage[normalem]{ulem}
\usepackage{dcolumn}
\usepackage{rotating}
\usepackage{hyperref}
\usepackage{multirow,bigdelim}  
\usepackage{lineno}
\usepackage[latin1]{inputenc}

\newcommand{\be}{\begin{equation}}
\newcommand{\ee}{\end{equation}}
\newcommand{\ba}{\begin{eqnarray}}
\newcommand{\ea}{\end{eqnarray}}
\newcommand{\capdef}{}
\newcommand{\mycaption}[2][\capdef]{\renewcommand{\capdef}{#2}
       \caption[#1]{{\footnotesize #2}}}
\makeatletter
\newcommand{\bi}{\begin{itemize}}
\newcommand{\ei}{\end{itemize}}

\newcommand{\ie}{{\it i.e.}}

\def\epsilon{\varepsilon}

\def\<{\langle}
\def\>{\rangle}

\def\lsim{\mathrel{\rlap{\lower4pt\hbox{\hskip1pt$\sim$}}
    \raise1pt\hbox{$<$}}}         
\def\gsim{\mathrel{\rlap{\lower4pt\hbox{\hskip1pt$\sim$}}
    \raise1pt\hbox{$>$}}}         

\begin{document}

\begin{titlepage}

\vspace*{-15mm}

\begin{flushright}
IP/BBSR/2020-3, TIFR/TH/20-19
\end{flushright}

\vspace*{0.8cm}

\begin{center}

{\bf\Large{From oscillation dip to oscillation valley 
in atmospheric neutrino experiments}} \\ [10mm]

{\bf Anil Kumar}$^{\, a,b,c,}$\footnote{E-mail: \texttt{anil.k@iopb.res.in (ORCID: 0000-0002-8367-8401)}}, 
{\bf Amina Khatun}$^{\, a,d,}$\footnote{E-mail: \texttt{amina.khatun@fmph.uniba.sk (ORCID: 0000-0003-3493-607X)}}, \\
{\bf Sanjib Kumar Agarwalla}$^{\, a,b,e,}$\footnote{E-mail: \texttt{sanjib@iopb.res.in (ORCID: 0000-0002-9714-8866)}}, 
{\bf Amol Dighe}$^{\, f,}$\footnote{E-mail: \texttt{amol@theory.tifr.res.in (ORCID: 0000-0001-6639-0951)}} \\
\vspace{8mm}
$^{a}$\,{\it Institute of Physics, Sachivalaya Marg, Sainik School Post, 
Bhubaneswar 751005, India} \\
\vspace{2mm}
$^{b}$\,{\it Homi Bhabha National Institute, Training School Complex, \\ 
Anushakti Nagar, Mumbai 400085, India} \\
\vspace{2mm}
$^{c}$\,{\it Applied Nuclear Physics Division, Saha Institute of Nuclear Physics, \\ Block AF, Sector 1, Bidhannagar, Kolkata 700064, India}\\
\vspace{2mm}
$^{d}$\,{\it Comenius University, Mlynsk\'{a} dolina F1, SK842 48 Bratislava, Slovakia} \\
\vspace{2mm}
$^{e}$\,{\it International Centre for Theoretical Physics, 
Strada Costiera 11, 34151 Trieste, Italy} \\
\vspace{2mm}
$^{f}$\,{\it Tata Institute of Fundamental Research, Homi Bhabha Road,  \\ Colaba, Mumbai 400005, India}

\end{center}

\vspace{8mm}

\begin{abstract}
\vspace{4mm}
\noindent 
Atmospheric neutrino experiments can show the ``oscillation dip'' feature
in data, due to their sensitivity over a large $L/E$ range. In experiments
that can distinguish between neutrinos and antineutrinos, like INO,
oscillation dips can be observed in both these channels separately.
We present the dip-identification algorithm employing a data-driven 
approach -- one that uses the asymmetry in the upward-going and downward-going events, binned in the reconstructed $L/E$ of muons -- to demonstrate the dip, 
which would confirm the oscillation hypothesis. We further propose, 
for the first time, the identification of an ``oscillation valley'' in the reconstructed 
($E_\mu$,$\,\cos\theta_\mu$) plane, feasible for detectors like ICAL 
having excellent muon energy and direction resolutions. We illustrate 
how this two-dimensional valley would offer a clear visual representation 
and test of the $L/E$ dependence, the alignment of the valley quantifying 
the atmospheric mass-squared difference. Owing to the charge identification capability of the ICAL detector at INO, we always present our results using $\mu^{-}$ and $\mu^{+}$ events separately. 
Taking into account the statistical 
fluctuations and systematic errors, and varying oscillation parameters over their currently allowed ranges, we estimate the precision 
to which atmospheric neutrino oscillation parameters would be determined with the 10-year simulated data at ICAL 
using our procedure.

\end{abstract}
\end{titlepage}

\setcounter{footnote}{0}

\section{Introduction and Motivation}
\label{introduction}

We have witnessed remarkable developments in neutrino physics
over the last two decades, driven by the astonishing discovery
of ``neutrino mass-induced flavor oscillation" 
suggesting that neutrinos have non-degenerate mass and they mix 
with each other~\cite{Pontecorvo:1967fh,Gribov:1968kq}. Due to
these two unique features, neutrinos change their flavor as they
move in space and time implying that leptonic flavors are not 
symmetries of Nature~\cite{Tanabashi:2018oca}. 
Such a path-breaking discovery of fundamental significance, 
recently recognized with the Nobel Prize~\cite{Nobel:2015}, has created 
enormous interest in the global neutrino community to measure 
the fundamental oscillation parameters with much better 
precision~\cite{Neutrino:Unbound}. Such measurements 
of neutrino oscillations parameters are essential to provide a stringent 
test of the standard three-flavor neutrino oscillation framework,
which has six fundamental parameters:
(i) the solar mass-squared difference, $\Delta m^2_{21}$
($\equiv m_2^2 - m_1^2 \approx 7.4 \times 10^{-5}$ eV$^2$),
(ii) the solar mixing angle, $\theta_{12} \approx 34^\circ$,
(iii) the atmospheric mass-squared difference, $|\Delta m^2_{32}|$
($\equiv |m_3^2 - m_2^2| \approx 2.5 \times 10^{-3}$ eV$^2$),
(iv) the atmospheric mixing angle, $\theta_{23} \approx 48^\circ$,
(v) the reactor mixing angle, $\theta_{13} \approx 8.6^\circ$, and
(vi) the Dirac CP phase, $\delta_{\rm CP}$, which at present 
lies in the third quadrant with large uncertainty.

It is quite remarkable to see that starting from almost no knowledge 
of the neutrino masses or lepton mixing parameters twenty years ago, 
we have been able to construct a robust, simple, three-flavor oscillation
framework which successfully explains most of the 
data~\cite{Capozzi:2020qhw,Esteban:2018azc,NuFIT,deSalas:2018bym}. 
Atmospheric neutrino experiments have contributed significantly 
to achieve this milestone~\cite{Learned:2019vcq,Kajita:2019bzu}
by providing an avenue to study neutrino oscillations over a wide range 
of energies ($E_\nu$ in the range of $\sim$100 MeV to a few hundreds of GeV) 
and baselines ($L_\nu$ in the range of a few km to more than 12,000 km) 
in the presence of Earth's matter with a density varying in the range 
of 0 to 10 g/cm$^3$. 

An important breakthrough in the saga of atmospheric neutrinos came 
in 1998 when the pioneering Super-Kamiokande (Super-K) experiment 
reported convincing evidence for neutrino oscillations in atmospheric neutrinos 
by observing the zenith angle (this angle is zero for vertically downward-going 
events) dependence of $\mu$-like and $e$-like events~\cite{Fukuda:1998mi}.
The data accumulated by the Super-K experiment, based on an exposure 
of 33 kt$\cdot$yr, showed a clear deficit of upward-going events in the 
zenith angle distributions of $\mu$-like events with a statistical significance 
of more than $6\sigma$, while the zenith angle distribution of $e$-like events 
did not show any significant up--down asymmetry. These crucial observations 
by the Super-K experiment were successfully interpreted in a two-flavor 
scenario assuming oscillation between $\nu_\mu$ and $\nu_\tau$, leading to
the disappearance of $\nu_\mu$. In this scenario, the survival probability of $\nu_\mu$ can be expressed in the
following simple fashion:
\begin{equation}
P(\nu_\mu \rightarrow \nu_\mu) = 1 - \sin^2 2\theta_{23} 
\cdot \sin^2\left({1.27 \cdot |\Delta m^2_{32}| \left(\mbox{eV}^2\right) 
	\cdot {L_\nu \left(\mbox{km}\right)\over E_\nu \left(\mbox{GeV}\right)}}\right).
\label{eq:2flavsurv}
\end{equation}
Due to the hierarchies in neutrino mass pattern 
($\Delta m^2_{21} \ll |\Delta m^2_{32}|$) and in 
mixing angles ($\theta_{13} \ll \theta_{12}, \theta_{23}$), 
the above simple two-flavor $\nu_\mu$ survival probability 
expression was sufficient to explain the following broad 
features of the Super-K atmospheric data, providing a 
solution of the long-standing atmospheric neutrino anomaly
in terms of ``neutrino mass-induced flavor oscillations". 
\begin{itemize}
	
	\item
	Around 50\% of upward-going muon neutrinos disappear after 
	traveling long distances. For very long baselines, the above
	$\nu_\mu$ survival probability expression (see Eq.~\ref{eq:2flavsurv})
	approaches $1 - (1/2)\sin^22\theta_{23}$ and the observed
	survival probability becomes close to 0.5, suggesting that the 
	mixing angles is close to the maximal value of 45$^\circ$.
	
	\item
	The disappearance of $\nu_\mu$ events starts for the zenith
	angle close to the horizon indicating that the oscillation length
	should be around $\sim$ 400 km for neutrinos having energies
	close to 1 GeV. This suggests that the atmospheric mass-squared
	difference, $|\Delta m^2_{32}|$ should be around $\simeq 10^{-3}$ 
	eV$^2$.
	
	\item
	There is no visible excess or deficit of electron neutrinos.
	It suggests that the oscillations of muon neutrinos mainly
	occur due to $\nu_\mu \rightarrow \nu_\tau$ transitions.
	
	\item
	The Super-K experiment was also the first experiment to
	observe the sinusoidal $L/E$ dependence of the $\nu_\mu$
	survival probability~\cite{Ashie2004}. A special analysis was
	performed by the Super-K collaboration, where they selected
	events with good resolution in $L/E$, largely excluding low-energy 
	and near-horizon events. Using this high-resolution event sample, 
	they took the ratio of the observed and expected event rates
	and the oscillation dip started to appear on the canvas around
	$L/E$ $\sim$ 100 km/GeV, with the deepest location of the dip
	around $L/E$ $\sim$ 500 km/GeV (see Fig.~4 in Ref.~\cite{Ashie2004}). 
	This study was quite useful to disfavor the other alternative models
	which showed different $L/E$ behaviors. Note that the  
	oscillation dip was also observed in the DeepCore data where  
	the ratio of observed events to unoscillated Monte Carlo (MC) events is plotted
	against the reconstructed $L/E$ of neutrino~\cite{Aartsen:2014yll}.
	
\end{itemize}

The proposed atmospheric neutrino detector Iron Calorimeter (ICAL) at the India-based Neutrino Observatory project aims to detect atmospheric neutrinos and antineutrinos separately over a wide range of energies and baselines. The performance of ICAL is optimized for the reconstructed muon energy ($E^{\rm rec}_\mu$) range of 1 GeV to 25 GeV, and reconstructed baseline ($L^{\rm rec}_\mu$) from 15 km to 12000 km except around the horizon. Thus, the range of reconstructed $L^{\rm rec}_\mu/E^{\rm rec}_\mu$ of detected muon to which ICAL is sensitive, is from 1 km/GeV to around $10^4$ km/GeV. Therefore, we expect to see an oscillation dip (or two) as in Super-K. In this paper, we investigate the capability of the ICAL detector to reconstruct the oscillation dip in the reconstructed  $L^{\rm rec}_\mu/E^{\rm rec}_\mu$  distribution of observed muon events. Moreover, it would be possible to identify the dip in $\mu^-$ and $\mu^+$ events separately, due to the muon charge identification capability of ICAL. 

In contrast to the studies in \cite{Ashie2004,Aartsen:2014yll} where the ratio between data and unoscillated MC as a function of $L/E$ was used, in this paper we use the ratio between upward-going and downward-going events for the analysis. We exploit the fact that the downward-going atmospheric neutrinos and antineutrinos in multi-GeV energy range do not oscillate, and up-down asymmetry reflects the features mainly due to oscillations in upward-going events in these energies. The magnetic field in the ICAL detector enables us to study these distributions in neutrino and antineutrino oscillation channels separately. In this paper, we also discuss the procedure to measure the atmospheric mass-squared difference $|\Delta m^2_{32}|$ and the mixing angle $\theta_{23}$ using the information from the oscillation dip, and provide the results.  
 
In this paper, we point out for the first time that there is an ``oscillation valley'' feature in the two-dimensional plane of   reconstructed muon observables ($E_\mu^\text{rec}$, $\cos\theta_\mu^\text{rec}$) for $\mu^-$ and $\mu^+$ events separately  at ICAL. The presence of valley-like feature in the oscillograms of  survival probability of $\nu_\mu$ and $\bar\nu_\mu$ in the plane of neutrino energy and direction is well known. But, it is not a priori obvious that the reconstructed energy and direction of muon will serve as a good proxy for the neutrino energy and direction, and will be able to reproduce this valley. In this work, we show that the reconstructed muon observables at ICAL preserve these features, which is a non-trivial statement about the fidelity of the detector. 

The valley-like feature would be recognizable at ICAL due to its sensitivity to muons having energies in the wide range of 1 GeV to 25 GeV and excellent angular resolution at these energies 
(the muon energy resolution at ICAL is around 10$\%$ to 15$\%$, whereas the direction resolution is less than $1^\circ$ for multi-GeV 
muons with directions away from the horizon~\cite{Chatterjee:2014vta}). We show,  using the complete migration matrices of ICAL for muon as obtained from GEANT4 simulation~\cite{Chatterjee:2014vta} that, due to the excellent energy and direction resolution, the oscillation valley in two-dimensional ($E_\mu^\text{rec},\cos\theta_\mu^\text{rec}$) plane would appear prominently at ICAL. We also demonstrate for the first time that the mass-squared difference may be determined using the alignment of the oscillation valley. 

There are studies to measure the sensitivity of the ICAL detector for measuring the atmospheric oscillation parameters $|\Delta m^2_{32}|$ and $\theta_{23}$, see Refs.\,\cite{Thakore:2013xqa,Devi:2014yaa,Kaur:2014rxa,Mohan:2016gxm,Rebin:2018fdl,Chacko:2019wwm}. 
ICAL would be able to determine these parameters separately in the neutrino and antineutrino channels~\cite{Kaur:2017dpd,Dar:2019mnk}. The difference between these analyses and our study is that  Refs.\,\cite{Thakore:2013xqa,Devi:2014yaa,Kaur:2014rxa,Mohan:2016gxm,Rebin:2018fdl,Chacko:2019wwm,Kaur:2017dpd,Dar:2019mnk} employ the $\chi^2$ method, while the focus of our study is to identify the oscillation dip and the oscillation valley, independently in $\mu^-$ and $\mu^+$ events, and determine the oscillation parameters from them. This is possible due to the wide range of energy and baselines reconstructed at ICAL with excellent detector properties. We also include statistical fluctuations, systematic uncertainties, and errors in other oscillation parameters while determine $|\Delta m^2_{32}|$ and $\theta_{23}$  with 10-year data at ICAL.

This paper is organized in the following fashion. In Sec.~\ref{sec:prob}, we present the survival probabilities of neutrinos and antineutrinos as one-dimensional functions of $L_\nu/E_\nu$ and as two-dimensional oscillograms in $(E_\nu, \cos\theta_\nu)$ plane, to pinpoint the oscillation dips and the oscillation valley, respectively. For exploring these features in atmospheric neutrino and antineutrino data separately, we simulate $\nu_\mu$ and $\bar\nu_\mu$ interactions at ICAL as discussed in  Sec.~\ref{sec:event-gen}. Next, in Sec.~\ref{sec:LbyE}, we formulate a data-driven variable, the ratio between upward-going and downward-going events (U/D), to use it as an observable for all the analyses in this paper. We propose a novel algorithm to identify the oscillation dip and to measure its ``location'' in reconstructed $L^{\rm rec}_\mu/E^{\rm rec}_\mu$ distributions. We present the  $90\%$ C.L. range for $|\Delta m^2_{32}|$ expected with 500 kt$\cdot$yr exposure of ICAL using the calibration curve between the location of the dip and $|\Delta m^2_{32}|$.  We also estimate the allowed range for $\sin^2\theta_{23}$ at 90$\%$ C.L. using the ratio of 
upward-going and downward-going events (U/D). Sec.~\ref{sec:2D_E-CT} is devoted to discuss our analysis of U/D distributions in  reconstructed ($E^{\rm rec}_\mu, \cos\theta^{\rm rec}_{\mu}$) plane. For identifying the oscillation valley in   reconstructed ($E^{\rm rec}_\mu, \cos\theta^{\rm rec}_{\mu}$) plane of these distributions and to measure the ``alignment'' of the valley, we suggest a unique algorithm and demonstrate its use. We present the $90\%$ C.L. range of $|\Delta m^2_{32}|$ using 500 kt$\cdot$yr exposure of ICAL with the help of the calibration curve between the alignment of the valley and $|\Delta m^2_{32}|$. In Sec.~\ref{sec:conclusions}, we summarize our findings and point out the utility of our novel approach in establishing the oscillation hypothesis in atmospheric neutrino experiments. In Appendix~\ref{app:5yr}, we explore the oscillation dip and oscillation valley with a relatively smaller exposure of 250 kt$\cdot$yr and show the 90$\%$ C.L. allowed ranges in $|\Delta m^2_{32}|$ from both the analysis procedures.

\section{Neutrino Oscillation Probabilities}
\label{sec:prob}

In this section, we discuss the survival probabilities of atmospheric
$\nu_\mu$ and $\bar\nu_\mu$, reaching the detector after their propagation
through the atmosphere and possibly the Earth. The oscillation probabilities
are functions of energy ($E_\nu$) and zenith angle ($\theta_\nu$) of neutrino.  
The net distance $L_\nu$ traversed by a neutrino (its ``baseline'')
is related to its zenith angle via
\be
L_\nu = \sqrt{(R+h)^2 - (R-d)^2\sin^2\theta_\nu} \,-\, (R-d)\cos\theta_\nu \, ,
\label{eq:zen-bl-rel}
\ee
where $R$, $h$, and $d$ denote the radius of Earth, the average height from
the surface at which neutrinos are produced, and the depth of
the detector underground, respectively. In this study, we take
$R = 6371$ km, $h = 15$ km, and $d=0$ km. Since $R \gg h \gg d$, the changes
in the survival probabilities with small variations in $h$ and $d$ are
negligible.

The survival probabilities may be represented in two ways: 
(i) as one-dimensional functions of  $L_\nu/E_\nu$, and
(ii) as two-dimensional oscillograms in the plane of energy and
zenith angle. Below we discuss these two representations, and
the major physics insights obtained via them.

\subsection{$L_\nu/E_\nu$ dependence of $\nu_\mu$ and $\bar\nu_\mu$
  survival probabilities}
\label{sec:1Dprob}

\begin{figure}[htb!]
  \centering
  \includegraphics[width=0.49\textwidth]{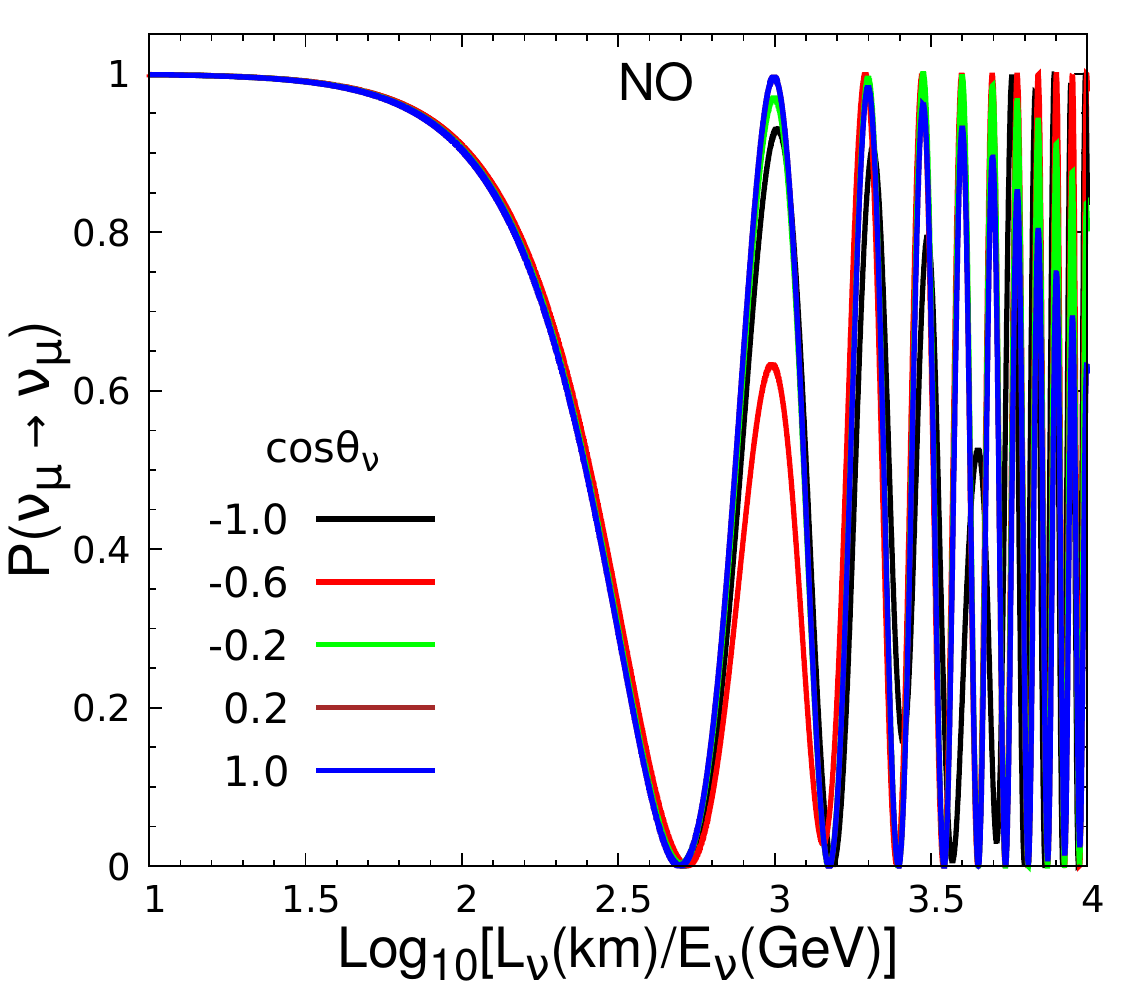}
  \includegraphics[width=0.49\textwidth]{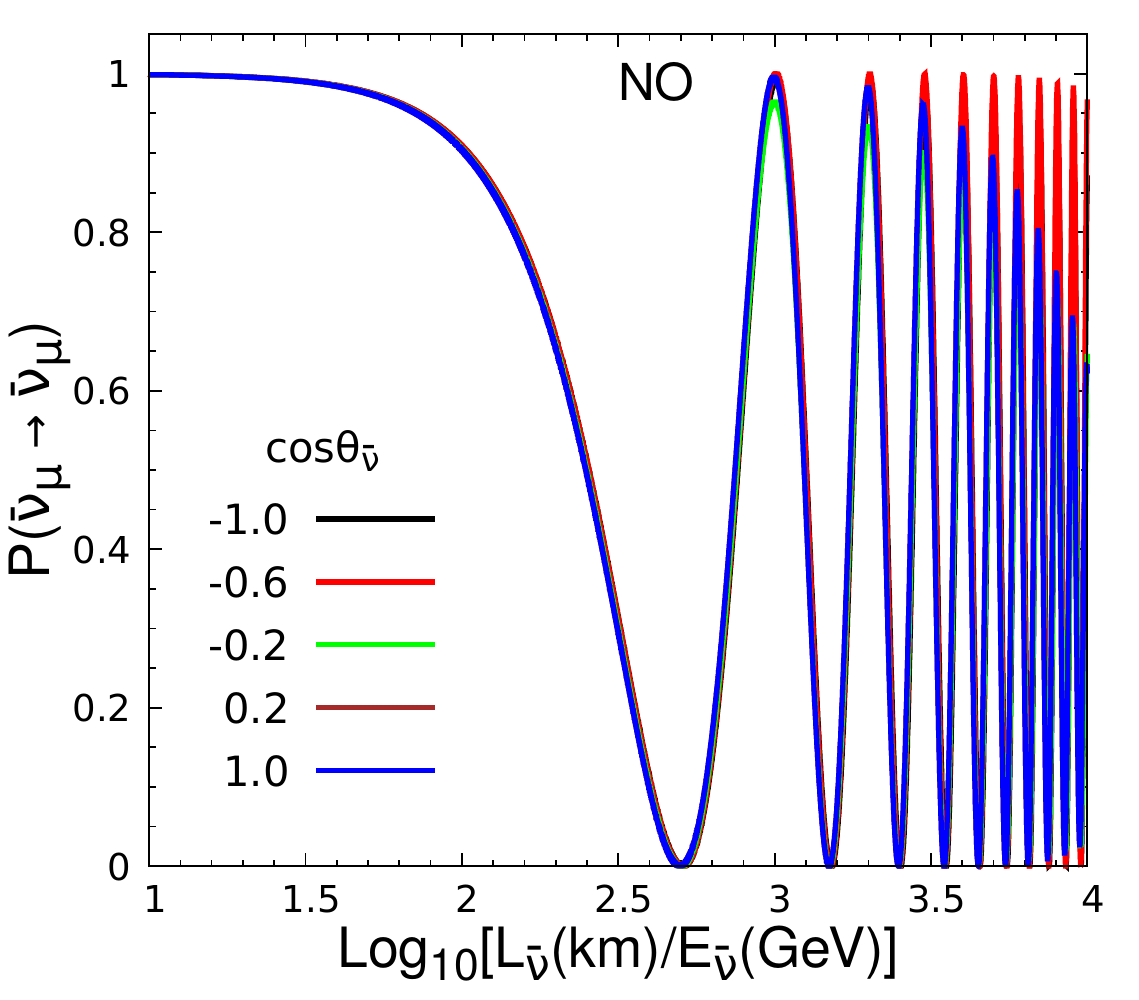}
  \mycaption{The survival probabilities of $\nu_\mu$ (left panel) and
    $\bar\nu_\mu$ (right panel) as functions of $L_\nu/E_\nu$, in the three-flavor neutrino oscillation 
    framework in the presence of Earth matter (PREM profile~\cite{PREM:1981}). Different colors represent different
    values of the neutrino zenith angle $\theta_\nu$ (or equivalently,
    the neutrino baseline $L_\nu$). We use the oscillation parameters given in Table~\ref{tab:osc-param-value}.}
  	\label{fig:1Dprob}
\end{figure}

Figure~\ref{fig:1Dprob} shows the survival probabilities for $\nu_\mu$
and $\bar\nu_\mu$ as functions of $\log_{10}[L_\nu/E_\nu]$
for different zenith angles for a benchmark set of oscillation parameters as shown in Table \ref{tab:osc-param-value}. 
It may be observed that at $\log_{10}[L_\nu/E_\nu] \approx 2.7$, the first
oscillation minimum appears for $\nu_\mu$ as well as for $\bar\nu_\mu$.
For $L_\nu/E_\nu$ values smaller than the location of the first oscillation minimum, the survival
probabilities for $\nu_\mu$ and $\bar\nu_\mu$ are almost equal, and the
overlapping nature of different $\cos\theta_\nu$ curves shows that there is no $\theta$-dependence beyond that coming from $L_\nu/E_\nu$.
This is an indication that the survival probabilities are dominated by vacuum
oscillations in this region.
However for higher $L_\nu/E_\nu$ values, Earth matter effects introduce
an additional $\theta$-dependence beyond that via $L_\nu/E_\nu$, which can be
prominently seen for $\nu_\mu$ survival probability in the case of NO.
Since matter effects on antineutrinos are not significant in NO,
the $\bar\nu_\mu$ survival probability does not show this feature.

\begin{table}[]
	\centering
	\begin{tabular}{|c|c|c|c|c|c|c|}
		\hline
		$\sin^2 2\theta_{12}$ & $\sin^2\theta_{23}$ & $\sin^2 2\theta_{13}$ & $|\Delta m^2_{32}|$ (eV$^2$) & $\Delta m^2_{21}$ (eV$^2$) & $\delta_{\rm CP}$ & Mass Ordering\\
		\hline
		0.855 & 0.5 & 0.0875 & $2.46\times 10^{-3}$ & $7.4\times10^{-5}$ & 0 & Normal (NO)\\
		\hline 
	\end{tabular}
	\caption{The values of the benchmark oscillation parameters used in this analysis. These values are in good agreement with the current global fits~\cite{Capozzi:2020qhw,Esteban:2018azc,NuFIT,deSalas:2018bym}. Normal mass ordering corresponds to $m_1 < m_2 < m_3$.}
	\label{tab:osc-param-value}
\end{table}

\subsection{Oscillograms in ($E_\nu$, $\cos\theta_\nu$) plane}

\begin{figure}[htb!]
  \includegraphics[width=0.49\textwidth]{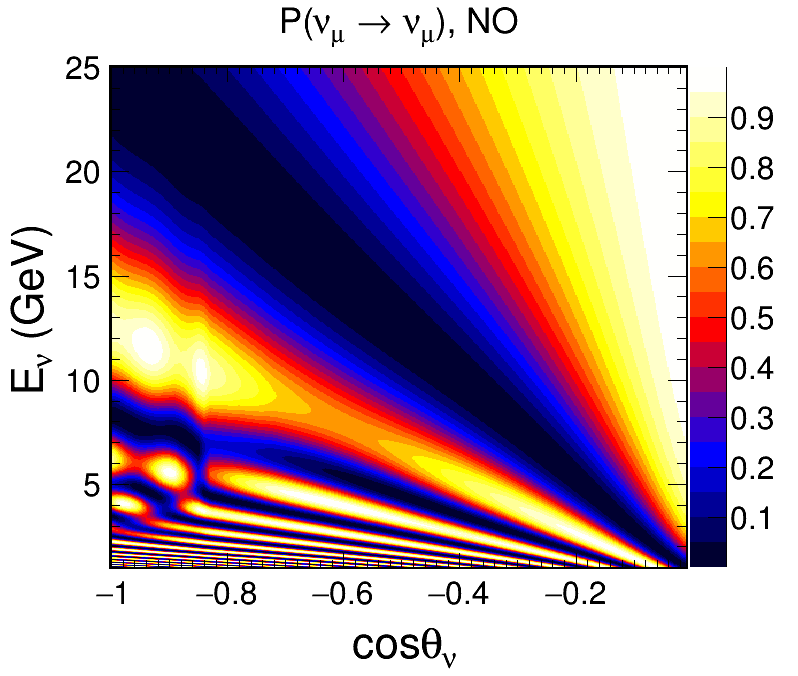}
  \includegraphics[width=0.49\textwidth]{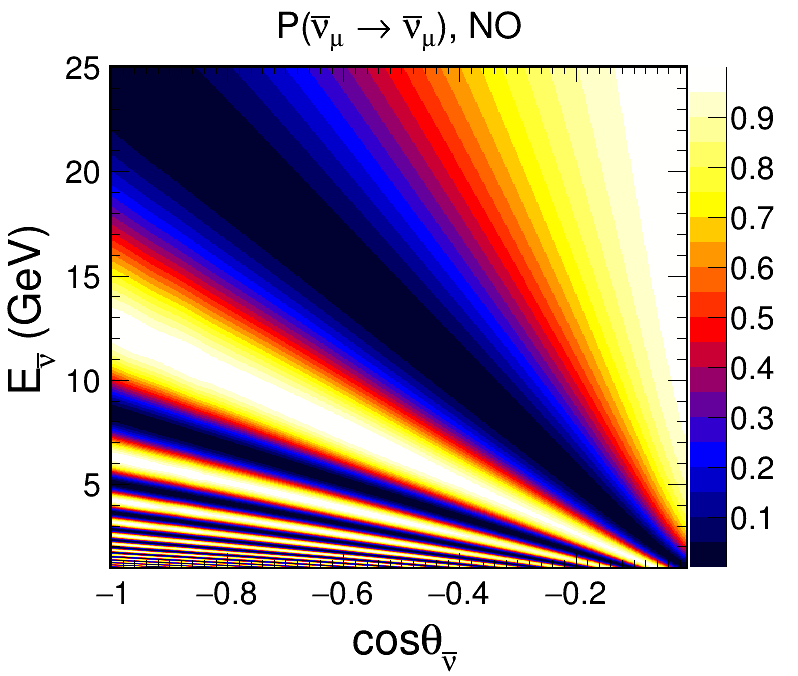}
  \mycaption{Oscillograms of survival probabilities of $\nu_\mu$ (left panel)
    and $\bar\nu_\mu$ (right panel) in the ($E_\nu$, $\cos\theta_\nu$) plane,
    using three-flavor neutrino oscillation framework in the presence of Earth matter
    (PREM profile~\cite{PREM:1981}).
    We use the oscillation parameters given in Table~\ref{tab:osc-param-value}.
  }
  	\label{fig:oscillogram}
\end{figure}

Figure~\ref{fig:oscillogram} shows the survival probabilities of $\nu_\mu$ and
$\bar\nu_\mu$ in the ($E_\nu$, $\cos\theta_\nu$) plane, with NO as the neutrino mass
ordering. 
In both the panels of Fig.~\ref{fig:oscillogram}, we can see the
prominent dark diagonal band, which represents the minimum survival
probability. This oscillation minimum region corresponds to the same
``oscillation dip'' as that earlier observed in
Fig.~\ref{fig:1Dprob} at $\log_{10}[L_\nu/E_\nu] \approx 2.7$.  Its nature in
neutrino and antineutrino survival probabilities is almost identical,
indicating that this region is dominated by vacuum oscillations. 
In this study, we will explore whether we can reconstruct this band of
oscillation minima (an ``oscillation valley'') from the atmospheric neutrino
data. The part of the oscillogram below the blue band corresponds to
lower energies and longer baselines, where the matter effects are
significant for neutrinos (in the NO scenario considered here). For neutrinos, the MSW resonance~\cite{Wolfenstein:1977ue, Mikheev:1986gs, Mikheev:1986wj} can be observed for $6 ~ \text{GeV} < E_{\nu} < 8 ~ \text{GeV}$ and $-0.7 < \cos\theta_{\nu} < -0.5$. The parametric resonance~\cite{Akhmedov:1998ui,Akhmedov:1998xq} or oscillation length resonance~\cite{Petcov:1998su,Petcov:199993} can be seen for $3 ~ \text{GeV} < E_{\nu} < 6 ~ \text{GeV}$ and $\cos\theta_{\nu} < -0.8$.
The region above the blue band corresponds to vacuum oscillations, and
is almost identical for $\nu_\mu$ and $\bar\nu_\mu$ survival probabilities. 

In the next sections, we demonstrate how these multi-GeV range features in the probability may be reflected in the expected event spectra of neutrinos and antineutrinos in an atmospheric neutrino experiment like ICAL.
   
\section{Event Generation at ICAL detector}
\label{sec:event-gen}

While the analysis in this paper can in principle be performed with any
atmospheric neutrino detector, we use the proposed ICAL
detector at the India-based Neutrino Observatory~\cite{INO,Kumar:2017sdq}
for our simulations.
There are two main reasons for this. (i) ICAL would be able to distinguish
neutrinos from antineutrinos, and hence independent analyses for these two
channels may be carried out. (ii) ICAL would be able to reconstruct the
energy as well as the direction of $\mu^-$ and $\mu^+$ (produced in the
charged-current interactions of $\nu_\mu$ and $\bar\nu_\mu$, respectively)
with a good precision, which will be crucial for the analysis. These are also the reasons due to which ICAL would be able to provide a direct measurement of neutrino mass ordering by observing the Earth's matter effects separately in neutrino and antineutrino channels in the multi-GeV energy range~\cite{Kumar:2017sdq}.

ICAL will be composed of 50 kt magnetized iron as the target, and glass
resistive plate chambers (RPCs) as the sensitive detector elements.
Iron plates of 5.6 cm thickness, and around 30000 glass RPCs, will be
stacked alternately in three modules, each of dimension
16 m (L) $\times$ 16 m (W) $\times$ 14.5 m (H). 
The magnetic field in the detector volume will be around 1.5 T.
Charged particles passing through the RPCs will ionize the gas inside them,
thereby inducing an electrical signal in the X- and Y- pickup strips
and providing the X-Y coordinates of the charged particle. 
The layer number provides the Z coordinate. In the charged-current (CC)
interactions of neutrinos, the muon, and hadrons in the final state will be
distinguished from their track-like and shower-like features respectively. The time resolution of less than 1 ns for each RPC layer~\cite{Dash:2014ifa, Bhatt:2016rek, Gaur:2017uaf} would enable ICAL to distinguish between upward-going and downward-going events with a high confidence.
The direction of bending of the charged muon track will enable the
identification of its charge, and hence the identification of whether
the original interacting particle was a neutrino or antineutrino.

To simulate the neutrino interactions in the ICAL detector, we use the
MC neutrino generator NUANCE~\cite{Casper:2002sd} and the neutrino
flux calculated for the INO site~\cite{Athar:2012it,PhysRevD.92.023004}.
In this context, it is worthwhile mentioning that the
studies of atmospheric neutrino oscillations~\cite{Kumar:2017sdq} for the ICAL detector have been performed
using the neutrino fluxes calculated for the Kamioka site. The neutrino fluxes vary from place to place due to the different geomagnetic
field of the Earth, and there are important differences in the fluxes at
Kamioka and at the proposed site of INO at Theni district of Tamil Nadu, India.
Theni is close to the region having the largest horizontal component
($\approx 40 ~\mu T$) of the Earth's magnetic field.
In comparison to this, the horizontal component of the geomagnetic field
at Kamioka is $\approx 30 ~ \mu T$.
As a result, the neutrino flux at the INO site is smaller by approximately
a factor of 3 at low energies. However, for neutrino energies around
10 GeV, the fluxes at INO and Kamioka sites are almost the
same~\cite{PhysRevD.92.023004}.
In this study, we use the neutrino fluxes given
in~\cite{Athar:2012it,PhysRevD.92.023004}, taking into account the 1 km
rock coverage of the mountain. Also, to take into account the effect of
solar modulation on the neutrino fluxes, we use the fluxes with high
solar activity (solar maximum) for half the exposure and the fluxes with
low solar activity (solar minimum) for the remaining half. 
These fluxes are given in the website~\cite{HondaFlux}.

In this work, we focus on the CC interactions of muon neutrinos and antineutrinos at the ICAL detector.
These may be a result of the unoscillated original $\nu_\mu / \bar\nu_\mu$, or original $\nu_e / \bar\nu_e$ which have oscillated to $\nu_\mu / \bar\nu_\mu$. The muon events from tau decay in the detector is small (around $2\%$ of total upward going muons from $\nu_\mu$ interactions~\cite{Pal:2014tre}), and these are mostly softer in energy and below the 1 GeV energy threshold of ICAL. In this work, we have not included this small contribution. 

The three-flavor neutrino oscillations in the presence of matter (PREM profile~\cite{PREM:1981}) are incorporated using the re-weighting algorithm as in~\cite{Ghosh:2012px,Thakore:2013xqa,Devi:2014yaa}. The events are then folded with the detector properties as obtained from the GEANT4 simulation for muons at ICAL~\cite{Chatterjee:2014vta}. The reconstructed energies ($E_{\mu}^{\rm rec}$) and zenith angles ($\theta_{\mu}^{\rm rec}$) of the charged muons are considered to be the observable quantities, and we do not reconstruct neutrino energy or its direction. We also use the observable $L_\mu^\text{rec}$, which is the effective baseline related to the reconstructed muon direction $\theta_\mu^\text{rec}$ in the following fashion, 
\begin{equation}
 L_\mu^{\rm rec} = \sqrt{(R+h)^2 - (R-d)^2\sin^2\theta_\mu^{\rm rec}} \,-\, (R-d)\cos\theta_\mu^{\rm rec} \,.
\label{eq:zen-bl-mu-rel}
\end{equation}
Note that $L_\mu^{\rm rec}$ is an observable associated with the reconstructed muon direction, and not to be related directly with the neutrino direction. 

In this paper, we use the ratios of upward-going and downward-going charged muons in various energy and direction bins. Note that the upward-going vs. downward-going events may lead to some ambiguity when the events are near-horizon, \ie, $-0.2 < \cos\theta_{\mu}^{\rm rec} < 0.2$ (or $73 ~ \text{km} < L_\mu^{\rm rec} < 2621 ~ \text{km}$). For these events, there is a significant change in the neutrino path length for a small variation in the estimated neutrino arrival direction. However, the detector response of ICAL to such events is anyway very poor, owing to the horizontally stacked structure, and it is observed that these events (corresponding to $\log_{10} [L_\mu^{\text{rec}}/E_\mu^{\text{rec}}]$ in the range of 1.5 -- 2.0) do not affect the analysis\footnote{Whenever we mention the value of $\log_{10} [L/E]$ in this paper, we take $L$ and $E$ in the units of km and GeV, respectively.}. Note that the Super-K collaboration selected only the ``high-resolution'' events in their $L/E$ analysis~\cite{Kajita:2014, Itow:2013zza}. In particular, their analysis rejected neutrino events near the horizon, as well as low-energy events where the large scattering angles would have led to large errors in the reconstruction of neutrino direction. The low efficiency of ICAL for the near-horizontal event, and our analysis threshold of 1 GeV for muons, automatically incorporates both these filters. However, a difference between our analysis and the $L/E$ analysis of Super-K~\cite{Kajita:2014} also needs to be pointed out. While the analysis in~\cite{Kajita:2014} is in terms of the inferred $L_\nu$ and $E_\nu$ of neutrinos, our analysis is in terms of the $L_\mu^\text{rec}$ and $E_\mu^\text{rec}$ of muons. The energy deposited at the detector by hadrons in the final state of neutrino interaction can be reconstructed at ICAL~\cite{Devi:2013wxa}, but we do not use the hadron energy information in this study. 
	
In Sections~\ref{sec:LbyE} and~\ref{sec:2D_E-CT}, we discuss the
event distributions as the one-dimensional functions of 
reconstructed $L_\mu^{\text{rec}}/E_\mu^{\text{rec}}$,
and in the two-dimensional reconstructed  ($E_\mu^{\rm rec}$, $\cos\theta_\mu^{\rm rec}$) plane, 
respectively.

\section{Oscillation dip in the $L_\mu^{\rm rec}/E_\mu^{\rm rec}$ distribution}
\label{sec:LbyE}

In this section, we analyze the expected event distributions of $\mu^-$
and $\mu^+$ as functions of reconstructed $L_\mu^{\text{rec}}/E_\mu^{\text{rec}}$,
with an MC event sample corresponding to 1000 years,
and simulated event sample corresponding to 10 years, for ICAL.
The 1000-year MC sample will correspond to the expected observations if
an unlimited amount of data were available, and analysis using this
sample will guide our algorithms. It will also lead to
the calibrations of oscillation parameters like $|\Delta m^2_{32}|$
and $\sin^2\theta_{23}$. On the other hand,
the analysis with the 10-year sample would give an idea of the
effects due to statistical fluctuations, an important consideration for
low-statistics experiments like those with atmospheric neutrinos.
To this end, we will analyze 100 independent sets of the 10-year samples.

 \begin{table}[htb!]
 \begin{center}
  \begin{tabular}{|l|c|c|}
   \hline
   & $\mu^-$ events & $\mu^+$ events \cr
   \hline
   U & 1654 & 740 \cr
   \hline
   D & 2960 & 1313 \cr
   \hline
   Total & 4614 & 2053\cr
   \hline
  \end{tabular}
  \mycaption{Total number of upward-going (U) and downward-going (D)
    $\mu^-$ and $\mu^+$ events expected at the 50 kt ICAL detector
    in 10 years (total exposure of 500 kt$\cdot$yr).
    We use the oscillation parameters given in Table~\ref{tab:osc-param-value}.}
 \label{tab:events}
  \end{center}
\end{table}

 The number of events expected in 10 years at the 50 kt ICAL
 are shown in Table~\ref{tab:events}. Here, U is the number of upward-going muon events 
($\cos \theta_\mu^{\rm rec} <0$), while D is the number of downward-going muon events ($\cos \theta_\mu^{\rm rec} >0$). The U/D ratio for reconstructed muons is defined as
\begin{equation}\label{eq:U/D_def}
\text{U/D} (E_\mu^\text{rec}, \cos\theta_\mu^{\rm rec}) \equiv
\frac{N(E_\mu^\text{rec}, -|\cos\theta_\mu^{\rm rec}|)}{N(E_\mu^\text{rec}, +|\cos\theta_\mu^{\rm rec}|)} \; ,
\end{equation}
where $N(E_\mu^\text{rec}, \cos\theta_\mu^{\rm rec})$ is the number of muon events with reconstructed energy $E_\mu^\text{rec}$ and 
the reconstructed zenith angle $\theta_\mu^{\rm rec}$. We associate the U/D ratio as a function of $\cos\theta_\mu^\text{rec}$ with the upward-going events, \ie, with $\cos\theta_\mu^\text{rec}<0$. We can also associate this U/D asymmetry with the $L_\mu^\text{rec}$ corresponding to these upward-going events, which is related to $\cos\theta_\mu^\text{rec}$ by Eq.~\ref{eq:zen-bl-mu-rel}. 

The ratio U/D  may be taken to be a proxy for the probability $P({\nu_\mu\rightarrow\nu_\mu})$ since
in the ICAL detector, around 98$\%$ events are contributed from the oscillation channel $\nu_\mu\rightarrow\nu_\mu$, and only $\sim 2\%$ of total events come from the $\nu_e\rightarrow\nu_\mu$ channel for the benchmark values of the oscillation parameters shown in Table~\ref{tab:osc-param-value}.
With the use of this ratio, the analysis becomes less dependent on the uncertainties in the absolute neutrino flux values. Below, we shall analyze the event distributions and U/D ratios for $\mu^-$ and $\mu^+$ events separately, as the functions of $L_\mu^{\text{rec}}/E_\mu^{\text{rec}}$.

\subsection{Events and U/D ratio using 1000-year Monte Carlo simulation }
\label{sec:LbyE-1000yrs}

\begin{figure}[htb!]
  \centering
  \includegraphics[width=0.49\textwidth]{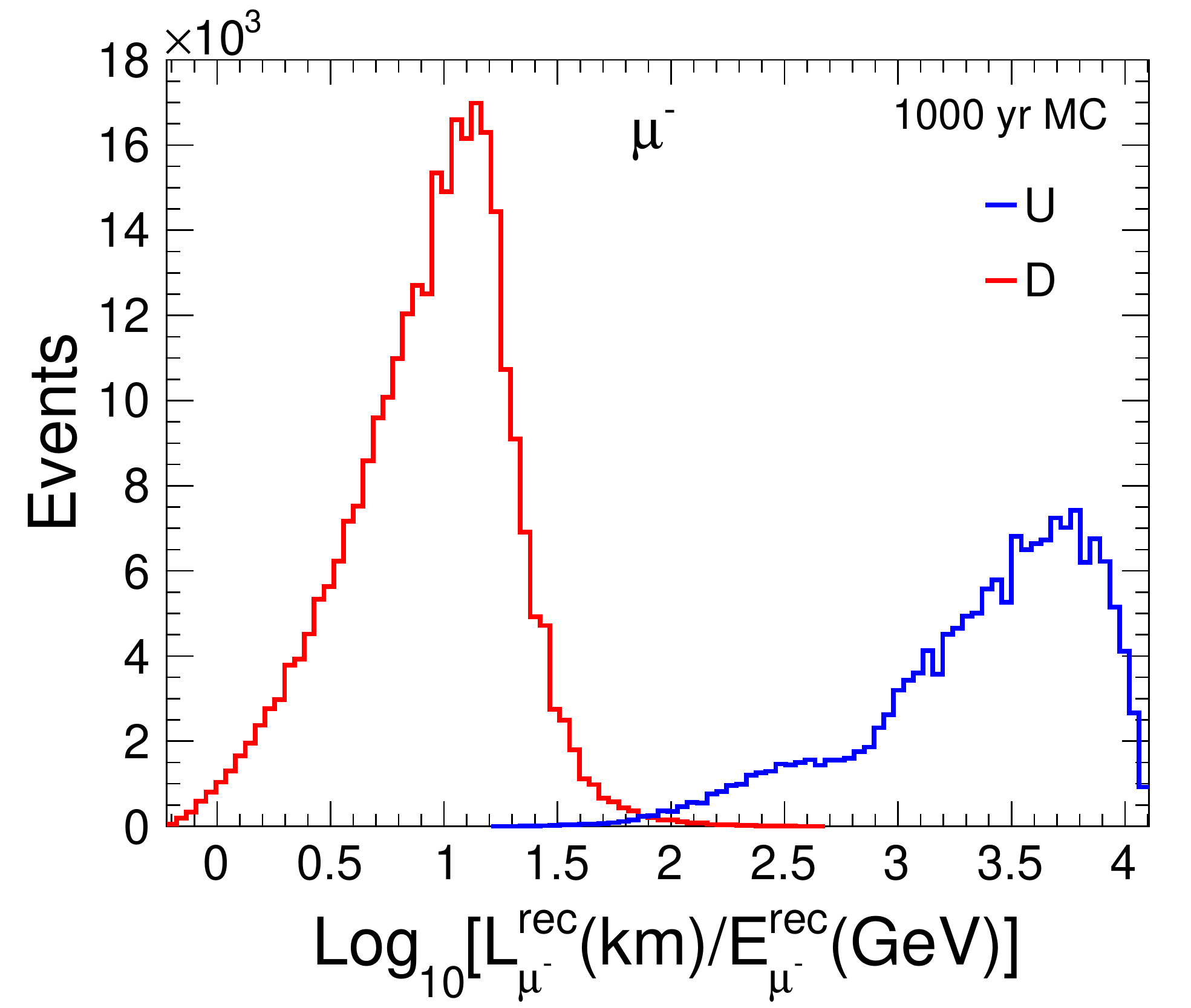}
  \includegraphics[width=0.49\textwidth]{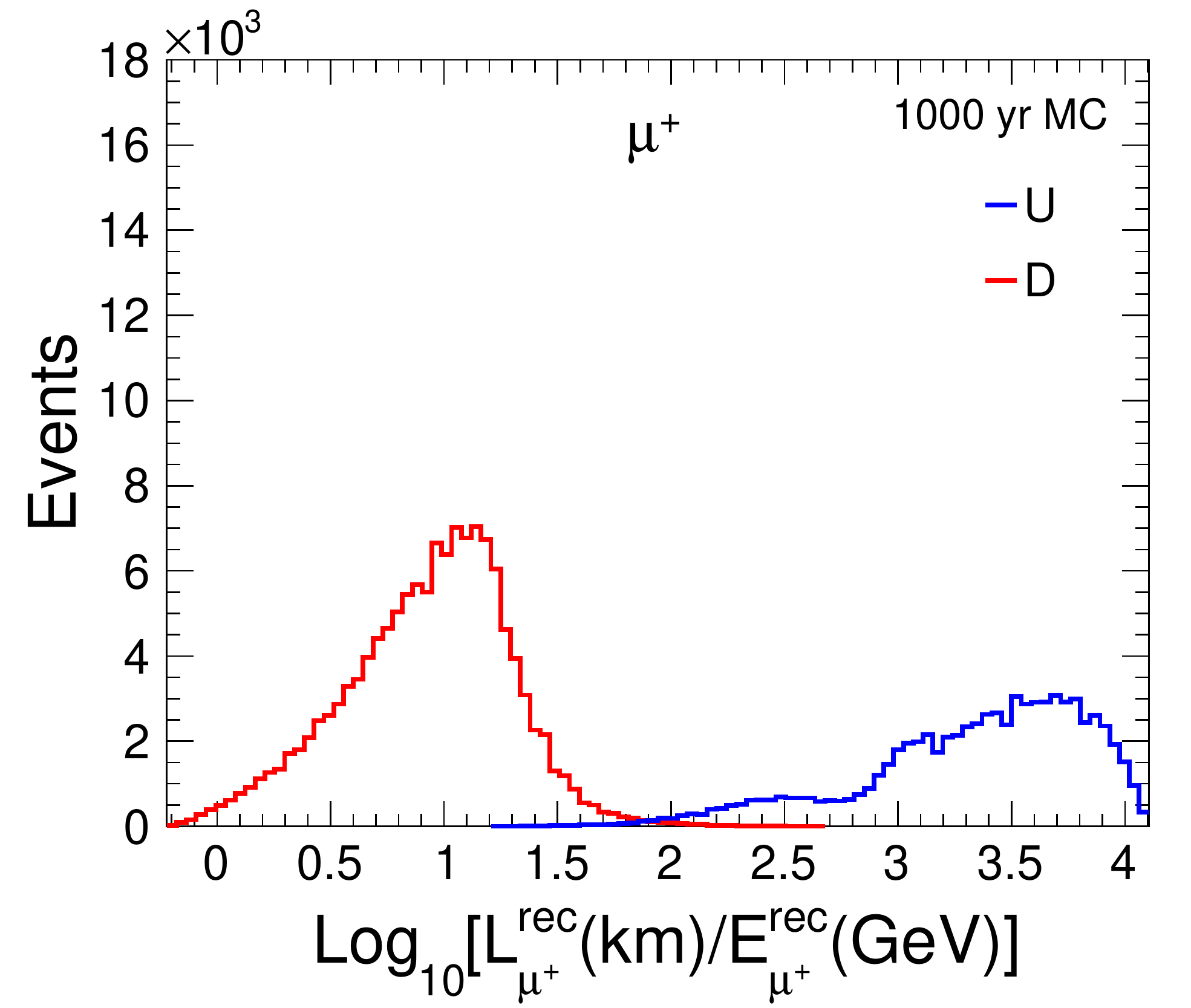}
  \includegraphics[width=0.49\textwidth]{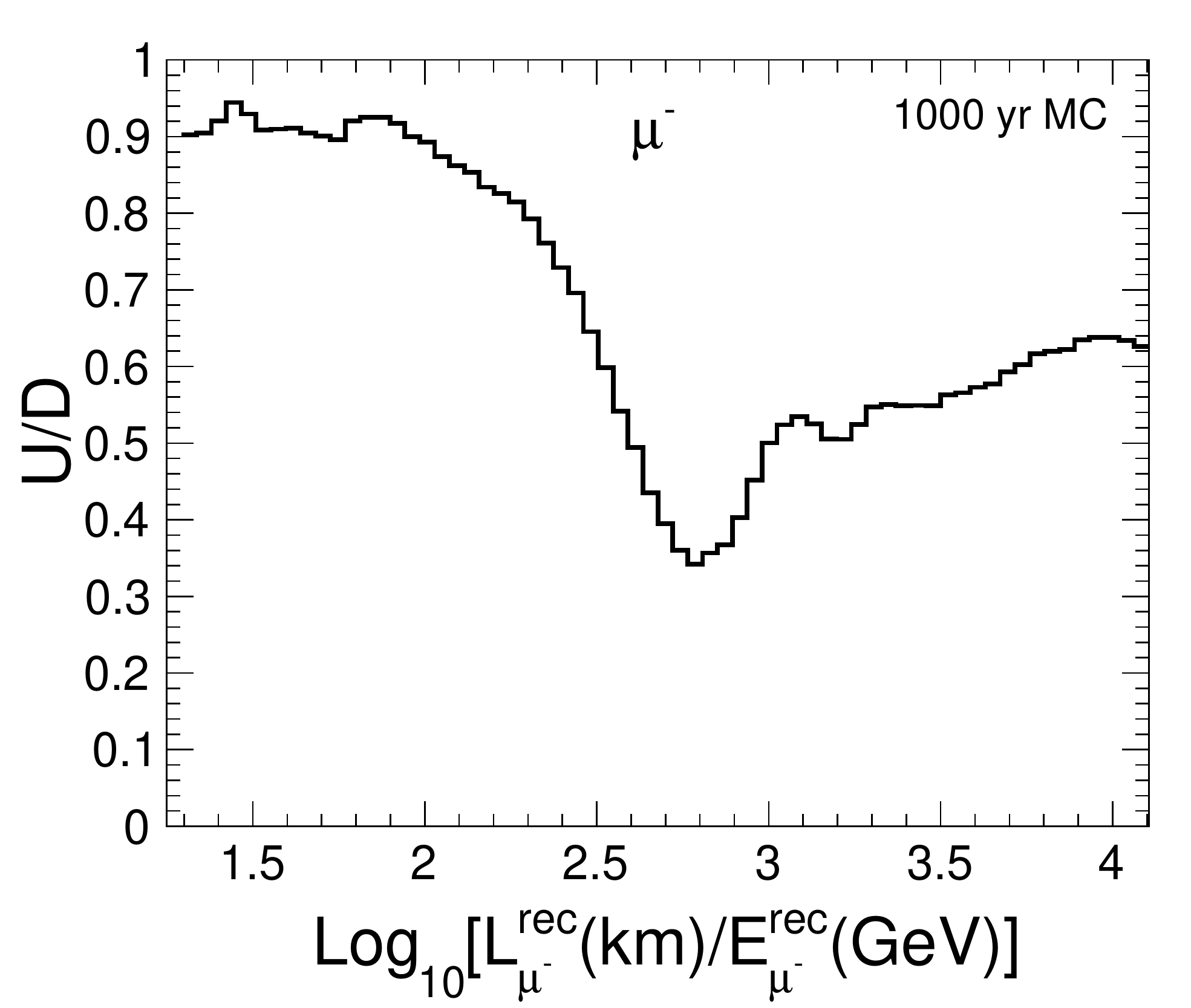}
  \includegraphics[width=0.49\textwidth]{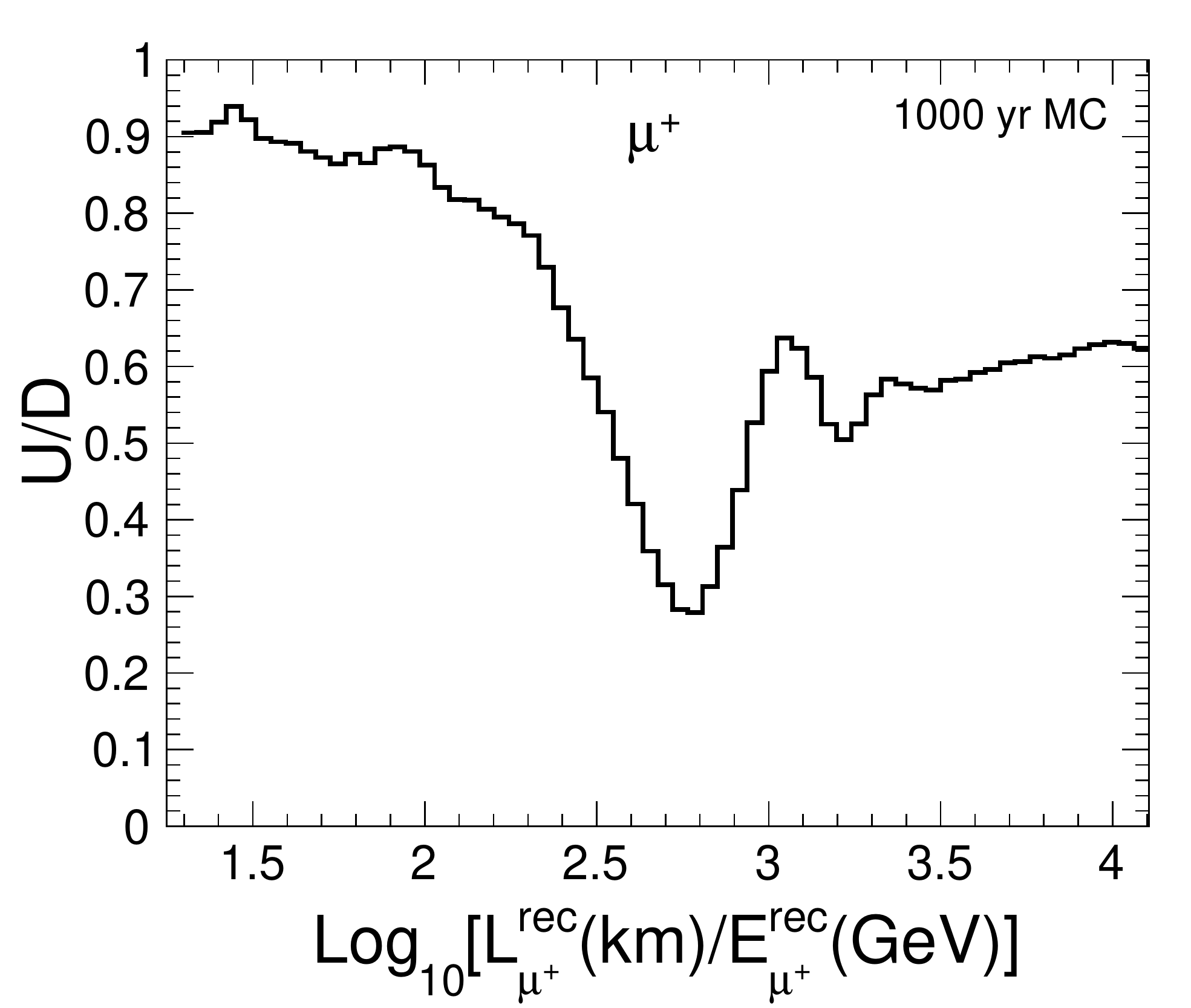}
  \mycaption{Upper panels: The expected $\log_{10}[L_\mu^\text{rec}/E_\mu^\text{rec}]$ distributions of
    $\mu^-$ (left panel) and $\mu^+$ (right panel) events. The upward-going and
    downward-going events are represented with blue and red lines respectively.
    Lower panels: The U/D ratios as functions of $\log_{10}[L_\mu^\text{rec}/E_\mu^\text{rec}]$ of $\mu^{-}$ (left panel)
    and $\mu^{+}$ (right panel). All of these results are obtained with the
    1000-year MC data sample.
    The $E_\mu^{\rm rec}$ lies in the range of 1 -- 25 GeV. We use the oscillation parameters given in Table~\ref{tab:osc-param-value}.
    } 
  \label{fig:LbyE-1000}
\end{figure}

The upper panels of Fig.~\ref{fig:LbyE-1000} present the distributions
of $\mu^-$ (left panel) and $\mu^+$ (right panel) events for the
1000-year MC sample. 
We take events with the reconstructed muon energy in the range
$(E_\mu^{\rm rec})_{\rm min} = 1$ GeV to $(E_\mu^{\rm rec})_{\rm max} = 25$ GeV.
Since $L_\mu^{\rm rec}$ is in the range of 15 km to 12757 km,
the minimum and maximum values of $\log_{10}[L_\mu^{\rm rec}/E_\mu^{\rm rec}]$
are $\log_{10}[(L_\mu^{\rm rec})_{\rm min}/(E_\mu^{\rm rec})_{\rm max}]$ = 0.22 and
$\log_{10}[(L_\mu^{\rm rec})_{\rm max}/(E_\mu^{\rm rec})_{\rm min}] = 4.1$, respectively. Note that for upward-going events, the minimum value of $\log_{10}[L_\mu^{\rm rec}/E_\mu^{\rm rec}]$ is 1.2, since $L_\mu^\text{rec}(\text{min})$ for upward-going events is 437 km.
We have binned the data in 100 bins, uniform in
$\log_{10}[L_\mu^{\rm rec}/E_\mu^{\rm rec}]$.
The number of upward-going events is clearly less than that of the downward-going
events, since the upward-going $\nu_\mu$ have traveled larger distances and
have had a larger chance of oscillating to the other neutrinos flavors. 
The number of $\mu^+$ events is less than that of $\mu^-$ events mainly because, at these energies, the cross-section of antineutrinos is smaller than that of neutrinos by a factor of approximately 2.

For the lower panels of Fig.~\ref{fig:LbyE-1000}, we divide the range
of upward-going muons ($\cos\theta_\mu^\text{rec} < 0$), \ie $\log_{10}[L_\mu^{\rm rec}/E_\mu^{\rm rec}] =$ 1.2 -- 4.1, in 66 uniform bins. 
A downward-going event with a given $\cos\theta_\mu^\text{rec} > 0$ value is then assigned to the bin corresponding to $-\cos\theta_\mu^\text{rec}$.
The U/D ratio in each bin is then calculated by dividing the number of
upward-going events by the number of downward-going events, and
plotted with the corresponding value of
$\log_{10}[L_\mu^{\rm rec}/E_\mu^{\rm rec}] $.
Two major dips in the U/D ratio may be observed in the figure,
in both the $\mu^-$ and $\mu^+$ channels. These dips occur at
$\log_{10}[L_\mu^{\rm rec}/E_\mu^{\rm rec}] \sim 2.75$ and $\sim$ 3.2, respectively.
In this study, we focus on the observation of the first dip.

Note that, while the two dips discussed above correspond to the first two
minima in the survival probability as shown in Fig.~\ref{fig:1Dprob}, they
are not exactly at the same location. This is because of the crucial
difference that, while Fig.~\ref{fig:1Dprob} uses the actual values of
energy and zenith angle of the atmospheric neutrino, 
Fig.~\ref{fig:LbyE-1000} uses the reconstructed values of energy and
zenith angle of the muon produced from the CC interaction
of that neutrino. In CC deep inelastic scattering of neutrino, hadrons in the final state take away some energy of the incoming neutrino, resulting in $E_\mu^\text{rec} < E_\nu$. As a result, the dips move towards higher values of $L_\mu^\text{rec}/E_\mu^\text{rec}$. Since the average inelasticities\footnote{Inelasticity of an event is defined as $y \equiv 1 - E_\mu/E_\nu$.} in the antineutrino events ($\langle y_{\bar\nu} \rangle \approx 0.3$) are smaller than that in the neutrino events ($\langle y_{\nu} \rangle \approx 0.45$) in the multi-GeV energy range~\cite{Devi:2014yaa}, the shift of the dip is smaller in the case of $\mu^+$ than $\mu^-$.

Note that the dips in Fig.~\ref{fig:LbyE-1000} are shallower and broader than those in Fig.~\ref{fig:1Dprob}, where the dips reach all the way to the bottom, that is $P(\nu_\mu \to \nu_\mu) \approx 0$, with $\sin^2 \theta_{23}=0.5$. The reason for this smearing lies in the difference between the momenta of the neutrino and muon, as well as the limitations on the muon momentum reconstruction in the detector. It may also be observed that the smearing is more in $\mu^-$ events as compared to the $\mu^+$ events in Fig.~\ref{fig:LbyE-1000}. This difference is due to the broader spread in the inelasticities of neutrino interactions than that in antineutrino interactions~\cite{Devi:2014yaa}. As a result of this spread, the dips in $\mu^-$ events get smeared more, and hence become shallower, as compared to the dips in $\mu^+$ events.

\subsection{Events and U/D ratio using 10-year simulated data} 
\label{sec:LbyE-10yrs}
\begin{figure}[htb!]
  \centering
  \includegraphics[width=0.49\textwidth]{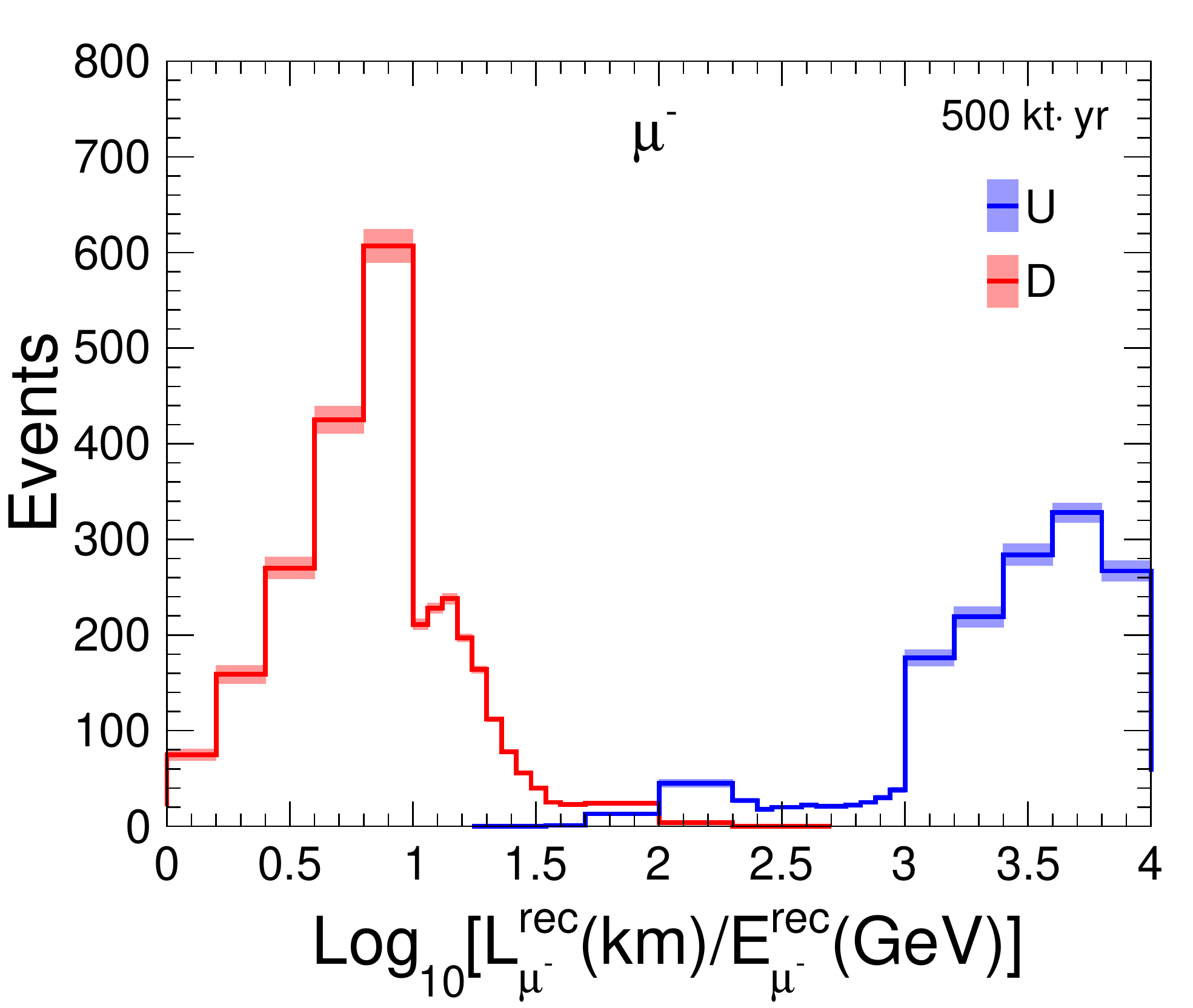}
  \includegraphics[width=0.49\textwidth]{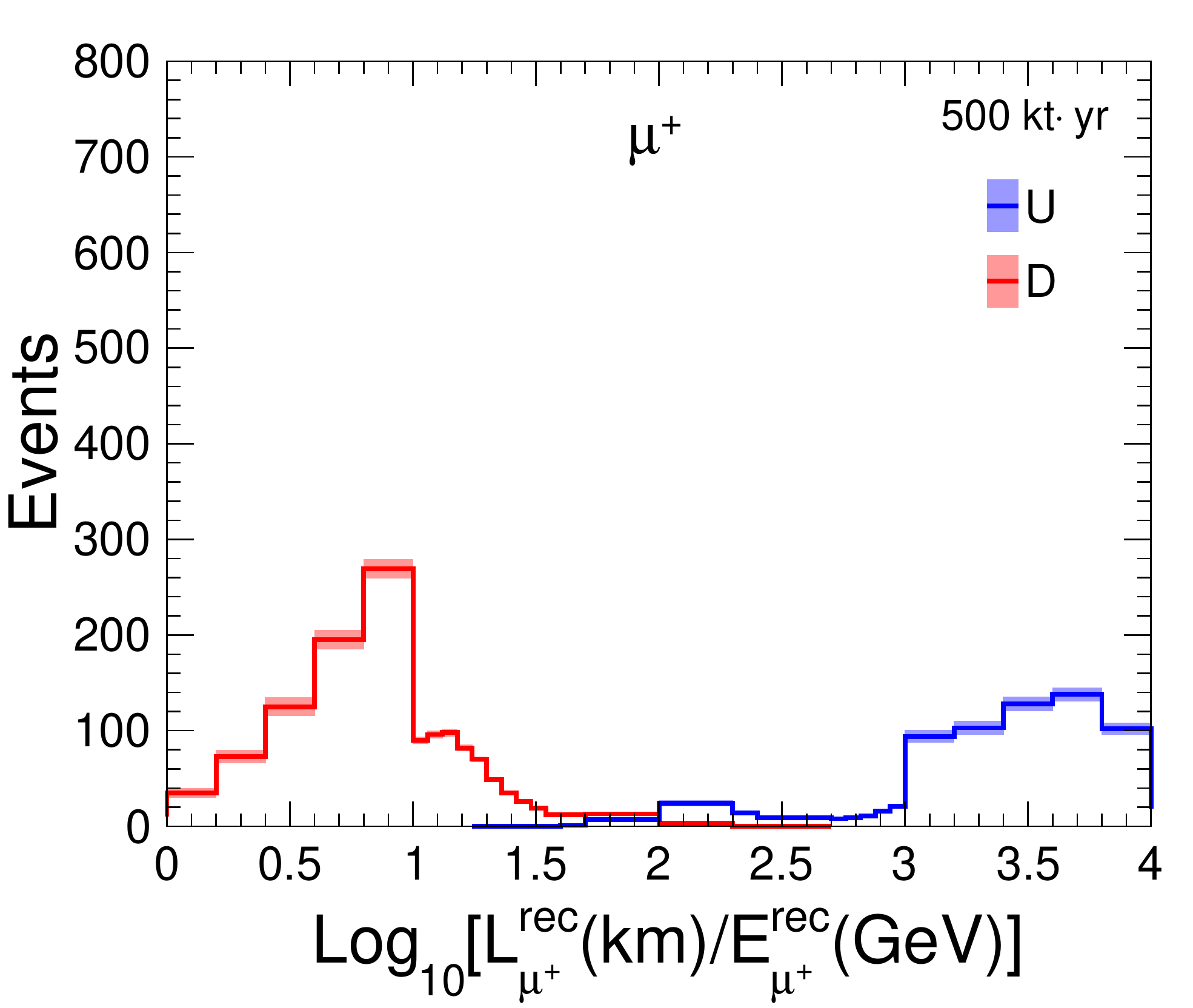}
  \includegraphics[width=0.49\textwidth]{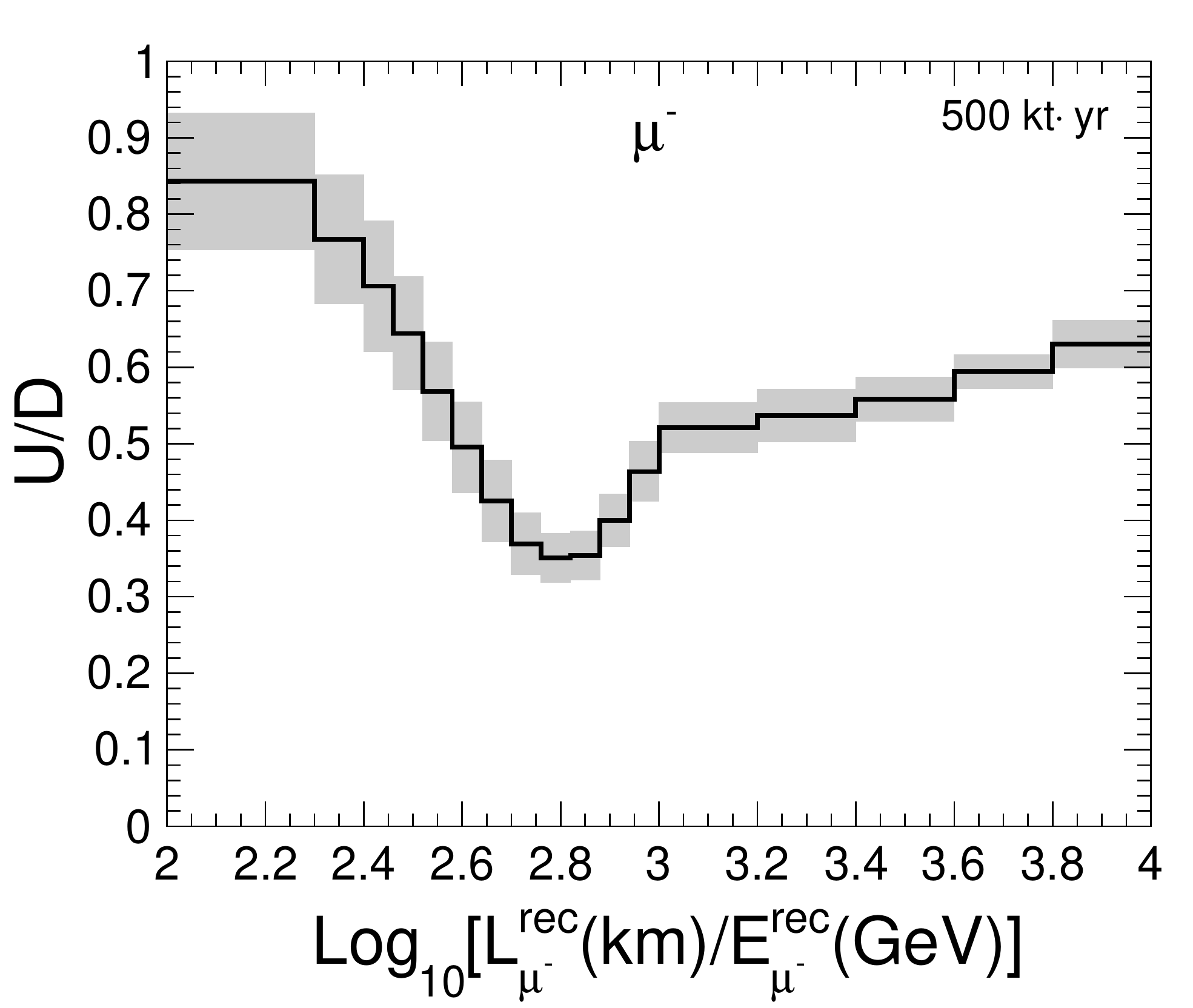}
  \includegraphics[width=0.49\textwidth]{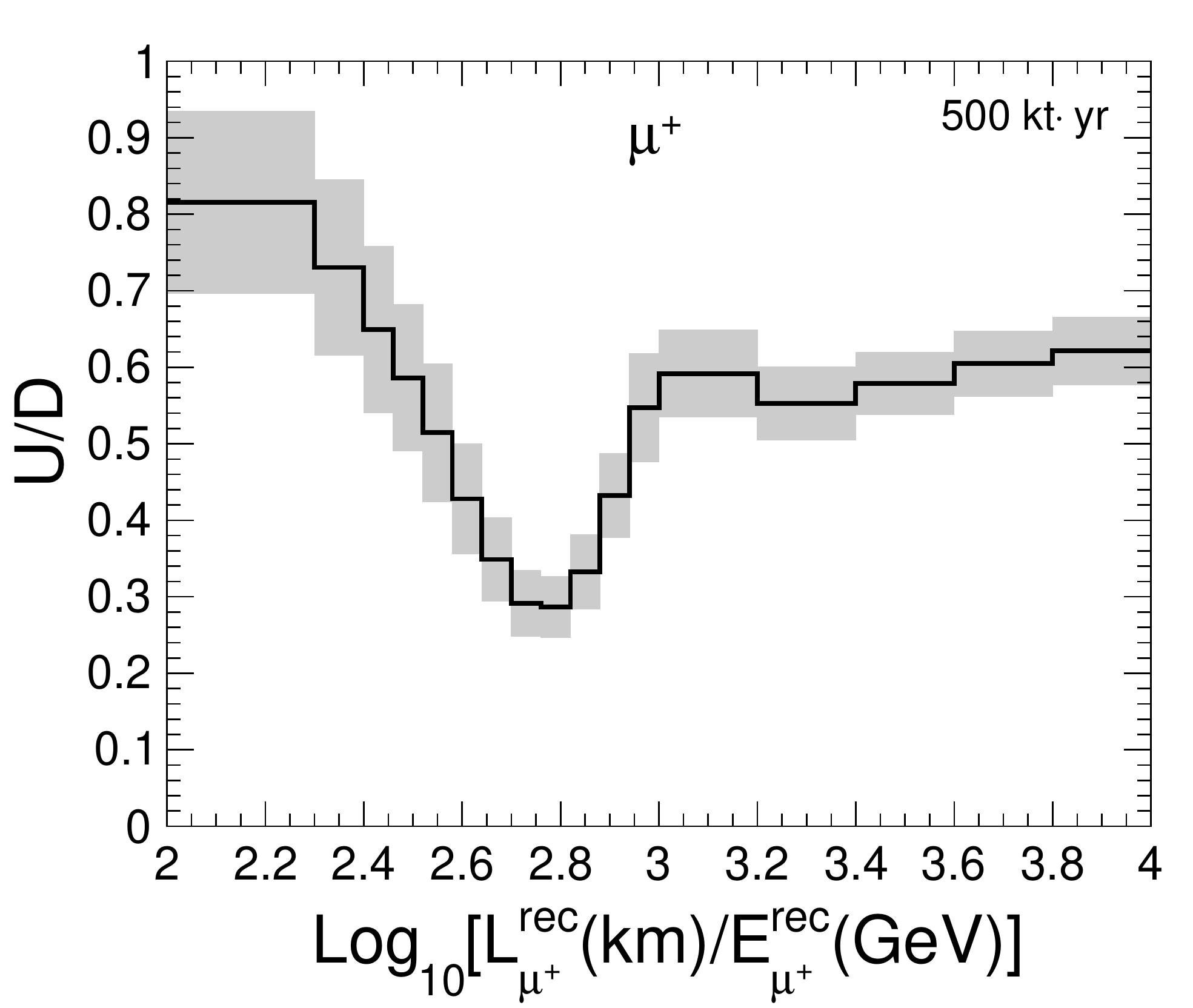}
  \mycaption{Upper panels: The expected $\log_{10} [L_\mu^{\text{rec}}/
      E_\mu^{\text{rec}}]$ distributions for $\mu^{-}$ (left panel) and
    $\mu^{+}$ (right panel) events. The red and blue lines are for
    downward-going and upward-going events, respectively. The shaded boxes
    represent statistical uncertainties. Lower panels: The U/D ratios
    as functions of  $\log_{10} [L_\mu^{\text{rec}}/E_\mu^{\text{rec}}]$.
    The height of shaded boxes represents expected uncertainties in
    the U/D ratio. The exposure is taken to be 10 years of 50 kt ICAL,
    \ie 500 kt$\cdot$yr. We use the oscillation parameters given in
    Table~\ref{tab:osc-param-value}. } 
	\label{fig:LbyE-10}
\end{figure}

In this section, we discuss the expected event distributions as functions of  reconstructed $L_\mu^{\rm rec}/E_\mu^{\rm rec}$ at the ICAL detector, with an exposure of 500 kt$\cdot$yr. In order to take into account the statistical fluctuations, which are expected to have a significant impact on the accuracy of the results, we take 100 independent sets, and calculate the mean as well as root-mean-square (rms) deviations of relevant  quantities.

We distribute the events in bins corresponding to the reconstructed   
$\log_{10}[L_\mu^{\rm rec}/E_\mu^{\rm rec}]$ values, as described in Sec.~\ref{sec:LbyE-1000yrs}. However, since
the number of events available here is much smaller than that for 1000 years, we choose a non-uniform binning scheme shown in Table~\ref{tab:binning-1D-10years}
such that typically, all the bins will have at least 10 down-going events. 
We have a total of 34 bins of
$\log_{10}[L_\mu^{\rm rec}/E_\mu^{\rm rec}]$ in the range of 0  to 4.
The distribution of the number of events in these bins is shown in the
top panels of Fig.~\ref{fig:LbyE-10}. The uncertainties due to
statistical fluctuations, calculated as the rms deviation obtained
from 100 independent simulated data sets, are also shown.
\begin{table}[htb!]
\centering
 \begin{tabular}{|c|c|c|c c|}
  \hline
  Observable & Range & Bin width & \multicolumn{2}{c|}{Number of bins} \\
  \hline 
  \multirow{7}{*}{ $\log_{10}\left[\frac{L_\mu^{\rm rec}(\text{km})}{E_\mu^{\rm rec}(\text{GeV})}\right]$} & [0, 1] & 0.2 & 5 & \rdelim\}{7}{7mm}[34] \cr 
   & [1, 1.6] &  0.06 & 10  & \cr
   & [1.6, 1.7]& 0.1& 1  & \cr
  & [1.7, 2.3]&0.3 & 2  & \cr
  & [2.3, 2.4]& 0.1& 1  & \cr
  & [2.4, 3.0] &0.06 & 10  &\cr
  & [3, 4] & 0.2 & 5  & \cr
  \hline
 \end{tabular}
\caption{The binning scheme adopted for $\log_{10}[L_\mu^{\rm rec}/E_\mu^{\rm rec}]$ for $\mu^-$ and $\mu^+$ events of 10-year simulated data.}
\label{tab:binning-1D-10years}
\end{table}

The lower panels of Fig.~\ref{fig:LbyE-10} show the U/D ratio in  reconstructed 
$\log_{10}[L_\mu^{\rm rec}/E_\mu^{\rm rec}]$ bins, using the same procedure
outlined in Sec.~\ref{sec:LbyE-1000yrs}. We choose the range of
$\log_{10}[L_\mu^{\rm rec}/E_\mu^{\rm rec}]$ to be in the range 2.0 -- 4.0,
which is the most interesting range. The fluctuations shown in the figure are the rms deviations obtained from the distributions of the U/D ratio in 100 independent simulated data sets.
It is observed that, while the first dip is quite prominent, the
second dip is lost in the statistical fluctuations, and due to the broad binning scheme that we have to employ owing to a small number of events.
While this does not rule out the identification of the second dip
using more efficient algorithms, in this work, we shall focus on the
first dip. Henceforth, when we refer to the dip position, we refer to the position of the first dip. It may be observed that the dip in the $\mu^+$ channel
is deeper than that in the $\mu^-$ channel similar to 1000-year MC sample as described in Sec.~\ref{sec:LbyE-1000yrs}.

\subsection{Identifying the dip with 10-year simulated data}
\label{sec:1dL/Efitting}

\begin{figure}[htb!]
	\centering
	\includegraphics[width=0.49\textwidth]{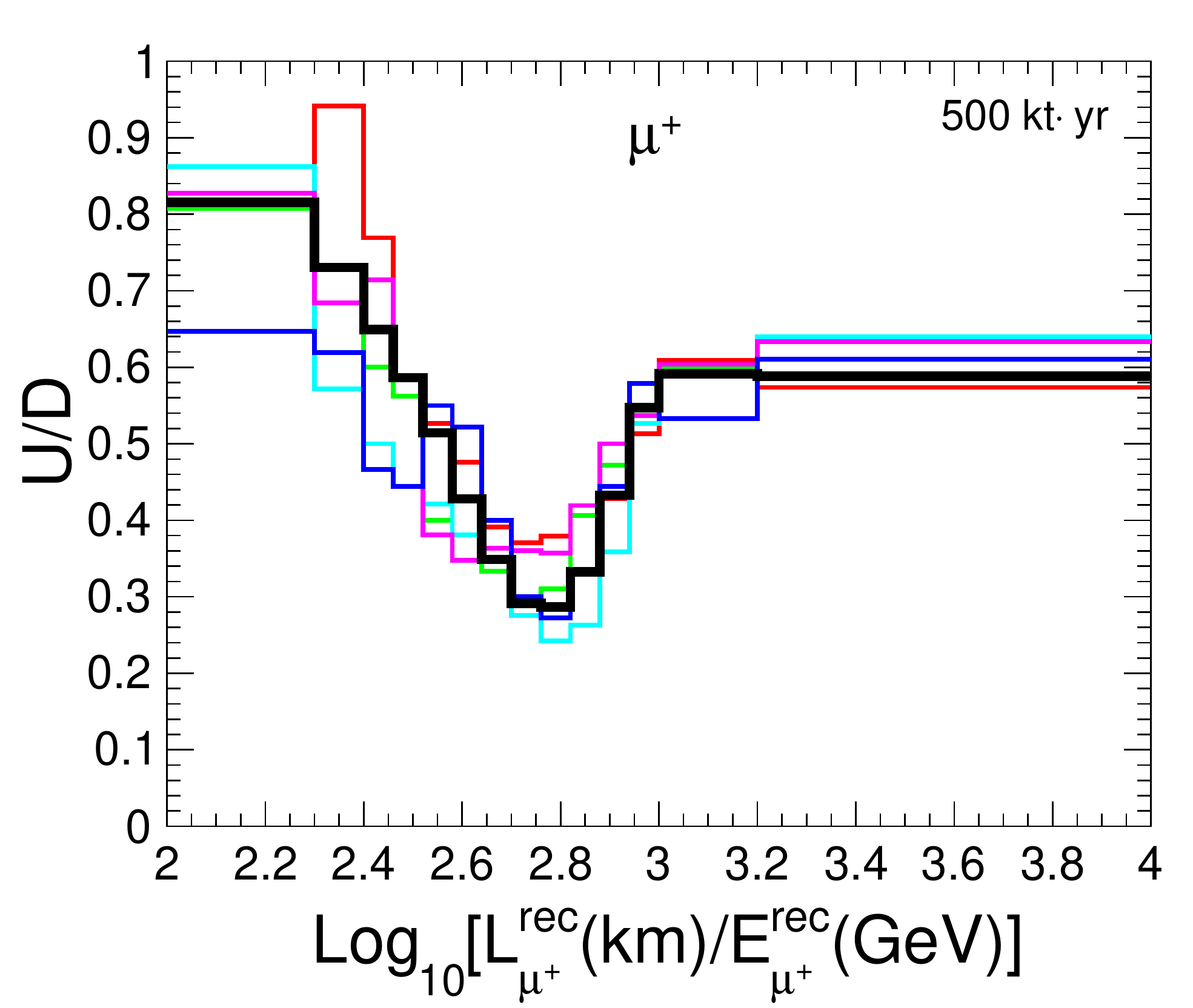}
	\includegraphics[width=0.49\textwidth]{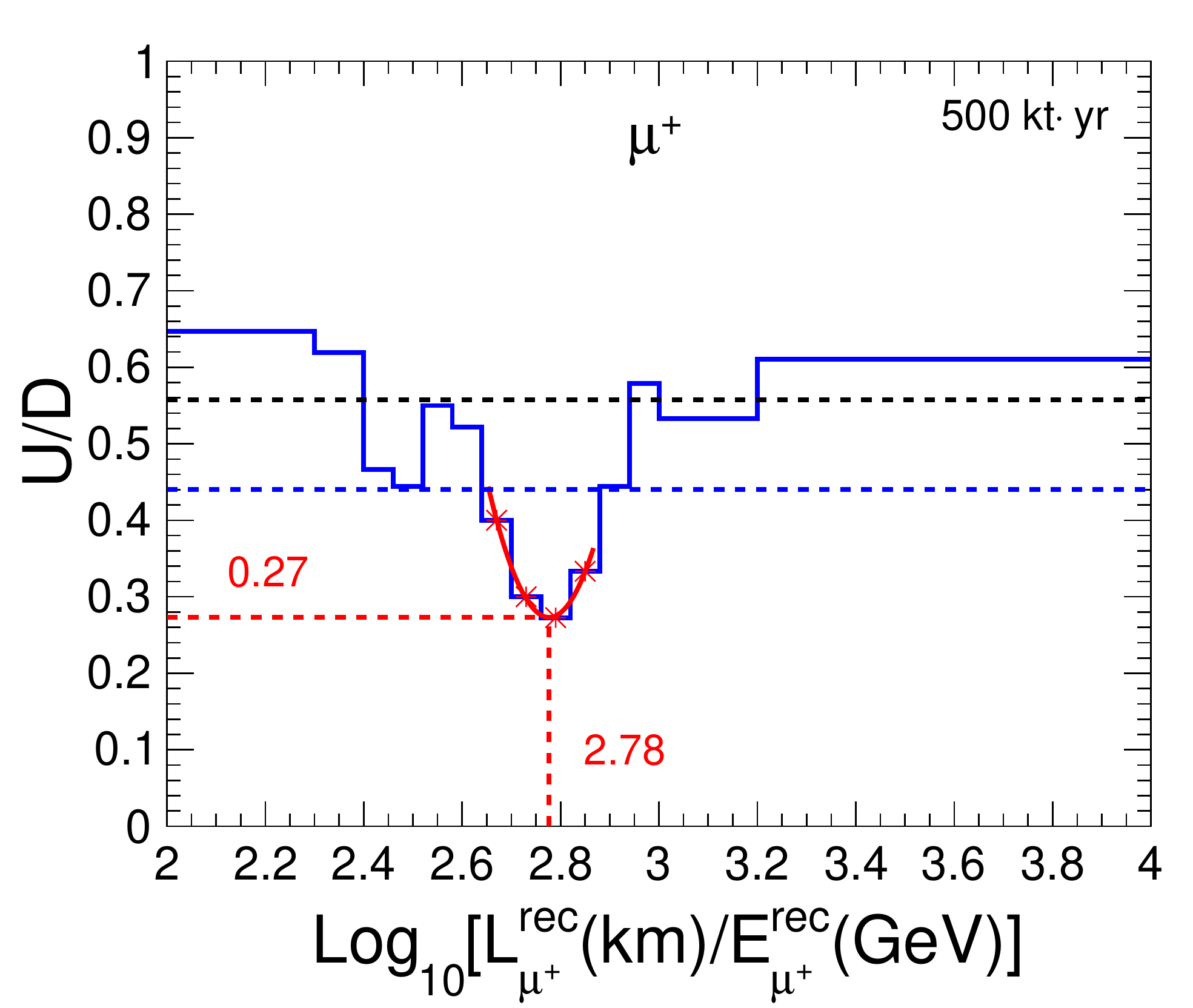}
	\mycaption{ Left panel: The U/D ratios as functions of
		$\log_{10} [L_\mu^{\text{rec}}/E_\mu^{\text{rec}}]$ for five independent simulated data sets
		with 500 kt$\cdot$yr exposure of ICAL are shown by thin colored lines. 
		The black solid line is the mean of 100 such data sets.
		Right panel: The red parabola represents the fit to the dip in
		one of the data sets (blue line), obtained after the dip-identification procedure described in the text. The black dashed line corresponds to
		the initial U/D ratio threshold for the algorithm, while the
		blue dashed line corresponds to the U/D ratio threshold once the dip
		has been identified. }
	\label{fig:LbyE-fit-procedure}
\end{figure}

The left panel of Fig.~\ref{fig:LbyE-fit-procedure} shows the U/D ratios
as functions of reconstructed $\log_{10} [L_\mu^{\text{rec}}/E_\mu^{\text{rec}}]$ of muons for five
independent simulated data sets with 500 kt$\cdot$yr exposure of ICAL. 
Clearly, the statistical fluctuations are significant, and may lead to
the misidentification of the dip position. An identification of the dip position should not correspond simply to the lowest value of the U/D ratio, but also be guided by the values of the U/D ratios in
surrounding bins. In order to achieve this, we use the dip-identification algorithm as described below. 

To start with, we consider the region corresponding to the range
$\log_{10} [L_\mu^{\text{rec}}/E_\mu^{\text{rec}}]$ = 3.2 -- 4.0 as a single bin,
since in this region, we expect the oscillations to be quite rapid,
leading to the U/D ratio averaging out to a constant value. 
Let the measured value of the U/D ratio in this bin be $R_0$.
From the simulations done with 100 independent sets, we have found that
the statistical fluctuations in data sets give rise to a rms deviation of
{\bf $\Delta R_0 \approx 0.02$} and \textbf{$0.03$} in this ratio for $\mu^-$ and $\mu^{+}$ respectively.
We take this fluctuation into account, and start with an initial
ratio threshold of $R_{\rm th} \equiv R_0 - 2 \Delta R_0$,
shown by the black dashed line in the right panel of
Fig.~\ref{fig:LbyE-fit-procedure}.
All the bins with measured U/D ratio less than $R_{\rm th}$ form the
initial candidates for the dip position.
However, all these bins need not be a part of the actual dip, due to
fluctuations.
In order to identify the actual bins surrounding the dip, we try to
find the cluster of consecutive bins, all of which have U/D ratio less
than that in all the other bins.
This is achieved by lowering the value of $R_{\rm th}$ till all the
bins with U/D ratio less than $R_{\rm th}$ are contiguous. This final
value of $R_{\rm th}$ is denoted by the blue dashed line in the right panel of
Fig.~\ref{fig:LbyE-fit-procedure}.

Once the dip-identification algorithm has thus identified a cluster of
contiguous bins as the dip region, we fit the U/D ratios in these bins
with a parabola as shown in the right panel of
Fig.~\ref{fig:LbyE-fit-procedure}.
The value of $\log_{10} [L_\mu^{\text{rec}}/E_\mu^{\text{rec}}]$
corresponding to the lowest U/D ratio obtained from this fit,
denoted by $x_{\rm min}$,
would be identified with the ``location'' of the dip.


\subsection{$|\Delta m^2_{32}|$ and $\theta_{23}$ from the dip and the U/D ratio}
\label{sec:Calib-LbyE}

\begin{figure}[htb!]
	\centering
	\includegraphics[width=0.49\textwidth]{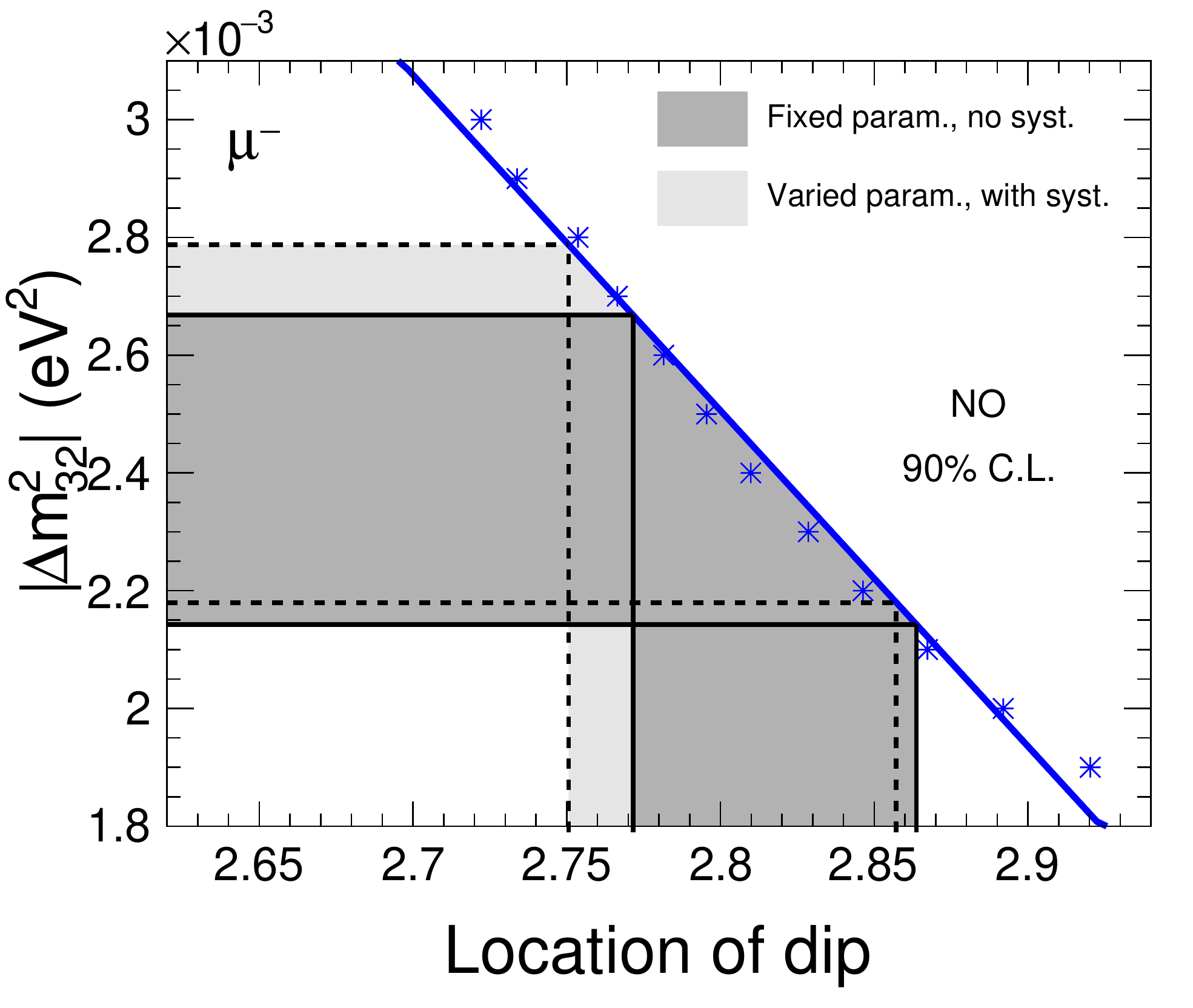}
	\includegraphics[width=0.49\textwidth]{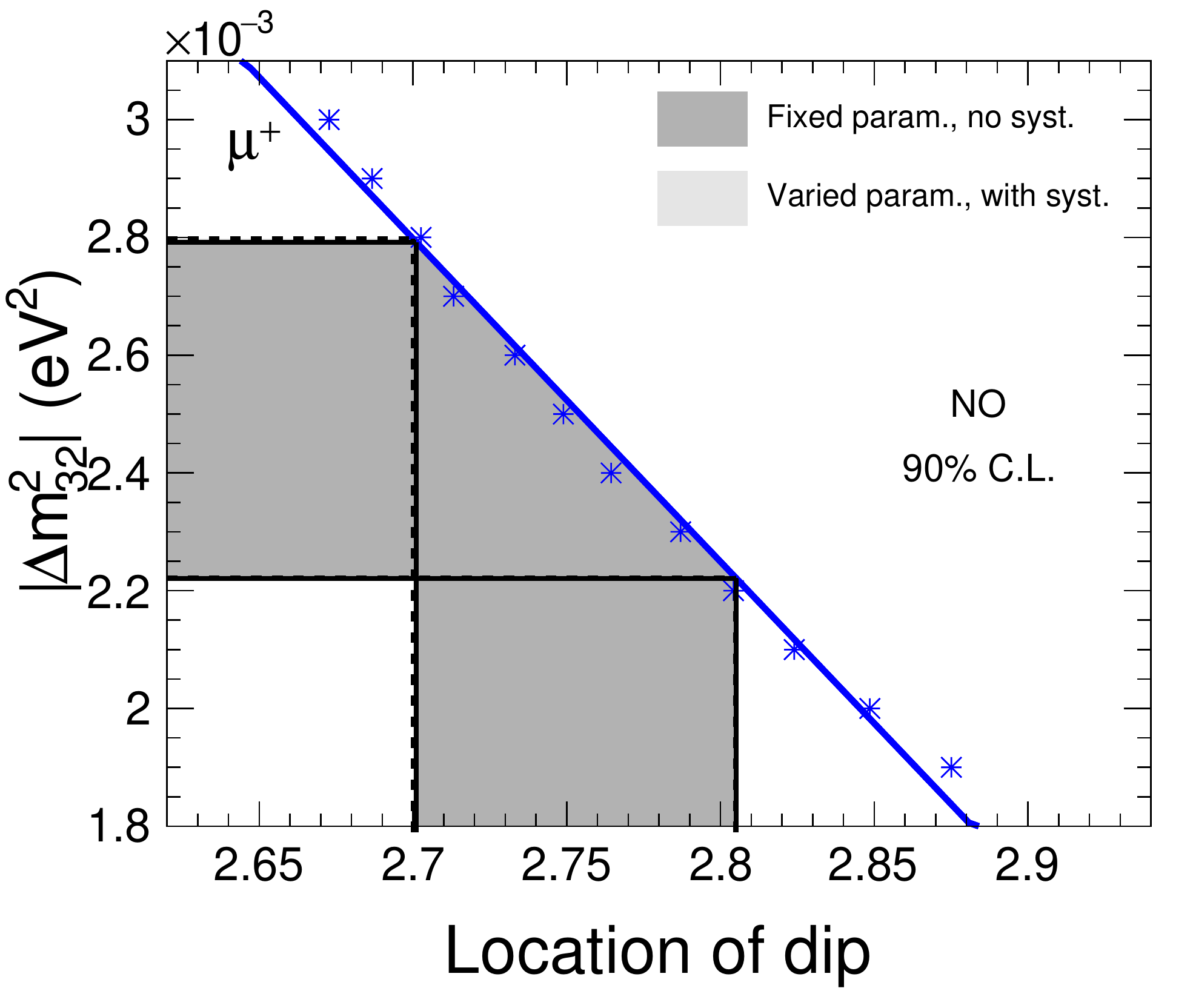}
	\includegraphics[width=0.49\textwidth]{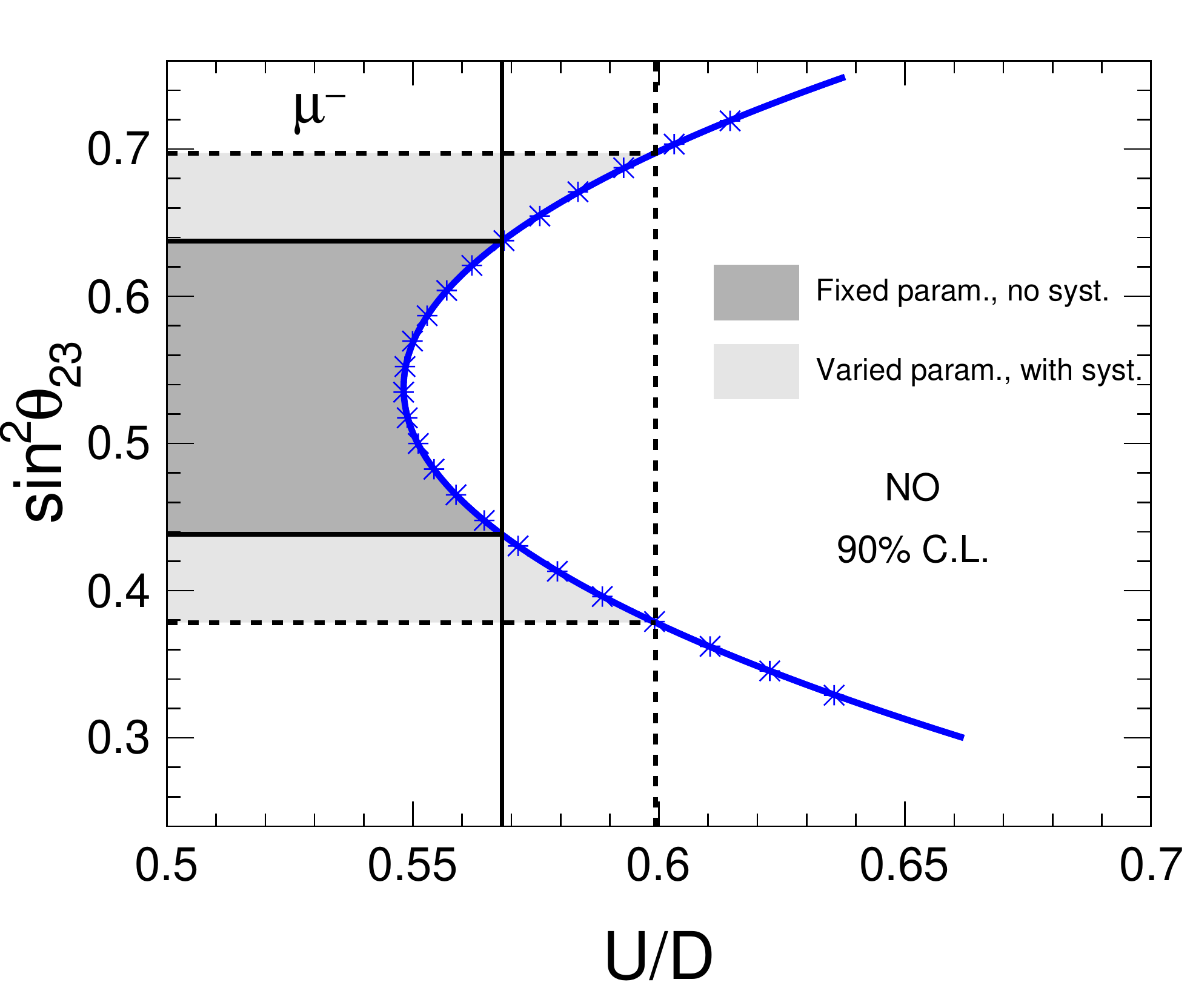}
	\includegraphics[width=0.49\textwidth]{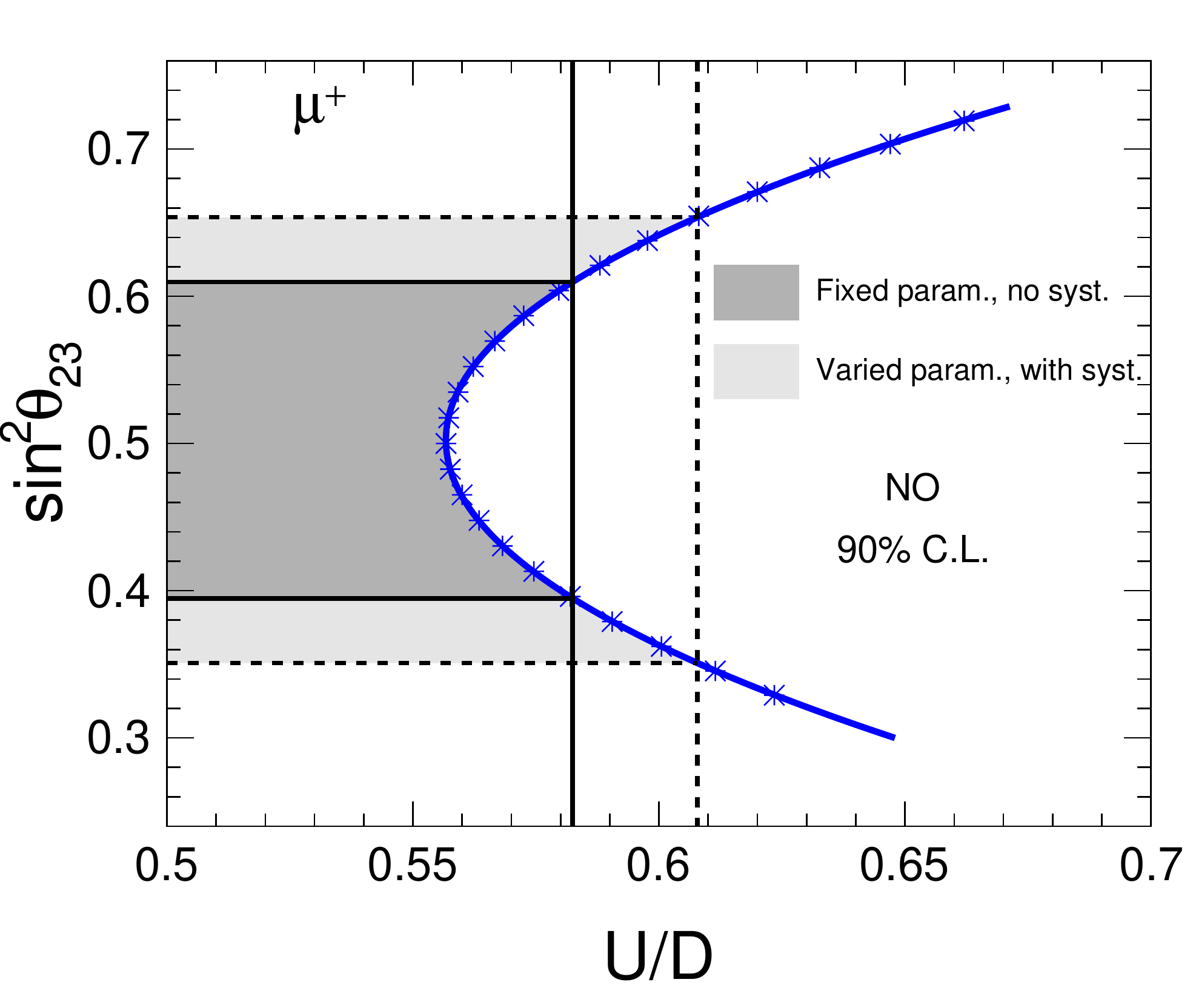}
	\mycaption{Upper panels: The blue points and the blue line correspond to
		the calibration of actual $|\Delta m^2_{32}|$ with the location of the
		dip, obtained by using the 1000-year MC data sample. The light (dark) gray bands represent the 90\% C.L. regions of the location of the dip, and hence the inferred $|\Delta m^2_{32}|$ through calibration, for $|\Delta m^2_{32}| \,(\text {true}) = 2.46\times 10^{-3}$ eV$^2$,  with (without) systematic errors and uncertainties in the other oscillation parameters.  
		Lower panels: The blue points and the blue line correspond to the
		calibration of actual $\sin^2\theta_{23}$ with the total U/D ratio
		across all bins. The light (dark) gray bands represent the 90\% C.L. allowed ranges of the U/D ratio, and hence the inferred $\sin^2\theta_{23}$
		through calibration, for $\sin^2\theta_{23} \,(\text {true}) = 0.5$,  with (without) systematic errors and uncertainties in the other oscillation parameters. For all the confidence regions, we use 10 years exposure of ICAL. The results obtained  from $\mu^-$ and $\mu^+$ events are shown in left and right panels, respectively. For the fixed-parameter analysis, we use the benchmark oscillation parameters given in Table~\ref{tab:osc-param-value}, while for the inclusion of systematic uncertainties and variation of oscillation parameters, we use the procedure described in Sec.~\ref{sec:Calib-LbyE}.
	} 
	\label{fig:Calib-LbyE}
\end{figure}

The first dip in the survival probability $P(\nu_\mu \to \nu_\mu)$, in the
approximation of two-flavor oscillations in vacuum, appears at 
$L_\nu (\text{km})/ E_\nu (\text{GeV})= \pi/(2.54\,|\Delta m^2_{32}|(\text{eV}^2))$ (see Eq.~\ref{eq:2flavsurv}). 
However as discussed in Sec.~\ref{sec:LbyE-1000yrs}, the location of the dip
$x_{\rm min}$ in the $L_\mu^{\rm rec}/E_\mu^{\rm rec}$ distribution 
would have a complicated dependence on the three flavor neutrino oscillation, matter effect, neutrino fluxes,
CC cross-sections, inelasticities of events, and detector response. 
However, we can obtain a calibration of $|\Delta m^2_{32}|$ with the
location of the dip, using the 1000-year MC sample. The binning scheme used is the same as in Sec.~\ref{sec:1dL/Efitting}. For obtaining the calibration curves, we keep the oscillation parameters (except for the one to be calibrated) fixed at benchmark values as given in Table 1.
The blue points in the upper panels of Fig.~\ref{fig:Calib-LbyE} represent
the values of $x_{\rm min}$ obtained from the U/D  ratio distribution with
different values of $|\Delta m^2_{32}|$. These points are observed to lie
close to a straight line, and we can draw a calibration curve that would
enable us to infer the actual value of $|\Delta m^2_{32}|$, given the
$x_{\rm min}$ value determined from the data. Note that this calibration
curve also has a dependence on the analysis procedure, including the
binning scheme and bin-identification algorithm.

While the calibration curve allows a one-to-one correspondence between
the actual $|\Delta m^2_{32}|$ value and the location of the dip for
sufficiently large data (like 1000 years of exposure), practically
speaking the data available is going to be limited. The statistical
fluctuations introduced due to this will lead to uncertainties in the
determination of $|\Delta m^2_{32}|$ via this method. To estimate these
uncertainties, we generate 100 independent simulated data sets with an exposure of 500 kt$\cdot$yr each, and determine the location of the dip in each of them. The distribution of the dip locations thus obtained allows us to determine 90\% C.L. allowed regions for the dip location, and hence
the 90\% C.L. allowed regions for the calibrated $|\Delta m^2_{32}|$
values. These allowed regions are represented in the upper panels of
Fig.~\ref{fig:Calib-LbyE} by dark gray bands, where all the oscillation parameters are kept fixed at benchmark values as given in Table~\ref{tab:osc-param-value} and no systematics are taken into account.  
The figure indicates that the 90$\%$ C.L. allowed range for
$|\Delta m^2_{32}|$ from $\mu^-$ events is (2.14 -- 2.67)$\times 10^{-3}$ eV$^2$, while that from $\mu^+$ data is (2.22 -- 2.79)$\times10^{-3}$ eV$^2$. 

We also investigate the effect on the measurements of $\Delta m^2_{32}$ and $\theta_{23}$ due to uncertainties in the other oscillation parameters. For this, we vary the values of the other oscillation parameters in 100 statistically independent unoscillated data sets. For each of these data sets, 20 random choices of oscillation parameters are taken following the Gaussian distributions:
\begin{align}
 \Delta m^2_{21} = (7.4 \pm 0.2)\times 10^{-5} \,\text{eV}^2, ~ \Delta m^2_{32} = (2.46\pm 0.03)\times 10^{-3} \,\text{eV}^2\,,\nonumber \\
 \sin^2 2\theta_{12} = 0.855 \pm 0.020,~ \sin^2 2\theta_{13} = 0.0875 \pm 0.0026, ~ \sin^2\theta_{23} = 0.50\pm 0.03,  
\end{align}
which are in accordance with the present neutrino global fit results~\cite{Esteban:2018azc}. This way, the variations of the results over a large number (2000) of different combinations of values of the other oscillation parameters are effectively included.  Note that the value of the oscillation parameter to be determined is kept fixed at benchmark value as given in Table~\ref{tab:osc-param-value}. We keep $\delta_\text{CP}$ fixed at zero because its effect on $\nu_\mu$ and $\bar\nu_\mu$ survival probabilities is negligible in the multi-GeV energy range.  

We do not see any major changes in the measurement of $|\Delta m^2_{32}|$ caused by uncertainties in the other oscillation parameters, viz. $\theta_{12}$, $\theta_{23}$, $\theta_{13}$, and $\Delta m^2_{21}$. This is expected, since (a) the mixing angle $\theta_{23}$ does not change the dip-location of $\nu_\mu$ and $\bar\nu_\mu$ survival probabilities, (b) the mixing angle $\theta_{13}$ is already precisely measured, (c) the solar oscillation parameters $\Delta m^2_{21}$ and $\theta_{12}$ have negligible impact on $\nu_\mu$ and $\bar\nu_\mu$ survival probabilities in the multi-GeV energy range.

We also check the impact of systematic uncertainties on the result. For this, we incorporate the standard systematic uncertainties used in $\chi^2$ analyses of ICAL~\cite{Kumar:2017sdq}. These are $(i)$ 20$\%$ error in flux normalization, $(ii)$ 10$\%$ error in cross-section, $(iii)$ 5$\%$ error in energy dependence of flux, $(iv)$ 5$\%$ error in zenith angle dependence of flux, and $(v)$ 5$\%$ overall systematics. The event number in each ($E_\mu^\text{rec}$, $\cos\theta_\mu^\text{rec}$) bin is modified in each of the 2000 simulated data sets as 
\begin{equation}
 N = N^{0} (1+\delta_1)(1+\delta_2)(E_\mu^\text{rec}/E_0)^{\delta_3} (1 + \delta_4 \cos\theta_\mu^\text{rec}) (1+\delta_5)\,, 
\end{equation}
with $E_0 = 2$ GeV and $N^0$ as the theoretically predicted event number. The parameters $\delta_1$, $\delta_2$, $\delta_3$, $\delta_4$, $\delta_5$ are random numbers generated following Gaussian distributions with mean zero and the $1\sigma$ widths as $20\%$, $10\%$, $5\%$, $5\%$, $5\%$, respectively, which correspond to the systematics mentioned above.

After incorporating systematic uncertainties and varying oscillation parameters, we get the 90$\%$ C.L. allowed range for $|\Delta m^2_{32}|$ from $\mu^-$ events as (2.18 -- 2.79)$\times 10^{-3}$ eV$^2$, and from $\mu^+$ data as (2.22 -- 2.80)$\times10^{-3}$ eV$^2$, for $|\Delta m^2_{32}| \,(\text {true}) = 2.46\times 10^{-3}$ eV$^2$. These results are shown with light gray bands in upper panels of Fig.~\ref{fig:Calib-LbyE}. Note that  the change in the measurement of $|\Delta m^2_{32}|$ from $\mu^+$ events after inclusion of systematic uncertainties and errors in other oscillation parameters is quite small, whereas that from $\mu^-$ events is larger. This is because the dip in $\mu^+$ data is quite sharp, owing to the lower inelasticity in $\bar\nu_\mu$ CC interactions.

We can follow the same procedure to determine the value of the mixing
angle $\theta_{23}$ from the data. In principle, this is related to
the depth of the dip, and such a calibration could be obtained
using the 1000-year MC events. However, the statistical fluctuations
in the 500 kt$\cdot$yr simulated data are observed to give rise to large
uncertainties. On the other hand, it is found that the
total U/D ratio of all events with $L_\mu^{\rm rec}/E_\mu^{\rm rec}$ in the
range of 2.2 -- 4.1 gives a much better estimate of $\theta_{23}$.
In the lower panels of Fig.~\ref{fig:Calib-LbyE}, we show in blue 
the calibration for $\sin^2\theta_{23}$ with the measured total U/D ratio,
obtained using the 1000-year MC sample for several
$\sin^2\theta_{23} \,(\text {true})$ values. 
We also show the 90\% C.L. allowed values of $\sin^2\theta_{23}$ inferred using the 100 independent simulated data sets with 500 kt$\cdot$yr exposure each, keeping all the oscillation parameters fixed at benchmark values as given in Table \ref{tab:osc-param-value}, with dark gray bands.
The expected allowed range for $\sin^2\theta_{23}$ at $90\%$ C.L. 
is (0.44 -- 0.64) from $\mu^-$ events, and (0.39 -- 0.61) from $\mu^+$ events. With the systematic uncertainties and the variation of  oscillation parameters $|\Delta m^2_{32}|$, $\theta_{13}$, $\theta_{12}$, and $\Delta m^2_{21}$ as mentioned above, we get the 90$\%$ C.L. allowed range for $\sin^2\theta_{23}$ from $\mu^-$ events as (0.38 -- 0.70), and from $\mu^+$ data as (0.35 -- 0.65), for $\sin^2\theta_{23} \,(\text {true}) = 0.5$. These results are shown with light gray bands in lower panels of Fig.~\ref{fig:Calib-LbyE}.

Note that we do not claim or demand that our analysis gives better results
than those obtained with the complete $\chi^2$ analysis done for ICAL in
Refs.~\cite{Thakore:2013xqa,Devi:2014yaa}. The $\chi^2$ analyses
take into account the complete event spectra,
while our focus is on the identification of the dip, a feature that
would reconfirm the oscillation paradigm for neutrinos.
The observation that the inferred $|\Delta m^2_{32}|$ range is comparable to the
range expected from the complete $\chi^2$ analysis points to the
conclusion that most of the information about the $|\Delta m^2_{32}|$
is concentrated in the location of the U/D dip. 

\section{Oscillation valley in the $(E_\mu^{\rm rec}, \cos\theta_\mu^{\rm rec})$ plane}
\label{sec:2D_E-CT}

The oscillation dip discussed in the last section was a dip in the
U/D ratio as a function of $L_\mu^{\rm rec}/E_\mu^{\rm rec}$. The
dependence on $L_\mu^{\rm rec}/E_\mu^{\rm rec}$ was motivated from the
approximate form of the neutrino oscillation probability (with two
flavors in vacuum). However, if the detector can reconstruct both
$L_\mu$ (hence $\cos\theta_\mu$) and $E_\mu$ accurately,
then one can go a step ahead and look at the distribution of the U/D ratio
in the $(E_\mu, \cos\theta_{\mu})$ plane. We perform such an analysis
in this section, and find that such a distribution can have many
interesting features with physical significance, which may be
identifiable with sufficient data. In particular, we point out
an ``oscillation valley'' corresponding to the dark diagonal band in Fig.~\ref{fig:oscillogram}, whose nature can provide stronger tests for the
oscillation hypothesis, albeit only with a large amount of data.
On the other hand, we show that the identification of the oscillation valley,
and the determination of $|\Delta m^2_{32}|$ based on its ``alignment'',
is possible with 500 kt$\cdot$yr of simulated data at ICAL.

\subsection{Events and U/D ratio using 1000-year Monte Carlo simulation}
\label{sec:2D-dist-1000yr}

\begin{figure}[htb!]
  \centering
  \includegraphics[width=0.49\textwidth]{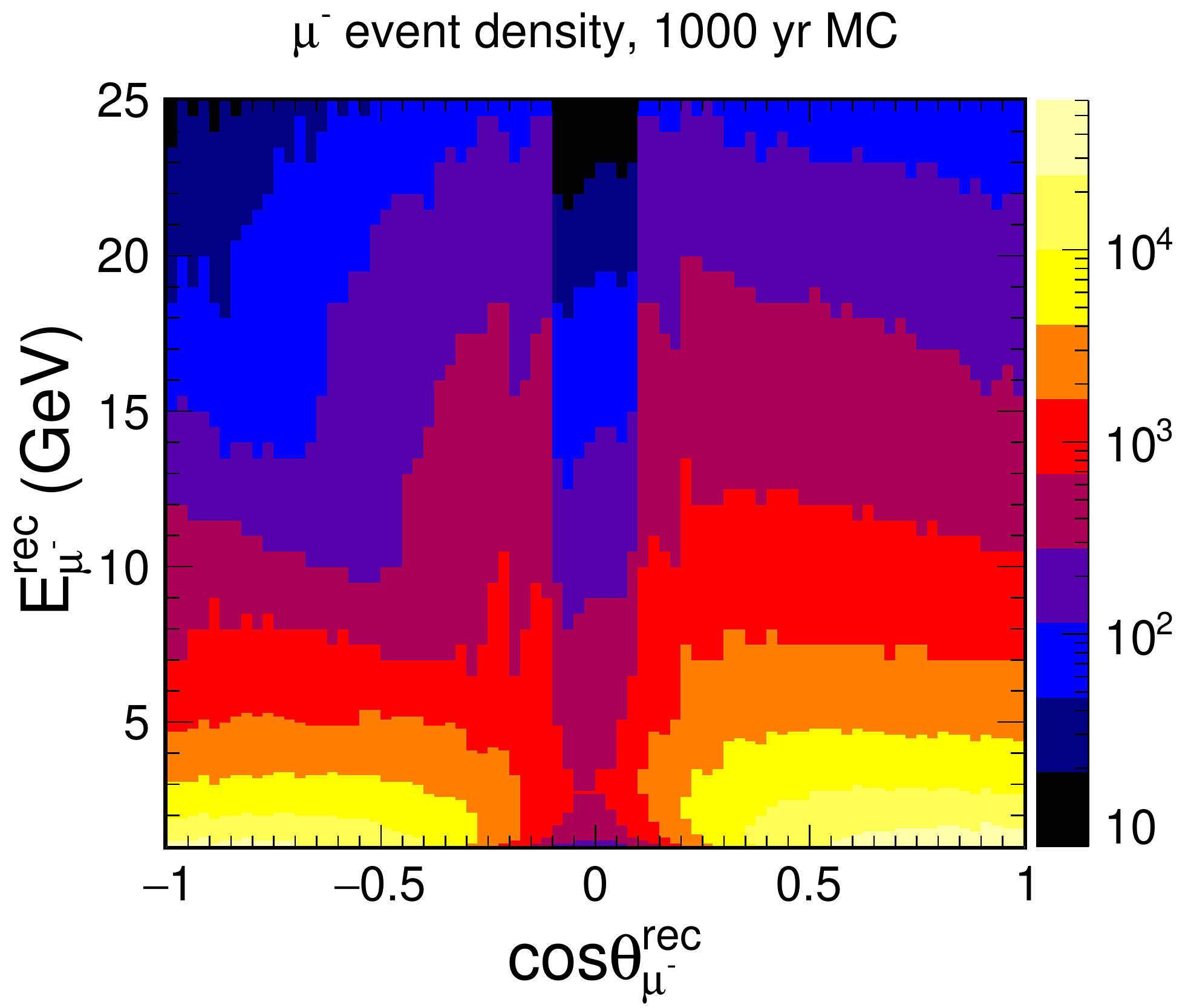}
  \includegraphics[width=0.49\textwidth]{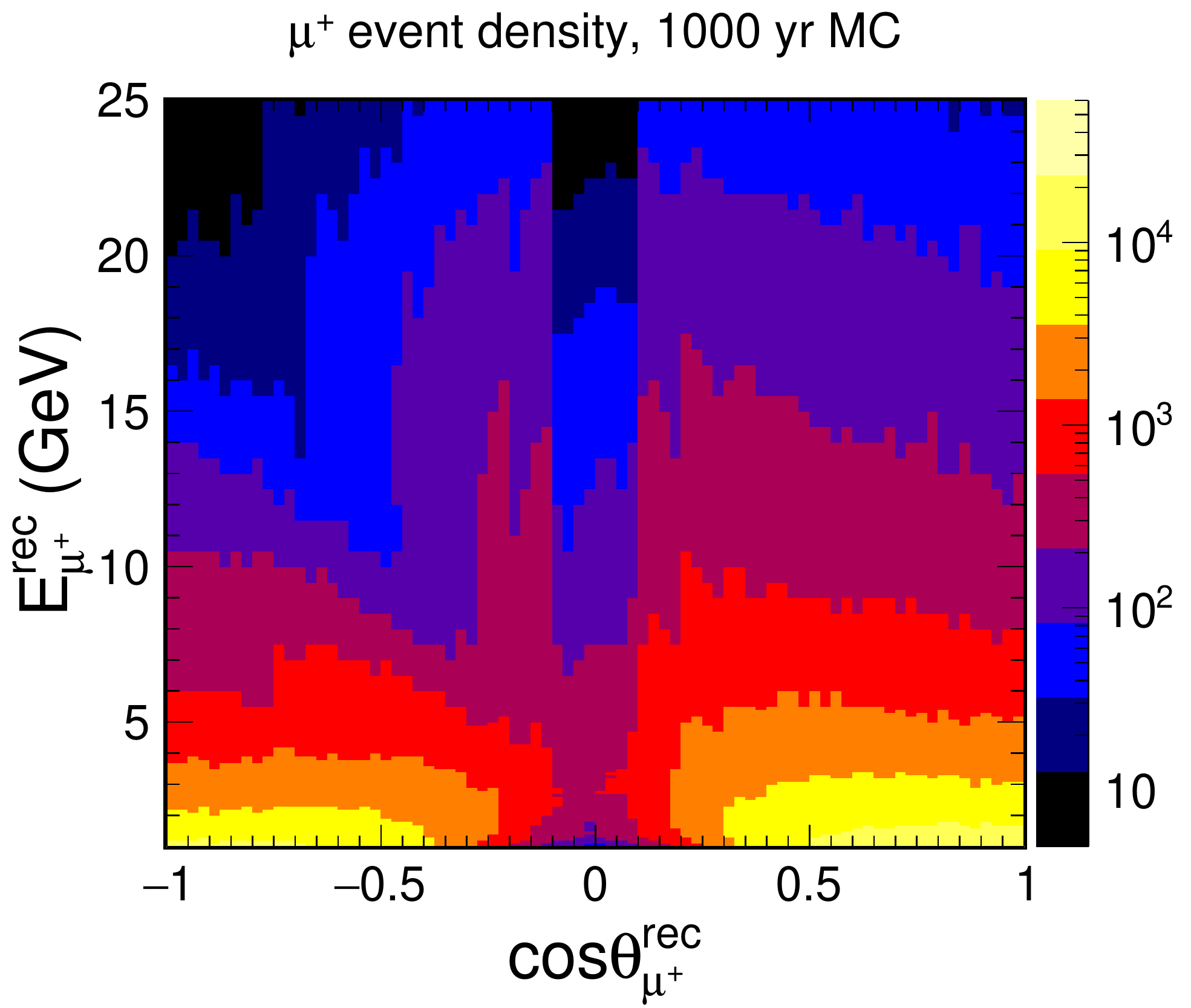}
  \includegraphics[width=0.49\textwidth]{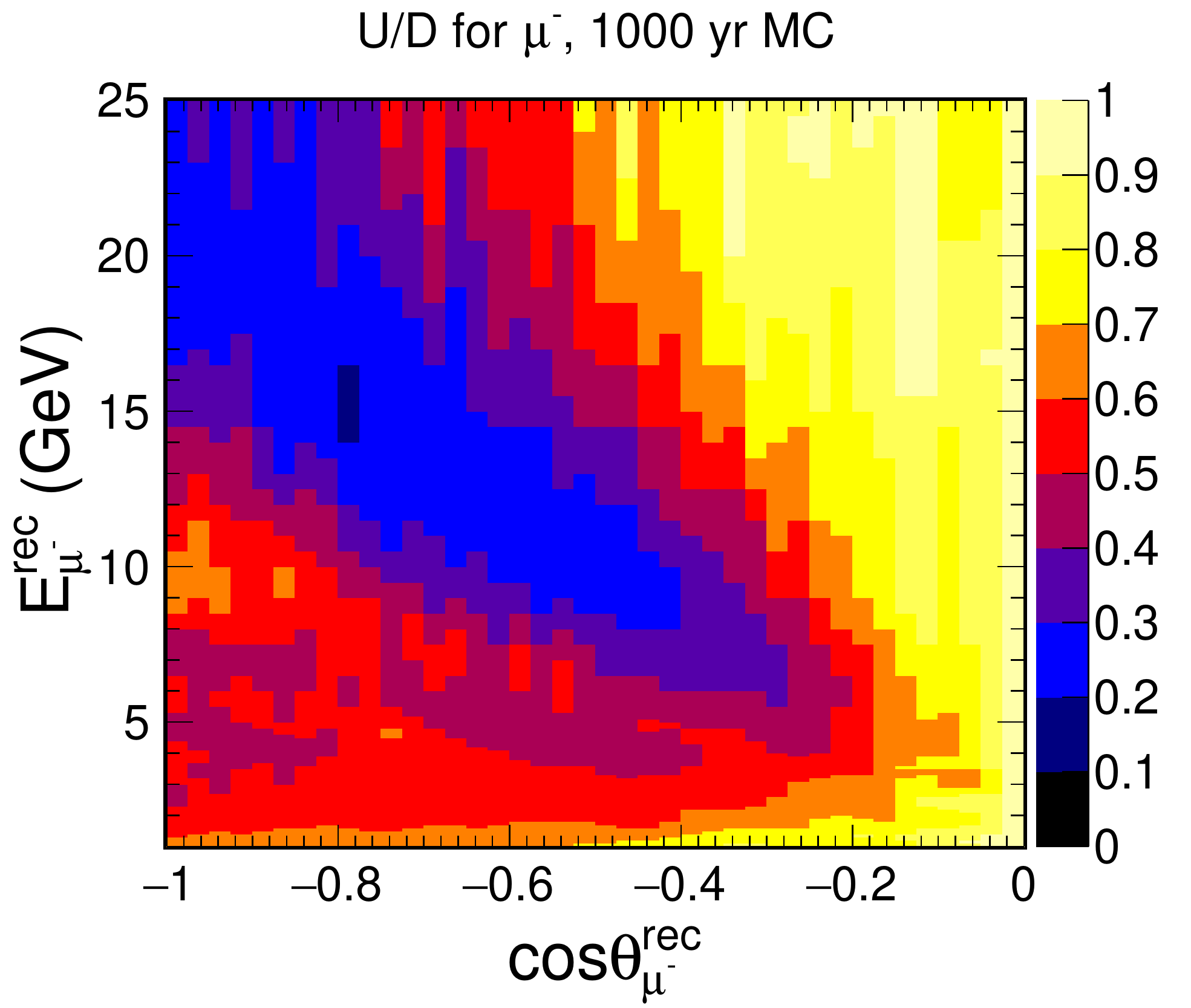}
  \includegraphics[width=0.49\textwidth]{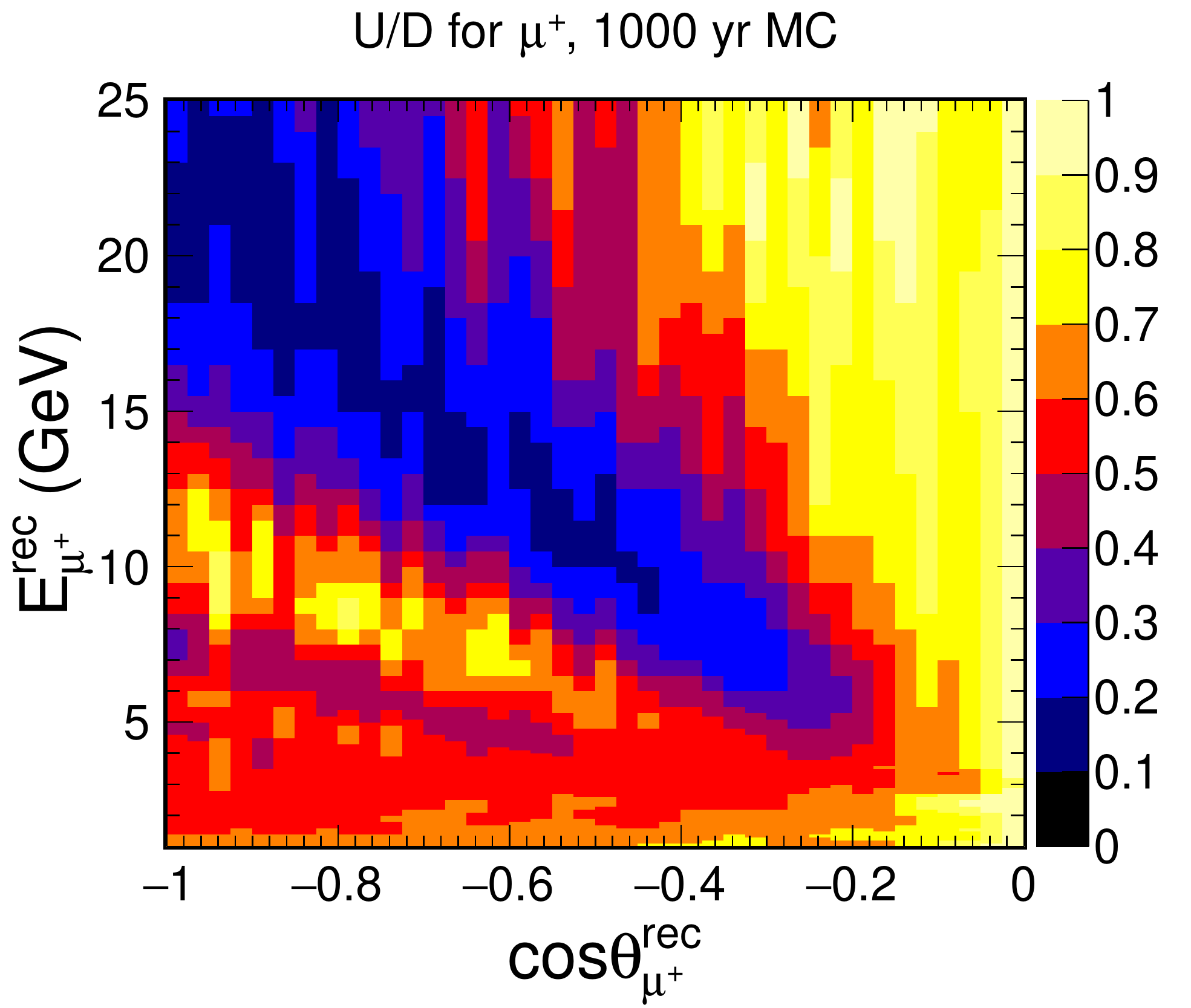}
  \mycaption{The distribution of event density (upper panels) and
    U/D ratio (lower panels), for $\mu^{-}$ (left panels) and
    $\mu^{+}$ (right panels), in the ($E_\mu^{\rm rec},\cos\theta_\mu^{\rm rec}$)
    plane, obtained using the 1000-year MC sample. We use the oscillation parameters given in Table~\ref{tab:osc-param-value}. Note that the event density in the
    upper panels has units of [GeV$^{-1}$sr$^{-1}$], and
    is obtained by dividing the number of events in each bin
    by ($2\pi~\times$ the bin area), \ie by
    ($2\pi\Delta E_{\mu}^{\rm rec} \Delta \cos\theta_{\mu}^{\rm rec}$).
    Here $\Delta E_{\mu}^{\rm rec}$ and  $\Delta \cos\theta_{\mu}^{\rm rec}$
    are the height and the width of the bin, respectively. Note that the U/D($E_\mu^\text{rec}, \cos\theta_\mu^\text{rec}$) is defined only for $\cos\theta_\mu^\text{rec} < 0$ (see Eq.~\ref{eq:U/D_def}).
  }
  \label{fig:2D-1000yr}
\end{figure}

In this section, we discuss the distributions of events and the U/D ratio,
in the plane of reconstructed  energy $E_{\mu}^{\rm rec}$ and zenith angle $\cos\theta_{\mu}^{\rm rec}$ of the $\mu^-$ and $\mu^+$ events,
for a 1000-year MC sample.
The main reason for considering such a huge exposure here is not to miss
those features, which survive in spite of our not using $E_\nu$ and
$\cos\theta_\nu$ directly, but which may disappear due to the limitation
of statistics.

The quantities shown in Fig.~\ref{fig:2D-1000yr} are binned based on the
reconstructed values of $E_{\mu}^{\rm rec}$ and $\cos\theta_{\mu}^{\rm rec}$.
The $\cos\theta_\mu^{\rm rec}$ range of -1.0 to 1.0 has been divided into
80 bins of equal width.
We have a total of 84 bins in $E_{\mu}^{\rm rec}$ in the range of 1 GeV to 25 GeV.
The bin width is not uniform in $E_\mu^{\rm rec}$, since at higher energies the number of events decreases rapidly. The first 45 $E_\mu^{\rm rec}$ bins,
in the range of 1.0 -- 5.5 GeV, have a width of 0.1 GeV each, and
the last 39 bins, in the range of 5.5 -- 25 GeV, have a width of 0.5 GeV each.
The upper panels of Fig.~\ref{fig:2D-1000yr}
present the distributions of $\mu^-$ (left panel) and $\mu^+$ (right panel)
event density.         
The abrupt lowering of the event density for
$-0.2<\cos\theta_{\mu}^{\rm rec}<0.2$ is due to the poor reconstruction
efficiency of the ICAL detector for the events coming near the
horizontal direction.
Note that this feature was absent in Fig.~\ref{fig:LbyE-1000}
  since it was distributed over a large $L_\mu^{\rm rec}/E_\mu^{\rm rec}$ range.

In lower panels of Fig.~\ref{fig:2D-1000yr}, we show the U/D ratio in each bin,
in the ($E_\mu^{\rm rec}, \cos\theta_\mu^{\rm rec}$) plane.  Note that the U/D($E_\mu^\text{rec}, \cos\theta_\mu^\text{rec}$) is defined only for $\cos\theta_\mu^\text{rec} < 0$ (See Eq.~\ref{eq:U/D_def}). This also makes the features
in Fig.~\ref{fig:2D-1000yr} resemble those in Fig.~\ref{fig:oscillogram},
thus enabling their understanding in terms of the survival probabilities
shown in the oscillograms. The lower panels may be observed to have many interesting features.
The most prominent one is of course the light/ dark blue band,
corresponding to U/D $\lesssim 0.25$, extending diagonally
from $(E_\mu^{\rm rec}, \cos\theta_{\mu}^{\rm rec}) \approx (5 ~{\rm GeV},-0.3)$
to $(E_\mu^{\rm rec}, \cos\theta_{\mu}^{\rm rec}) \approx (25 ~{\rm GeV},-1.0)$.
This is the band with the lowest values of U/D ratio in this plane,
and we shall henceforth refer to it as the ``oscillation valley''.
Clearly, this valley is deeper for $\mu^+$, since the number of dark blue
bins with U/D $< 0.15$ is seen to be larger for $\mu^+$. This is
consistent with the observation of a deeper oscillation dip in $\mu^+$,
as discussed in Sec.~\ref{sec:LbyE-1000yrs}. The main reason for this is the
smaller inelasticity of antineutrino CC events. 
The differences observed between the left and right panel may be attributed to the Earth matter effects, which affect the survival probabilities of neutrinos and antineutrinos differently for a given neutrino mass ordering.

\subsection{Events and U/D ratio using 10-year simulated data}
\label{sec:2D-dist-10yr}

\begin{figure}[htb!]
  \centering
  \includegraphics[width=0.49\textwidth]{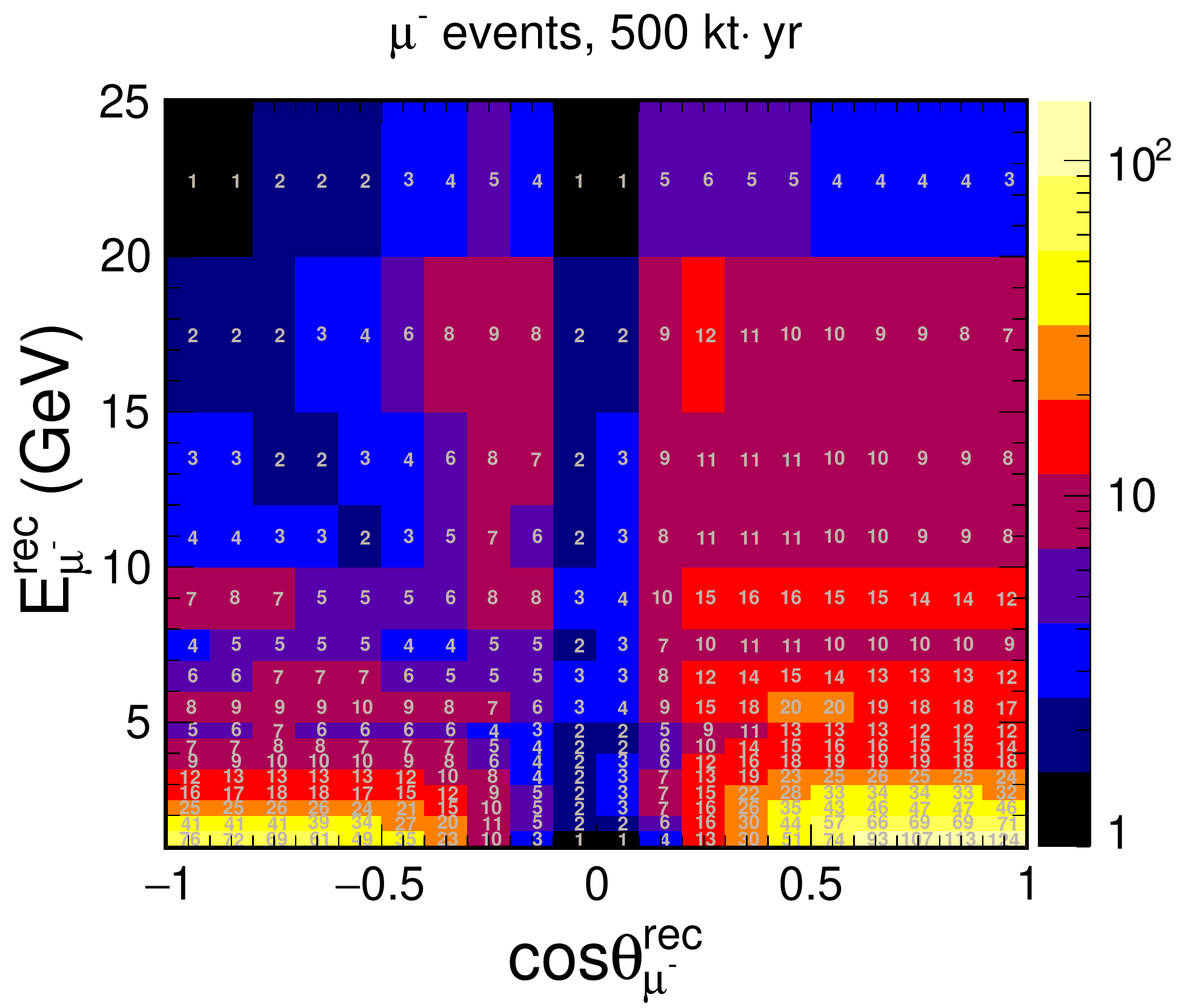}
  \includegraphics[width=0.49\textwidth]{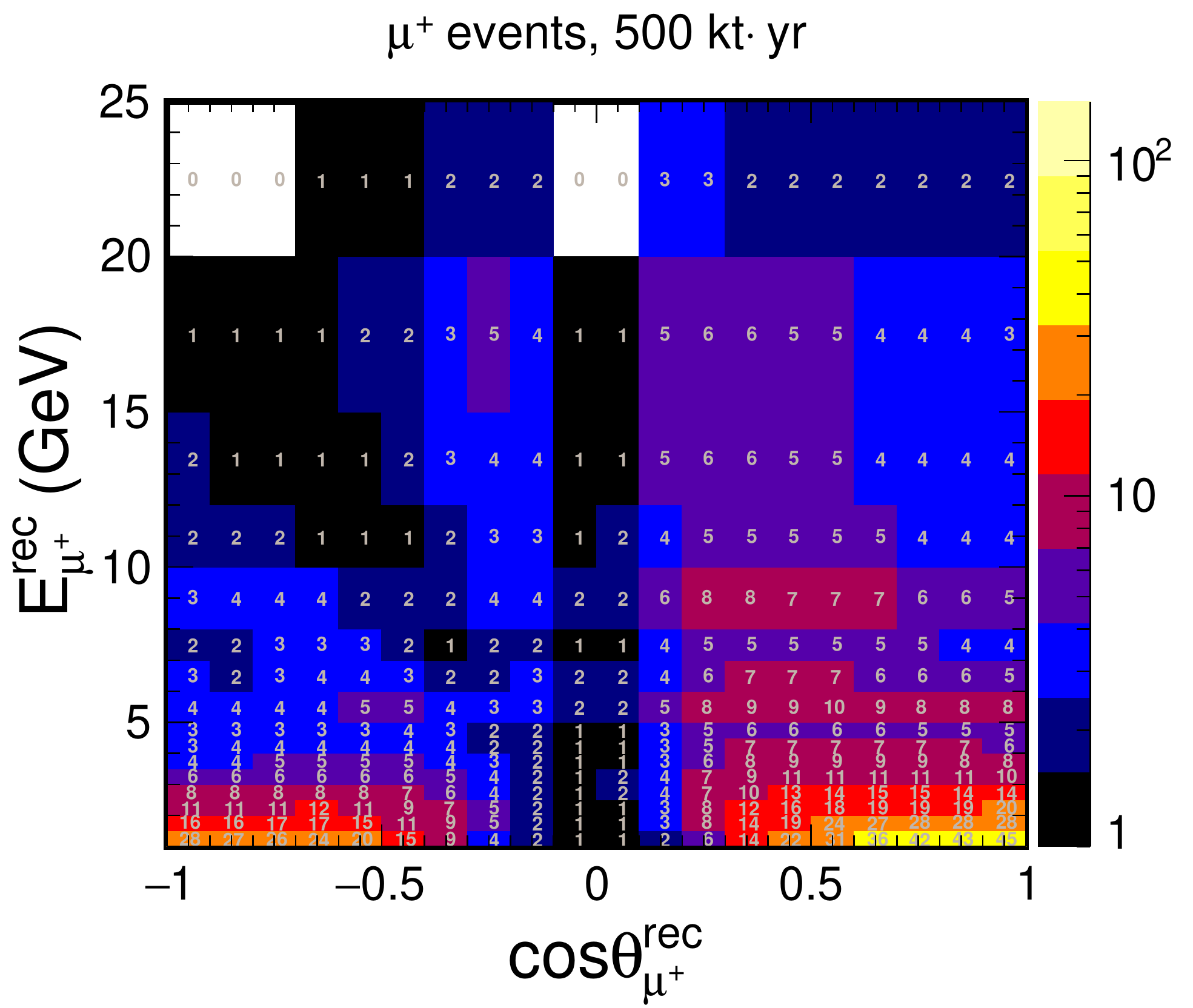}
  \includegraphics[width=0.49\textwidth]{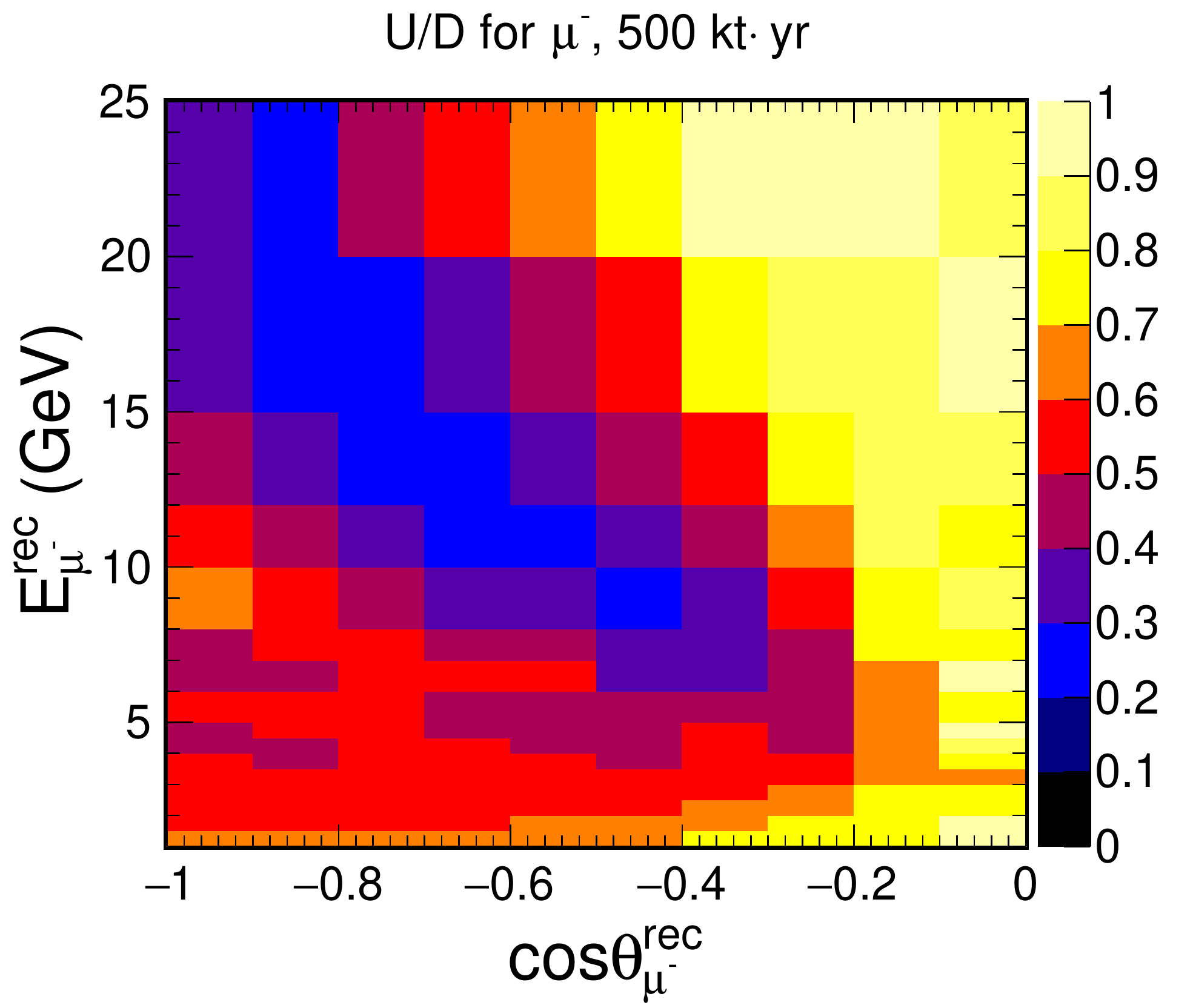}
  \includegraphics[width=0.49\textwidth]{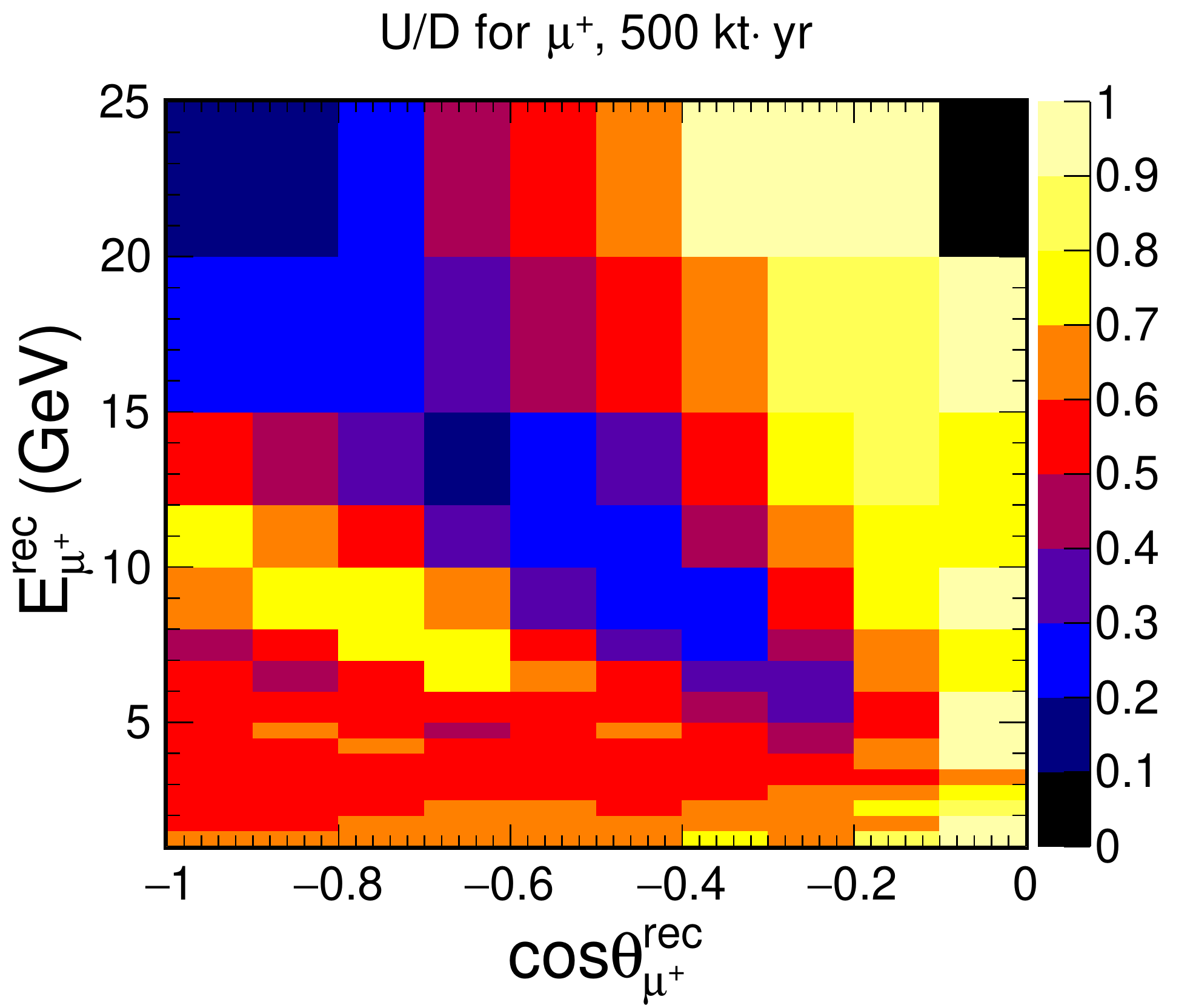}
  \mycaption{The distribution of mean number of events (upper panels)
    and mean U/D ratio (lower panels) for $\mu^{-}$ (left panels) and
    $\mu^{+}$ (right panels) in the plane of $\cos\theta_{\mu}^{\rm rec}$
    and $E_{\mu}^{\rm rec}$. The mean is calculated from 100 independent
    data sets of exposure 500 kt$\cdot$yr each of ICAL.
    The number of events written in white are rounded to the
    nearest integer. 
   We use the oscillation parameters given in Table~\ref{tab:osc-param-value}.  Note that the U/D($E_\mu^\text{rec},  \cos\theta_\mu^\text{rec}$) is defined only for $\cos\theta_\mu^\text{rec} < 0$ (see Eq. \ref{eq:U/D_def}).}
	\label{fig:2D-10}
\end{figure}

In the upper panels of Fig.~\ref{fig:2D-10}, we show the expected event
distributions of $\mu^-$ (left panel) and $\mu^+$ (right panel) events,
for 500 kt$\cdot$yr exposure of ICAL, in the plane of reconstructed muon energy
($E_{\mu}^{\rm rec}$) and reconstructed muon
zenith angle ($\cos\theta_{\mu}^{\rm rec}$).
The $\cos\theta_{\mu}^{\rm rec}$ range of -1.0 to 1.0 is divided into
20 uniform bins of width 0.1 each. 
The binning in $E_{\mu}^{\rm rec}$ is non-uniform, and is such that
there are at least a few number of events in each bin (except in the largest
energy bins, where this may not be possible due to the smaller neutrino
flux at higher energies). 
The $E_{\mu}^{\rm rec}$ range of 1 -- 25 GeV has been divided into 16 non-uniform bins, as given in Table \ref{tab:binning-2D-10years}. 
\begin{table}[htb!]
\centering
 \begin{tabular}{|c|c|c|c c|}
  \hline
  Observable & Range & Bin width & \multicolumn{2}{c|}{Number of bins} \\
  \hline 
  \multirow{4}{*}{ $E_\mu^{\rm rec}$ } & [1, 5] &  0.5 & 8 & \rdelim\}{5}{7mm}[16] \cr 
   & [5, 8] & 1 & 3  & \cr
 \multirow{2}{*}{ (GeV) }  & [8, 12] & 2 & 2  & \cr
  & [12, 15] & 3 & 1  & \cr
  & [15, 25] & 5 & 2  &\cr
  \hline 
 $\cos\theta_\mu^{\rm rec}$ & [-1.0, 1.0] & 0.1 & 20  & \cr
  \hline
 \end{tabular}
\caption{The binning scheme considered for $E_\mu^{\rm rec}$ and $\cos\theta_\mu^{\rm rec}$ for $\mu^-$ and $\mu^+$ events of 10-year simulated data while we show event distribution and U/D plots in the ($E_\mu^\text{rec}$, $\cos\theta_\mu^{\text{rec}}$) plane. This binning scheme is used in analysis of oscillation valley as well. }
\label{tab:binning-2D-10years}
\end{table}

In order to take care of the statistical fluctuations that will clearly be
significant with 500 kt$\cdot$yr data, we generate 100 independent data sets,
each with this exposure. In the upper panels of Fig.~\ref{fig:2D-10},
we show the mean number of events obtained from these 100 data sets. 
Similarly, in the lower panels of Fig.~\ref{fig:2D-10}, we calculate the
U/D ratios in each bin for the above 100 sets, and present their mean values in
the figure, using appropriate colors. Note that the U/D($E_\mu^\text{rec}, \cos\theta_\mu^\text{rec}$) is defined only for $\cos\theta_\mu^\text{rec} < 0$ (see Eq. \ref{eq:U/D_def}).

Clearly, while many of the features corresponding to matter effects
  seem to get diluted due to the coarser bins and statistical fluctuations,
the oscillation valley along the diagonal survives.
In the following sections, we shall provide a procedure for identifying
this oscillation valley and extracting information about the
oscillation parameters from it.

\subsection{Identifying the valley with 10-year simulated data}
\label{sec:2D-fit-procedure}
\begin{figure}[htb!]
  \centering
  \includegraphics[width=0.49\textwidth]{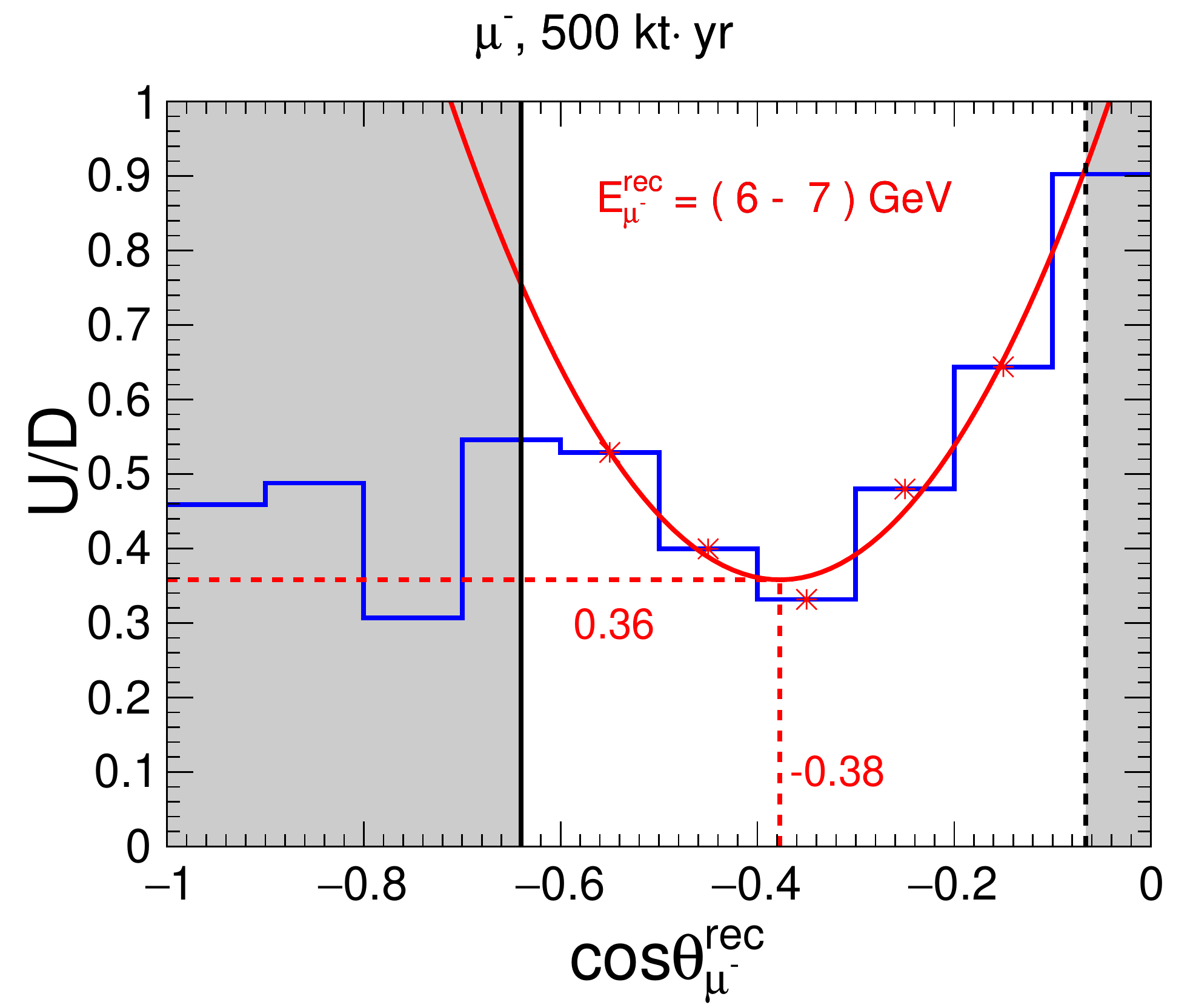}
  \includegraphics[width=0.49\textwidth]{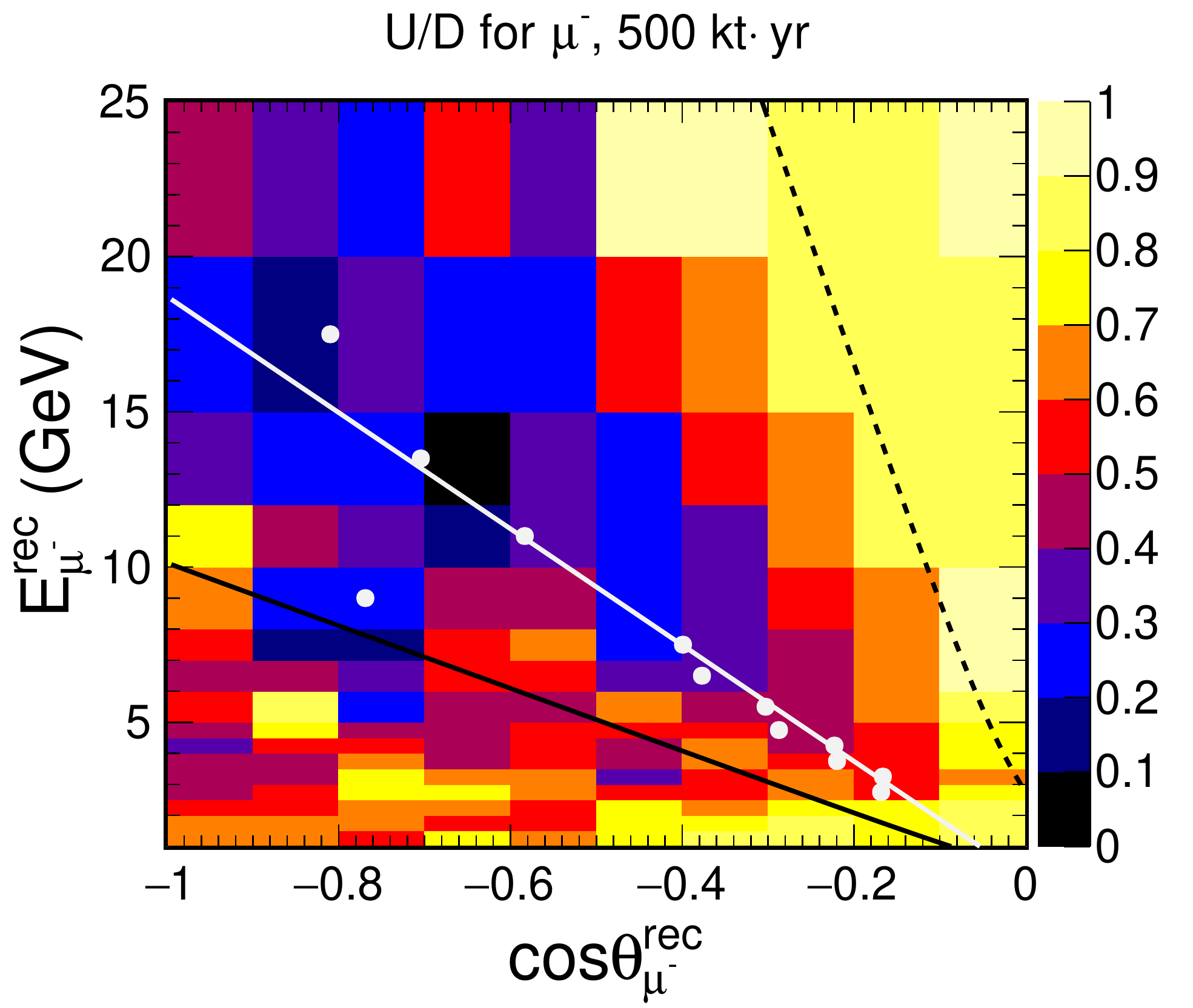}
  \mycaption{Demonstration of the algorithm for identifying the
    valley and determining its alignment.
    Left panel: The blue line represents the U/D ratio as a function of
    ($\cos\theta_\mu^{\rm rec}$) for a fixed energy bin. The gray areas
    are those with $\log_{10} [L_\mu^{\text{rec}}/E_\mu^{\text{rec}}] < 2.2$
    and $> 3.1$. The red curve represents the parabolic fit to the
    U/D ratios with
    $2.2 < \log_{10} [L_\mu^{\text{rec}}/E_\mu^{\text{rec}}] < 3.1$. 
    Right panel: The colored background shows the U/D ratios for each bin in
    the $(E_\mu^{\rm rec}, \cos\theta_\mu^{\rm rec})$ plane, for
    a particular 500 kt$\cdot$yr data set.
    The black dashed and solid lines show the cuts
    corresponding to $\log_{10} [L_\mu^{\text{rec}}/E_\mu^{\text{rec}}]= 2.2$
    and $3.1$, respectively.
    White points represent the locations of minima for
    the fitted parabolas corresponding to each horizontal energy ``strip''.
    The white line is a fit to the white points with appropriate weights
    as described in the text.  Note that the U/D($E_\mu^\text{rec}, \cos\theta_\mu^\text{rec}$) is defined only for $\cos\theta_\mu^\text{rec} < 0$ (see Eq. \ref{eq:U/D_def}).
  }
  \label{fig:2D-fit-procedure}	
\end{figure}

We now describe the methodology adopted for the identification of the
oscillation valley in the $(E_\mu^{\rm rec}, \cos\theta_\mu^{\rm rec})$ plane,
and the determination of its alignment.
To this end, we use the binning described in sec.~\ref{sec:2D-dist-10yr}
for the 500 kt$\cdot$yr data set.

\begin{itemize}
  
\item Since the events with $\log_{10} [L_\mu^{\text{rec}}/E_\mu^{\text{rec}}] > 3.1 $ are susceptible to significant matter effect and rapid oscillations,
the information from these bins cannot be directly connected to the dip corresponding to the vacuum oscillations. As far as the region $\log_{10} [L_\mu^{\text{rec}}/E_\mu^{\text{rec}}] < 2.2$ is concerned, the oscillation are not developed yet.
Therefore, we only choose the region
  $2.2 < \log_{10} [L_\mu^{\text{rec}}/E_\mu^{\text{rec}}] < 3.1$ for our analysis,
  where vacuum oscillations are expected to dominate, and to be resolvable.

\item We then analyze each horizontal ``energy strip'' in this region, \ie
  the U/D ratio for all $\cos\theta_\mu^{\rm rec}$,
  as shown in the left panel of Fig.~\ref{fig:2D-fit-procedure}. 
  We perform a parabolic fit to the points in this strip that are in the
  allowed region as specified above.
  The equation of the parabola to be fitted is $y = y_E + a_E (x-x_E)^2$,
  where $y$ is the U/D ratio while $x$ is the value of
  $\cos\theta_\mu^{\rm rec}$.
  The parameters $(x_E,y_E)$ correspond to coordinates ($\cos\theta_\mu^\text{rec}$, U/D) at the bottom of the parabola,
  while the parameter $a_E$ describes the ``width'' of the parabola.
  Higher the value of $a_E$, more prominent the dip in that energy strip.

\item Once the parabolic fits in all energy strips are obtained,
  all the points ($E_\mu^\text{rec} = E,~\cos\theta_\mu^\text{rec} = x_E$) are plotted on the
  $(E_\mu^{\rm rec}, \cos\theta_\mu^{\rm rec})$ plane,
  as shown in the right panel of Fig.~\ref{fig:2D-fit-procedure}. 
  A straight line, passing through the origin $(E_\mu^\text{rec}=0,~\cos\theta_\mu^\text{rec} = 0)$, is fitted to
  these points. For this fitting, each of the points is given a weight
  of $\sqrt{a_E N_E}$, where $N_E$ is the number of downward-going
  events  in the analysis region (between the solid and dashed black lines in the right panel Fig.~\ref{fig:2D-fit-procedure}) of the energy strip. The points lying outside the analysis region are not taken into account in the fit. 
  The best-fit line is thus obtained of the form $y = m\cdot x$, where $y$ represents
  $E_\mu^{\rm rec}$, and $x$ represents $\cos\theta_\mu^{\rm rec}$.
  The slope $m$ of this line quantifies the ``alignment'' of the
  valley. 
\end{itemize}

Note that the above procedure is motivated by our expectation that
the valley would correspond to the value of $L_\mu^{\rm rec}/E_\mu^{\rm rec}$
which would correspond to the dip in the one-dimensional $L/E$ analysis 
described in Sec.~\ref{sec:1dL/Efitting}.
Our analysis is rather simple, focused only on the identification of the valley and on determining its alignment.
With more data and a refined analysis, one maybe
able to use the goodness of the
fit to the line, or the intercepts of the line on either axes,
for testing the oscillation framework in more detail.
We show here that 500 kt$\cdot$yr of exposure would be enough
to locate the valley and quantify its alignment even with our simplified treatment.
In the following section, we shall also show that the value of
$|\Delta m^2_{32}|$ may be calibrated using the alignment of the valley.

\subsection{$|\Delta m^2_{32}|$ from the alignment of the valley}
\label{sec:Calib-valley}

\begin{figure}[htb!]
  \centering
  \includegraphics[width=0.49\textwidth]{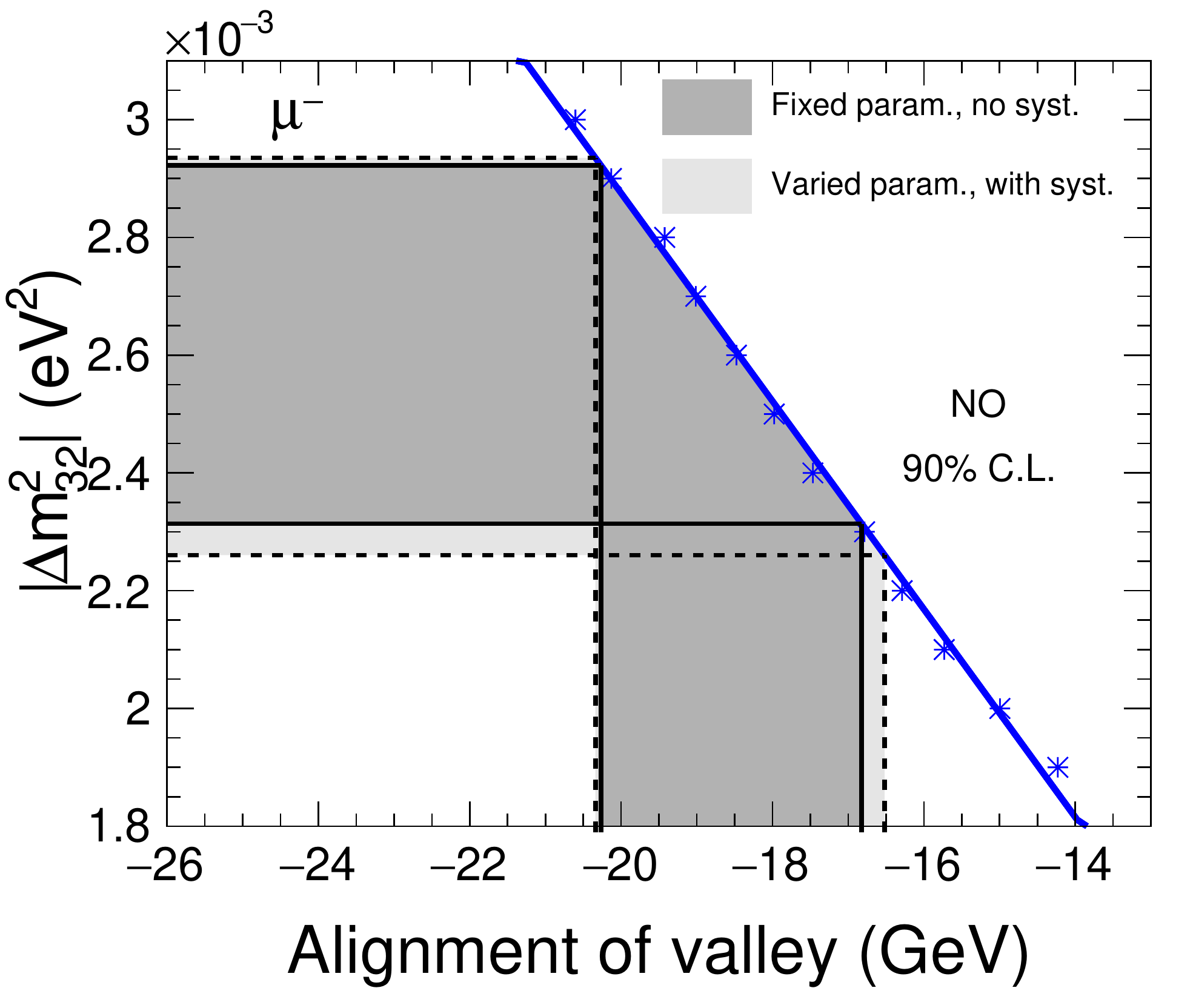}
  \includegraphics[width=0.49\textwidth]{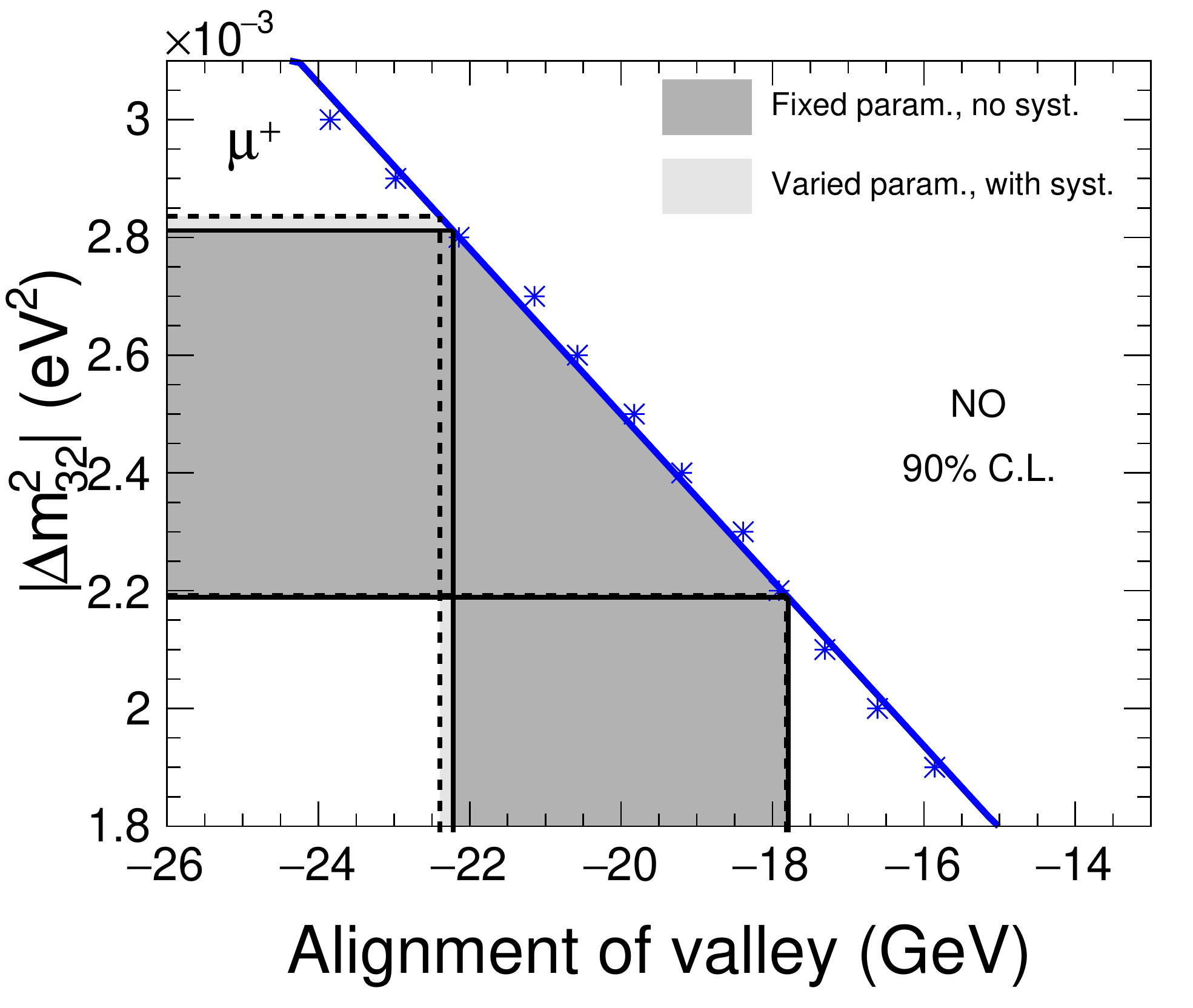}
  \mycaption{The blue points and the blue line correspond to
    the calibration of actual $|\Delta m^2_{32}|$ with the alignment of
    the valley, obtained by using the 1000-year MC data sample. The light (dark) gray bands represent the 90\% C.L. regions of the alignment
    of the valley, and hence the inferred $|\Delta m^2_{32}|$ through
    calibration, with (without) systematic uncertainties and error in other oscillation parameters. Here, we take $|\Delta m^2_{32}| \,(\text {true}) = 2.46\times 10^{-3}$ eV$^2$ and 10 years exposure of ICAL.  The negative values of the alignments of the valley correspond to the negative slope of the best-fit line in the right panel of Fig.~\ref{fig:2D-fit-procedure}. The results obtained  from $\mu^-$ and $\mu^+$ events are shown in left and right panels, respectively. For the fixed-parameter analysis, we use the benchmark oscillation parameters given in Table~\ref{tab:osc-param-value}, while for the inclusion of systematic uncertainties and variation of oscillation parameters, we use the procedure described in Sec.~\ref{sec:Calib-LbyE}. 
    }
  \label{fig:Calib-2D}
\end{figure}

For the same reasons described in Sec.~\ref{sec:Calib-LbyE},
we expect the alignment of the valley, as determined through the
procedure outlined in the previous section, to lead to the value of $|\Delta m^2_{32}|$.
Indeed, for upward-going events, $L_\mu^{\rm rec}$ is approximately
proportional to $\cos \theta_\mu^{\rm rec}$, so that the alignment
of the valley, \ie $E_\mu^{\rm rec} (\text{GeV})/\cos\theta_\mu^{\rm rec}$,
is approximately  $(5.08/\pi)\,|\Delta m^2_{32}| (\text{eV}^2)\, R(\text{km})$.

We calibrate the actual value of $|\Delta m^2_{32}|$ with the alignment
of the valley using the same procedure as in Sec.~\ref{sec:Calib-LbyE}. The 1000-year MC samples, with different true values of
$|\Delta m^2_{32}|$, are used to locate the blue calibration points,
and hence the blue calibration curve, in Fig.~\ref{fig:Calib-2D}. The statistical fluctuations with 500 kt$\cdot$yr data are taken into account 
by generating 100 independent data sets. The dark gray bands in Fig.~\ref{fig:Calib-2D} present the 90\% C.L. allowed values for the alignment of
the valley, and hence the 90\% C.L. allowed values of the calibrated
$|\Delta m^2_{32}|$, with all the oscillation parameters  fixed at benchmark values as given in Table.~\ref{tab:osc-param-value}. 
The figure indicates that the 90$\%$ C.L. allowed range for
$|\Delta m^2_{32}|$ from $\mu^-$ events is (2.31 -- 2.92)$\times 10^{-3}$ eV$^2$,
while that from $\mu^+$ data is (2.19 -- 2.81)$\times10^{-3}$ eV$^2$, for $|\Delta m^2_{32}| \,(\text {true}) = 2.46\times 10^{-3}$ eV$^2$.  
After incorporating the systematics uncertainties and varying the oscillation parameters $\theta_{12}$, $\theta_{23}$, $\theta_{13}$, and $\Delta m^2_{21}$ as discussed in Sec.~\ref{sec:Calib-LbyE}, we get the 90$\%$ C.L. allowed range for $|\Delta m^2_{32}|$ from $\mu^-$ events as (2.26 -- 2.94)$\times 10^{-3}$ eV$^2$, and from $\mu^+$ data as (2.19 -- 2.84)$\times10^{-3}$ eV$^2$. These results are shown with light gray bands in  Fig.~\ref{fig:Calib-2D}. 

So far, our analysis has been based upon 500 kt$\cdot$yr of simulated data. However, ICAL should be able to identify the oscillation dip/oscillation valley even with half the exposure, and may provide some early measurement of $|\Delta m^2_{32}|$ with data in the multi-GeV range. We show some results with 250 kt$\cdot$yr exposure in Appendix \ref{app:5yr}.

\section{Summary and Concluding Remarks}
\label{sec:conclusions}

Atmospheric neutrino experiments access a large range of energies
and baselines for neutrino oscillations, hence they are capable of 
testing basic features of the neutrino flavor conversion framework
over a large parameter space. In this work, 
we study the potential of the ICAL detector to visualize the $L/E$
dependence of survival probabilities in neutrino and antineutrino
channels separately, by distinguishing $\mu^-$ and $\mu^+$ events.
For this, we use the reconstructed
muon energy and zenith angle, as opposed to the inferred neutrino
energy and zenith angle. Further, we consider the ratio of upward-going
(U) and downward-going (D) events, instead of the ratio of observed
oscillated events and simulated unoscillated events. 
The reconstructed $L_\mu^{\rm rec}/E_\mu^{\rm rec}$ distributions of the U/D ratio for 1000-yr MC indicate that the feature of the first oscillation
dip in this ratio in muon neutrinos (and antineutrinos) is still preserved
in the detected $\mu^-$ and $\mu^+$.
We demonstrate that this first oscillation dip can be clearly identified
at ICAL with an exposure of 500 kt$\cdot$yr, \ie 10 years running of the 50 kt ICAL.

We develop a novel dip-identification algorithm to identify the
oscillation dip and measure the value of $|\Delta m^2_{32}|$,
using the information from the first oscillation dip in the
$L_\mu^{\rm rec}/E_\mu^{\rm rec}$ distributions of the U/D ratio, in
$\mu^-$ and $\mu^+$ events separately.
This algorithm finds a contiguous set of $L_\mu^{\rm rec}/E_\mu^{\rm rec}$
bins with the lowest U/D values, and uses the combined information
in all of them to determine the dip location precisely.
The value of $|\Delta m^2_{32}|$ is then calibrated against the location
of the dip using the 1000-yr MC, and the statistical uncertainties expected
with the 10-year data are estimated using simulations of 100 independent
data sets. The expected 90\% C.L. ranges of $|\Delta m^2_{32}|$ obtained after taking into account the systematic uncertainties and varying the other oscillation parameters over their currently allowed ranges, are summarized in Table~\ref{tab:result:summary}.   
We also calibrate $\theta_{23}$, however, here simply the ratio of the
total number of U and D events is found to be a better choice to calibrate
against, rather than the depth of the dip.
We find that, when the true value of $\sin^2\theta_{23}$
is 0.50, the U/D ratio would determine the 90\% C.L.
range of its value to be (0.38 -- 0.70) with $\mu^-$ events,
and (0.35 -- 0.65) with $\mu^+$ events, taking into account systematic uncertainties and errors in the other oscillation parameters.

\begin{table}[htb!]
	\begin{center}
		\begin{tabular}{|c|c|c|}
			\hline
			\multirow{2}{*}{Reconstructed observables} &
			\multicolumn{2}{c|}{$90\%$ C.L. range}   \\
			\cline{2-3}
			& $\mu^-$ & $\mu^+$ \\
			\hline
			& & \cr
			$L_\mu^{\rm rec}/E_\mu^{\rm rec}$ & $(2.18~{\rm -}~2.79)\times 10^{-3}$ eV$^2$
			& $(2.22 ~{\rm -}~ 2.80)\times10^{-3}$ eV$^2$ \cr
			& &  \cr
			\hline
			\hline
			& & \cr
			$(E^{\rm rec}_\mu$, $\cos\theta^{\rm rec}_\mu)$   & $(2.26 ~{\rm -}~ 2.94)\times 10^{-3}$ eV$^2$ & $(2.19 ~{\rm -}~ 2.84)\times 10^{-3}$ eV$^2$  \cr
			& & \cr
			\hline
		\end{tabular}
		\caption{The expected $90\%$ C.L. allowed ranges for
			$|\Delta m^2_{32}|$, from the analyses using $L_\mu^{\rm rec}/E_\mu^{\rm rec}$
			and $(E_\mu^{\rm rec}, \, \cos\theta_\mu^{\rm rec})$ distributions of the
			U/D ratio, with 500 kt$\cdot$yr exposure. The true value of
			$|\Delta m^2_{32}|$ is taken to be $2.46 \times 10^{-3}$ eV$^2$, and the other oscillation parameters $\theta_{12}$, $\theta_{23}$, $\theta_{13}$, and $\Delta m^2_{21}$ are varied over their currently allowed ranges, and systematics are taken into account.}
		\label{tab:result:summary}
	\end{center}
\end{table}

In this paper, we point out for the first time that the identification
of an ``oscillation valley'' feature is possible, in the distribution
of the U/D ratio in the plane of
$(E_\mu^{\rm rec}, \, \cos \theta_\mu^{\rm rec}$).
The oscillation dip observed in the one-dimensional
$L_\mu^{\rm rec}/E_\mu^{\rm rec}$ analysis
above now appears as an oscillation valley in the two-dimensional
$(E_\mu^{\rm rec},\cos\theta_\mu^{\rm rec}$) plane.
We go on to develop an algorithm for finding the alignment of the valley,
\ie the slope of the best-fit line to the valley 
in the $(E_\mu^{\rm rec},\cos\theta_\mu^{\rm rec}$) plane, and show that it
serves as a good proxy for $|\Delta m^2_{32}|$.
As in the oscillation dip analysis, we calibrate $|\Delta m^2_{32}|$
against the alignment of the valley using 1000-yr MC data,
and estimate the uncertainties with 10-year data using 100
independent data sets. 
The expected 90\% C. L. ranges of $|\Delta m^2_{32}|$ with systematic uncertainties and errors in other oscillation parameters are summarized 
in Table~\ref{tab:result:summary}. 

We also show the expected results for 90\% C.L. ranges for
$|\Delta m^2_{32}|$, obtained with half the data (5 years data,
or an exposure of 250 kt$\cdot$yr), in Appendix~\ref{app:5yr}.
This shows that preliminary results with the oscillation dip and oscillation valley
analyses may already start appearing within the first few years of
ICAL. Note that the aim of this exercise is not to compare the precisions
achieved by different methods -- indeed, this study does not aim to
compete with the conventional $\chi^2$ method for precision.
Our aim is to present an approach (``oscillation dip'') that has
proved useful in the past for establishing the oscillation hypothesis
and rejecting some alternative hypotheses, and to go one step
beyond it, by performing the ``oscillation valley'' analysis in
the two-dimensional $(E_\mu^{\rm rec}, \, \cos \theta_\mu^{\rm rec}$) plane.
An important feature of our analysis is that the reconstruction of dip and valley and their identification processes are data-driven. Of course, to retrieve the information on oscillation parameters using the calibration curve, we use MC simulations.

Note that the precisions on $|\Delta m^2_{32}|$, obtainable from our dip and valley analyses, are very similar to each other. This is expected, since the L/E dependence is assumed and the two-dimensional analysis offers no advantages for the determination of $|\Delta m^2_{32}|$ in the context of the SM.  However, our focus has been on confirming that such two-dimensional reconstruction of the valley survives in the detector environment, \ie, the ICAL characteristics are sufficient to reconstruct the valley feature. This would also act as the fidelity test for the detector. The power of the valley analysis would become evident when L/E dependence is not assumed, where it would open up avenues for testing the
oscillation framework from more angles -- for example, by looking
for effects of new physics on the shape, alignment, width, or
depth of the valley. In the context of non-standard interactions (NSI) of neutrinos, this has been addressed in~\cite{kumar2021new}.

Though we have presented our results for ICAL, the same procedure can be
used for any present or upcoming atmospheric neutrino experiment
that has access to a large range of energies and baselines.
An atmospheric neutrino experiment can perform the analysis based on the oscillation dip as well as the oscillation valley. Of course, if the detector cannot identify the muon charge, the data from neutrino
and antineutrino channels will have to be combined, which may
dilute some of the observable features. However, the principle
of performing the oscillation valley analysis in the
$(E_\mu^{\rm rec}, \, \cos \theta_\mu^{\rm rec}$) plane -- going one step
ahead of the $L_\mu^{\rm rec}/E_\mu^{\rm rec}$ analysis -- may be adopted
in any atmospheric neutrino experiment. 
Note that while the fixed-baseline experiments may in principle be able
to identify the $L_\nu/E_\nu$ oscillation dip, they do not have access to the
$(E_\nu, \, \cos \theta_\nu$) plane, and hence they cannot perform the oscillation valley analysis.

Even then, certain features of ICAL make it uniquely suited for
such an analysis. It will likely be the first detector to identify
the oscillation dip separately for the neutrino and antineutrino channels,
which will also be a test for new physics that may affect the neutrino and antineutrino differently.  
Further, as far as the oscillation valley analysis goes,
a crucial requirement for the capability to identify the alignment
of the valley is enough number of points to determine
the best-fit line as shown in the right panel of
Fig.~\ref{fig:2D-fit-procedure}. This needs a large energy range to
which the detector is sensitive (large number of y-bins),
and an excellent angular resolution (large number of x-bins).
ICAL is thus uniquely suited for the oscillation valley analysis,
for providing an orthogonal approach to establish the nature of neutrino
oscillations, and hence for making the neutrino oscillation picture
more robust.

\section{Acknowledgements}

This work is performed by the members of the INO-ICAL collaboration 
to suggest a new approach to establish the oscillation hypothesis using 
atmospheric neutrino experiments. We thank Vivek Datar, S. Uma Sankar, 
and Srubabati Goswami for their useful and constructive comments on our work. 
We would like to thank the Department of Atomic Energy (DAE), Govt. of India for 
financial support. S.K.A. is supported by the DST/INSPIRE Research Grant 
[IFA-PH-12] from the Department of Science and Technology (DST), India, and 
the Young Scientist Project [INSA/SP/YSP/144/2017/1578] from the 
Indian National Science Academy (INSA). S.K.A. acknowledges 
the financial support from the Swarnajayanti Fellowship Research Grant 
(No. DST/SJF/PSA-05/2019-20) provided by the DST, Govt. of India and 
the Research Grant (File no. SB/SJF/2020-21/21) provided by the 
Science and Engineering Research Board (SERB) under the 
Swarnajayanti Fellowship by the DST, Govt. of India. A.D. acknowledges 
partial support from the European Union's Horizon 2020 research and 
innovation programme under the Marie-Sklodowska-Curie grant agreement 
Nos. 674896 and 690575.

\begin{appendix}

\section{Oscillation dip and valley using 5-year simulated data}
\label{app:5yr}

\begin{figure}[htb!]
	\centering
	\includegraphics[width=0.49\textwidth]{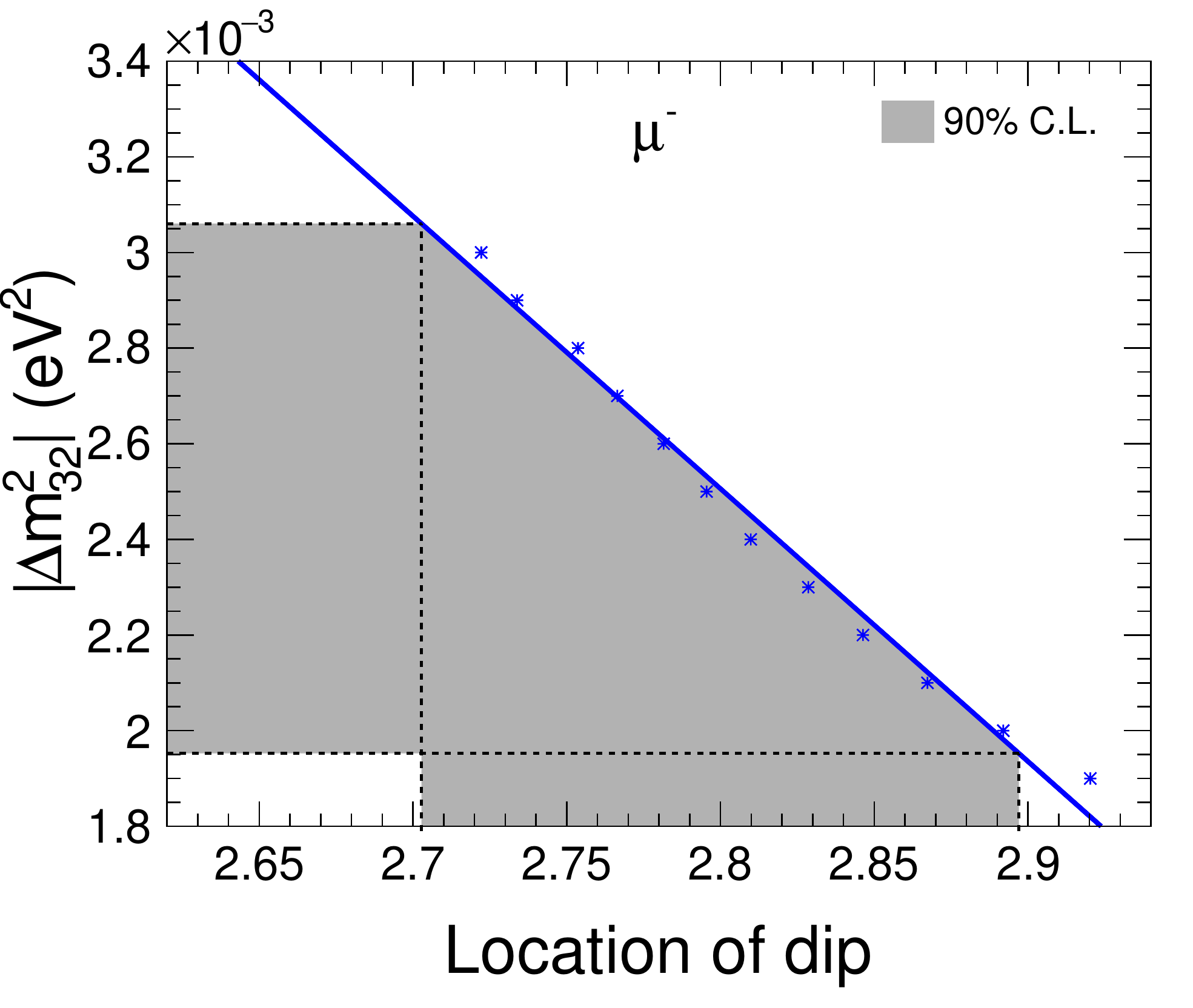}
	\includegraphics[width=0.49\textwidth]{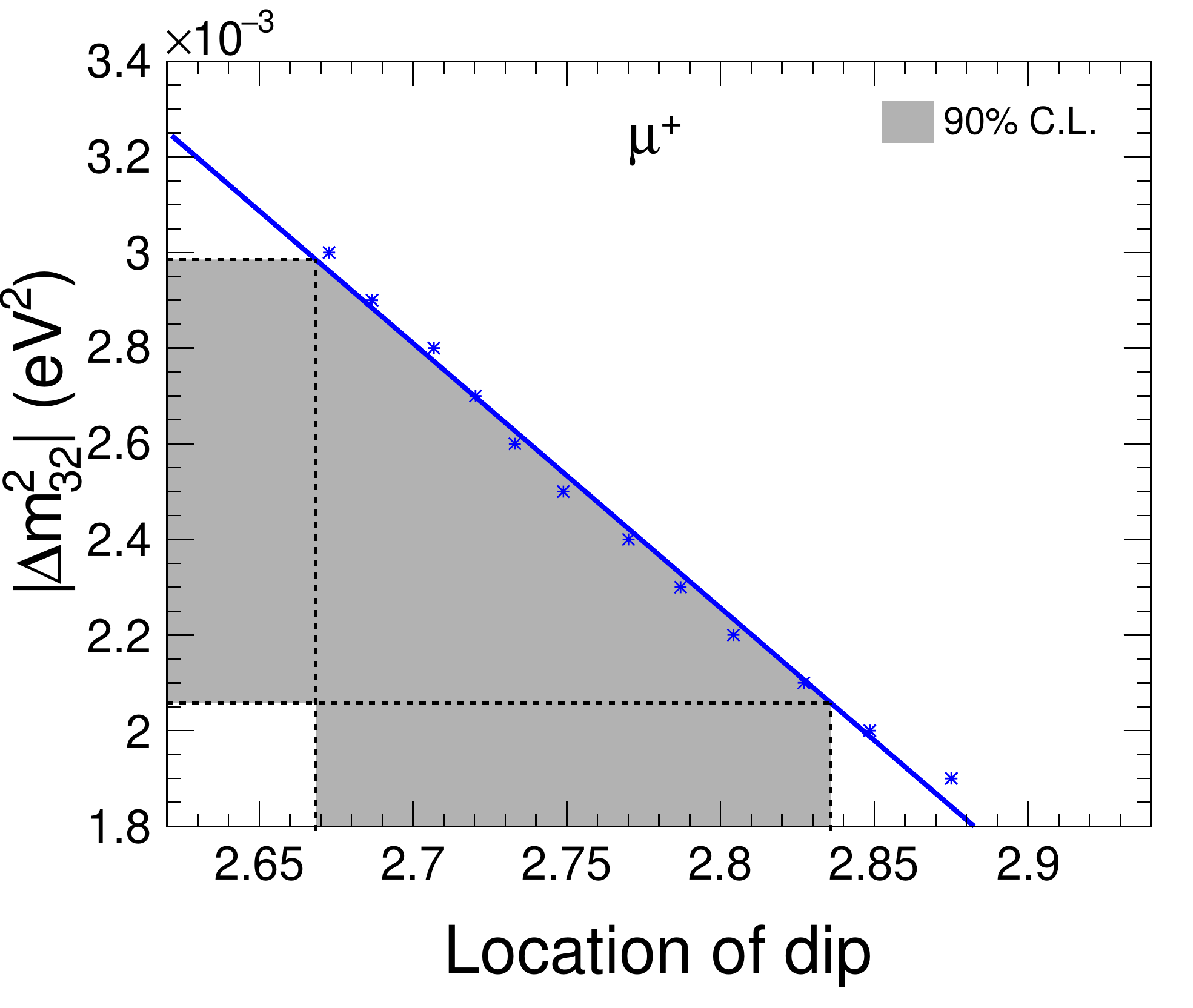}
	\includegraphics[width=0.49\textwidth]{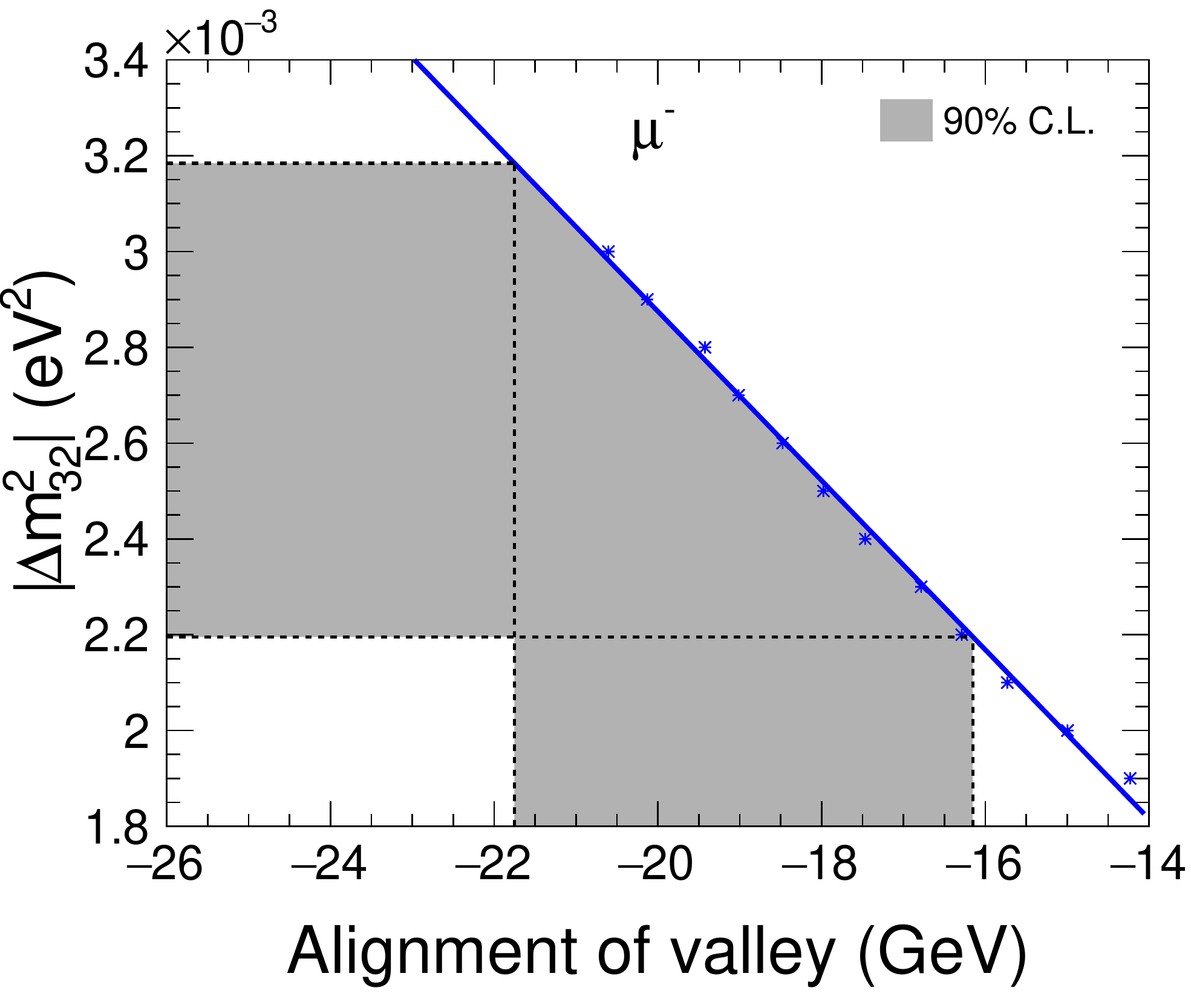}
	\includegraphics[width=0.49\textwidth]{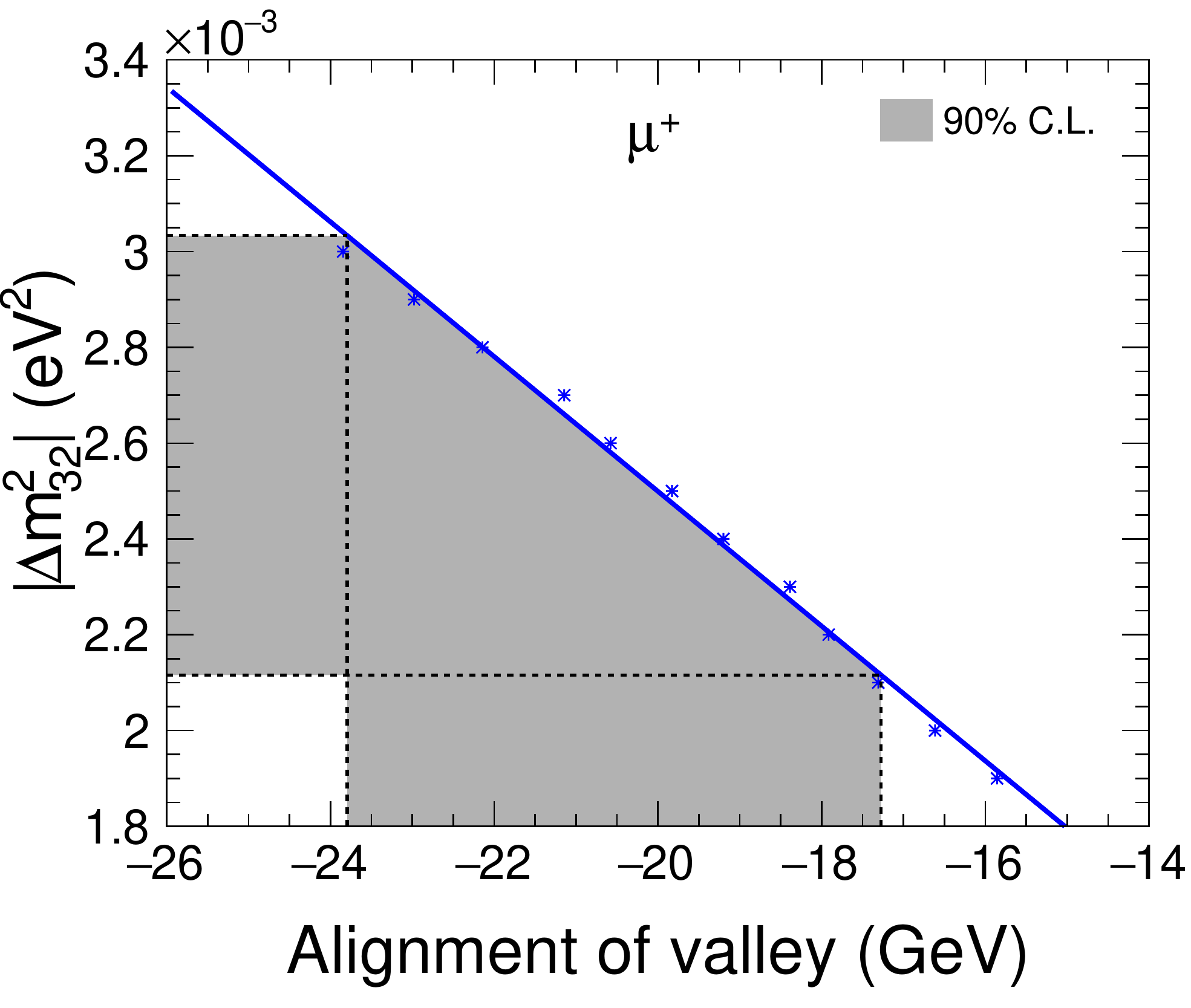}
	\mycaption{Upper panels: The blue points and the blue line correspond to
		the calibration of actual $|\Delta m^2_{32}|$ with the location of the dip,
		obtained by using the 1000-year MC data sample.
		The gray bands represent the 90\% C.L. regions of the location of the dip, and hence the inferred $|\Delta m^2_{32}|$ through
		calibration, for
		$|\Delta m^2_{32}| \,(\text {true}) = 2.46\times 10^{-3}$ eV$^2$,
		with 5 years of exposure.
		Lower panels: The same results, by using the alignment of the valley. The results obtained  from $\mu^-$ and $\mu^+$ events are shown in left and right panels of both the panels, respectively.
	}
	\label{fig:Calib-5yr}
\end{figure}

While in this work we have presented the expected results with
an exposure of 500 kt$\cdot$yr, \ie with 10 years running of the 50 kt
ICAL, it is worthwhile noticing that the identification of the
oscillation dip and the oscillation valley, as well as the
determination of the dip position and valley alignment, is possible
even with half the data size. In this appendix,  we show
in Fig.~\ref{fig:Calib-5yr} the
results of the calibration of $|\Delta m^2_{32}|$ with the dip position
and valley alignment, with 250 kt$\cdot$yr exposure of ICAL. We have adopted the same analysis as used for 500 kt$\cdot$yr with the same binning schemes as given in Tables \ref{tab:binning-1D-10years} and \ref{tab:binning-2D-10years}.
These could be some of the earliest significant neutrino oscillation 
  results from ICAL.

\end{appendix}

\bibliographystyle{JHEP}
\bibliography{ICAL-Oscillation-References}

\end{document}